# Physics of SrTiO$_3$-Based Heterostructures and Nanostructures: A Review

Yun-Yi Pai[1,2], Anthony Tylan-Tyler[1,2], Patrick Irvin[1,2] and Jeremy Levy[1,2]

[1]Department of Physics and Astronomy, University of Pittsburgh, Pittsburgh, PA  15260 USA

[2]Pittsburgh Quantum Institute, Pittsburgh, PA  15260 USA

# 1  Overview

## 1.1  Introduction

This review provides a summary of the rich physics expressed within SrTiO$_3$-based heterostructures and nanostructures.  The intended audience is researchers who are working in the field of oxides, but also those with different backgrounds (e.g., semiconductor nanostructures).

Strontium titanate-based heterostructures and nanostructures intersect two major areas in condensed matter and materials physics—the rich field of complex oxides and the physics of semiconductor interfaces and nanostructures.  The initial goal was to extend techniques of material growth with unit-cell precision, through advanced thin film techniques, to the relatively unexplored realm of complex oxides. Methods for creating high-mobility compound semiconductors were largely developed in the 1970s, and refinements resulted in major advances such as the discovery of the integer [1] and fractional [2] quantum Hall effects.  As noted in the perspective by Heber [3], the extension of these thin-film growth techniques to complex oxides represented a major new frontier.  Harold Hwang, working at Bell Laboratories and managed by Horst Stormer, undertook a program to create atomically precise oxide heterostructures, with the hope that new physics would emerge from these interfaces.

Hwang was not disappointed.  A milestone in the field was the demonstration of LaTiO$_3$/SrTiO$_3$ heterostructures in which charge transfer was observed to occur from single atomic planes.  The annular dark field images, produced by David Muller, were essential to providing convincing proof that these structures could be produced with unit-cell precision.  Also essential to the advances of SrTiO$_3$-based heterostructure growth was the development of techniques capable of controlling the surface termination. In particular, as will be discussed below, TiO$_2$-terminated SrTiO$_3$ is particularly desirable for conductive interfaces.  This control was first established in 1994 by Masashi Kawasaki and collaborators [4]. The development of new growth techniques—most notably, the invention of laser ablation or pulsed-laser deposition (PLD), by T. "Venky" Venkatesan—enabled growth of a wide variety of oxide thin films derived from bulk precursors.  *In situ* techniques used to monitor the growth of oxide thin films—most notably RHEED—were adapted by Guus J. H. M. Rijnders and collaborators [5] to the high-pressure growth environment.

In 2004, Ohtomo and Hwang reported a high-mobility electron gas at the interface between SrTiO$_3$ and LaAlO$_3$ [6].  Both materials are band insulators (unlike LaTiO$_3$ which is a Mott insulator), and LaAlO$_3$ has a relatively large bandgap (5.6 eV).  The existence of a conductive interface, with a mobility exceeding $10^4$ cm$^2$/Vs, was remarkable. The conductivity of the interface was found to depend crucially on the SrTiO$_3$ termination: for TiO$_2$-terminated SrTiO$_3$, the interface was conductive, while for SrO-terminated SrTiO$_3$, the interface was insulating. Quantum oscillations were also reported for this interface by a number of groups [7-14], another signature of high-mobility quantum transport.

The 2004 paper by Ohtomo and Hwang is a landmark because it had delivered on the promise of novel emergent behavior stemming from an oxide heterointerface. And yet (as with all significant advances), the work also raised many questions about the origin of the conductance, whether it was truly two-dimensional, and what other emergent properties would be found. Questions about the role of oxygen vacancies, interdiffusion, stoichiometry, ferroelastic and polar degrees of freedom have played a large role in subsequent unfolding of the literature. But it is worth appreciating the intellectual contributions of pioneering research that is driven by a far-reaching vision that is still playing out today.

### 1.1.1 Oxide growth techniques are rooted in search for high-$T_c$ superconductors

The community of researchers who made early contributions to the field of oxide heterostructures came in large part from research in high-temperature superconductors. These materials--all of them oxides and many of them with perovskite structures--have the largest superconducting transition temperatures known. SrTiO$_3$ is a "superconducting semiconductor", i.e., it becomes superconducting at exceptionally low electron densities (~$10^{17}$ cm$^{-3}$). The fact that it becomes superconducting at all is truly remarkable, and the existence of superconductivity in SrTiO$_3$ has fascinated researchers ever since it was first reported in 1964 by Schooley et al [15]. The 1987 Nobel Prize lecture, co-authored by Bednorz and Muller, describes the early search for novel superconducting compounds within the context of SrTiO$_3$ and LaAlO$_3$: *"The key material, pure SrTiO$_3$, could even be turned into a superconductor if it were reduced, i.e., if oxygen were partially removed from its lattice… The transition temperature of 0.3 K, however, was too low to create large excitement in the world of superconductivity research. Nevertheless, it was interesting that superconductivity occurred at all, because the carrier densities were so low compared to superconducting NbO, which has carrier densities like a normal metal."* [16].

### 1.1.2 First reports of interface conductivity

The early years following the report of Ohtomo and Hwang (2004) [6] were dominated by discussion and some dissention about the nature and even existence of 2D transport. Many groups emphasized the importance (and perhaps dominance) of oxygen vacancies. The existence of such vacancies was in fact clear from the very high electron densities reported, higher than 0.5 electrons/unit cell predicted by the polar discontinuity at the interface [6]. Additionally, the quantum oscillations did not show the correct angular dependence expected for a clean 2D semiconductor interface.

It was not until 2006 that experiments clearly began to show evidence for quasi-2D interfacial conductivity. A sharp insulator-to-metal transition was found for LaAlO$_3$/SrTiO$_3$ films grown by pulsed laser deposition and subsequently annealed [17], establishing a length scale (4 unit cells or 1.6 nm) that (at room temperature) appeared to characterize the depth of the electron liquid. Growth conditions were established in which bulk oxygen vacancies produced during growth were minimized by thermal annealing in oxygen-rich environment [17]. Scanning cross-sectional conductive-AFM measurements [18] (Figure 1.2) clearly showed that annealing produced an insulating SrTiO$_3$ substrate, leaving only the interface conductive. Nanostructures created by conductive-AFM lithography [19] provided a strong indication that the LaAlO$_3$/SrTiO$_3$ interface conductivity was localized well within 2 nm.

## 1.2 2D physics

The 2D physics explored at oxide interfaces coexists with a large body of research in conventional compound semiconductor heterostructures (see [20] for a recent review), binary oxide heterostructures

(e.g., ZnO/(Zn,Mg)O) [21-23], and graphene [24-26] and related Van der Waals heterostructures and devices ([27] for recent review). Given the early controversies surrounding the existence of 2D transport at the LaAlO$_3$/SrTiO$_3$ interface, it is important to keep in mind the accepted definition of a 2D material (from semiconductor heterostructures) [28]. All materials in nature exist in three dimensions. The term "2D heterostructure" is commonly used to describe structures in which the envelope wavefunction for electrons (or holes), when separated into a wavefunction along the growth ($z$) direction and in-plane ($x$-$y$) directions, exists in the lowest (non-degenerate) energy z-subband [28]. These envelope wavefunctions can extend many lattice constants away from the heterostructure interface.

## 1.3 Emergent and inherited properties of oxide heterostructures and nanostructures

One recurring theme of oxide heterostructures is that new and "emergent" properties may exist. The most celebrated example is the emergent conductivity at the interface between two insulators (SrTiO$_3$ and LaAlO$_3$). Emergent magnetism has also been reported for the same interface, when neither SrTiO$_3$ nor LaAlO$_3$ are separately magnetic [29]. Other properties are perhaps better characterized in terms of inheritance. Interfacial superconductivity [30,31] is an inherited property, since bulk n-type SrTiO$_3$ is a well-known superconducting semiconductor [15]. The properties of the low-dimensional superconducting state are unusual and many of them may be related to the carrier confinement. Similarly, new effects emerge when the dimensionality is further reduced. In one-dimensional (1D) nanostructures, it has been established from a number of measurements that mobility is greatly enhanced, and is ballistic at low temperatures over many micrometers [32,33].

## 1.4 Outline

In this review, we present an overview of what is known about the physics of SrTiO$_3$-based heterostructures. After reviewing the relevant properties of SrTiO$_3$ itself, we will then discuss the basics of SrTiO$_3$-based heterostructures, how they can be grown, and how devices are typically fabricated. Next, we will cover the physics of these heterostructures, including their phase diagram and coupling between various degrees of freedom. We then cover the rich landscape of quantum transport phenomena, as well as the devices that elicit them.

Before proceeding, we point the reader to several earlier reviews on oxide interfaces [3,34-48]. Bhattacharya et al. [46] review magnetic phenomena at oxide interfaces. There are several reviews that focus on the LaAlO$_3$/SrTiO$_3$ system [42,49-53], and there is a recent review by Pai et al. [54] on magnetism at the LaAlO$_3$/SrTiO$_3$ interface [54]. Pentcheva and Pickett [55], Chen, Kolpak and Ismail-Beigi [50] and Bjaalie et al. [45] have written reviews of this system from the perspective of first-principles theory. We would also like to refer readers to the reviews of related field such as Fiebig et al. [56] on multiferroic materials, Catalan et al. [57] on the domain walls of those multiferroic materials, Hellman et al. [58] on interface-related magnetic phenomena, Salje [59] on ferroelastic materials, and Salje et al. [60] on ferroelastic domain boundaries.

# 2 Relevant properties of SrTiO$_3$

In this section, we will briefly discuss the properties of SrTiO$_3$, as well as new findings on the interplay between the phases of bulk SrTiO$_3$ [61]. Reasearch into the properties of STO and related materials constitutes a large field, with a long history and depth far beyond the scope of this rather short review. We are only able to cover a subset of properties that we deem the most relevant for SrTiO$_3$-based oxide heterointerfaces. We refer readers to the renowned work by Lines and Glass [61], and recent work by

Tilley [62]. In addition, the recent reports on emergent magnetism found related to oxygen vacancies are discussed in Sec. 4.6.

## 2.1 Structural properties and transitions

At room temperature, SrTiO$_3$ is an idealized cubic structure (space group $Pm\bar{3}m$) with lattice parameter $a = 0.3905$ nm [62], as shown in Figure 2.1 a. Under stress or change of temperature, the SrTiO$_3$ lattice can deform from the ideal cubic structure. The most relevant distortions for SrTiO$_3$ are:

(1) Ferroelectric-type (FE) displacement: polar displacement between cations and anions, resulting in a built-in polarization (Figure 2.1 b)
(2) Antiferrodistortive (AFD) rotation: antiphase rotation of neighboring oxygen TiO$_6$ octahedra (Figure 2.1 c).

Generally, for each perovskite $ABO_3$, the Goldschmidt tolerance factor ($t = \frac{r_A+r_O}{\sqrt{2}(r_B+r_O)}$ [63], where $r_A$, $r_B$, and $r_O$ are the ionic radii of A-site cation, B-site cation, and oxygen, respectively) provides a reasonably accurate guideline in identifying whether FE displacement or AFD rotation is more likely to occur [64]. Perovskites with larger Goldschmidt tolerance factors ($t > 1$), such as $t = 1.063$ for BaTiO$_3$, favor FE displacement and are often found to be ferroelectric. On the other hand, perovskites with smaller tolerance factors ($t < 1$), such as $t = 0.946$ for CaTiO$_3$, favor AFD rotation. The tolerance factor for SrTiO$_3$ is found to be very close to unity ($t = 1.00$ [65-67]). It is not surprising that both FE and AFD are relevant in the phase diagram of SrTiO$_3$. However, whether they compete or cooperate with each other is not well established [68-70].

As temperature decreases, bulk SrTiO$_3$ undergoes a cubic-to-tetragonal, ferroelastic, AFD phase transition around $T_{AFD} = 100 - 110$ K [71-76] (Figure 2.1 d) (For ferroelectric perovskites such as PbTiO$_3$, the cubic-to-tetragonal structural transition is due to the FE displacement.) This AFD transition has a rotation angle (Glazer classification [77] $a^0a^0c^-$ ) α = 2.1°; the resulting distortion changes the $c/a$ ratio (1.00056) [72], reducing the symmetry of the crystal structure from cubic to tetragonal. The $R_{25}$ phonon condenses at 105 K on the Brillouin zone boundary [75]. Ferroelastic domains with different rotation axes appear below the AFD transition temperature [76]. These domains disappear as the sample is warmed above $T_{AFD}$ and typically form a different domain pattern upon temperature cycling [76]. The cubic-to-tetragonal transition can also be triggered by stress and pressure [78-86] and can be affected by both defects [87] and doping [70]. Inhomogeneous transport associated with ferroelastic domains is discussed in Sec. 5.2.

## 2.2 Ferroelectricity, Paraelectricity, and Quantum Paraelectricity

SrTiO$_3$ is one of the few known systems to exhibit *quantum paraelectricity* [88,89]—a phase in which quantum fluctuations between degenerate lower symmetry configurations suppress ferroelectric ordering. At room temperature, the static dielectric constant of SrTiO$_3$ is already large ($\varepsilon_r \sim 300$). As the temperature decreases, $\varepsilon_r$ increases, saturating at $T \sim 4$ K, and can be as large as $\varepsilon_r \sim 30{,}000$, depending on the strain, field, and sample quality. The polar transverse optical soft mode for SrTiO$_3$ scales approximately as $\omega_{TO}^2 \propto (T - T_c)$ for $T > T_c$, with $T_c = 35$ K [88,89]. For a regular ferroelectric material, the phonon frequency intercepts zero at $T_c$ and becomes imaginary [90]. However, for SrTiO$_3$, as $T$ approaches $T_c$ from above, $\omega_{TO}^2$ starts to deviate from $(T - T_c)$ and eventually levels off at low temperatures [91].

While the bulk of SrTiO₃ is paraelectric, the surface can be ferroelectric. Using low-energy electron diffraction (LEED), Bickel et al. [92] found that the surface layer exhibits significant puckering at $T = 120$ K (oxygen ions extending away from the surface). Later theoretical work by Ravikumar et al. [93] suggests that the surface can be ferroelectric if it is SrO-terminated (but not if it is TiO₂-terminated). The structure of SrTiO₃ in the polar ferroelectric phase is in fact still under debate [94]. Bussmann-Holder et al. [95] argue that the ferroelectric phase does not possess true long-range order, and instead consists of coexisting ordered-disordered micrograins. Besides bulk ferroelectricity, ferroelectricity in SrTiO₃ films [96-99], and flexoelectricity [100] have also been reported.

There are several experimental parameters that are known to drive SrTiO₃ from a paraelectric to a ferroelectric phase (or otherwise impact the ferroelectric transition):

(1) *Electric field*. By applying an electric field greater than $2 \times 10^5$ V/m, SrTiO₃ starts to become ferroelectric at $T \sim 20$ K [101]. The dielectric constant $\varepsilon_r$ decreases rapidly with an applied electric field [102]. The electric field also induces Raman-active odd parity phonons [103,104].

(2) *Stress or strain*. The ferroelectric phase can be triggered by stress on the order of 0.5 GPa at $T = 2$ K [105,106]. The stress forces the domains with polarization parallel to the stress direction to switch to the perpendicular direction. Notably, the sample must be warmed above $T_{AFD} \sim 110$ K to remove the remnant ferroelectric polarization. This observation indicates a possible interplay between the AFD transition and the ferroelectric phase. Thanks to the advances of thin-film growth technologies and development of suitable substrates [107], SrTiO₃ films can be grown with substrate-derived strain. Room-temperature ferroelectricity has been reported for SrTiO₃ films grown on lattice-mismatched DyScO₃ [97], and later at even higher temperature $T = 400$ K for SrTiO₃ films grown on silicon [108].

(3) *Cation substitution*. Substituting the Sr atoms with other cations such as Ca, Pb, or Ba can induce ferroelectricity. The critical substitution fraction $x_c$ is different for each material: $x_c = 0.0018$ for $Sr_{1-x}Ca_xTiO_3$ [109,110], $x_c = 0.002$ for $Sr_{1-x}Pb_xTiO_3$ [111], $x_c = 0.035$ for $Sr_{1-x}Ba_xTiO_3$ [112].

(4) *Isotope substitution.* Itoh et al. [113] reported that, by substituting ¹⁶O with heavier ¹⁸O, at $x_c = 0.33$ [114], SrTiO₃ becomes ferroelectric at $T_c = 23$ K. This is because the heavier ¹⁸O atoms suppress the quantum fluctuations. Stucky et al. [115] found that this isotope substitution raises the superconducting transition temperature of SrTiO₃, indicating that the superconductivity and the ferroelectric phase of SrTiO₃ may be related (See Sec. 2.5).

(5) *Defects and defect centers*. Several reports suggest Sr-vacancy-related defects or oxygen vacancies may be responsible for observed ferroelectricity in SrTiO₃ films, for example, Sr-O-O [116], $V_{Sr}$-$I_{Ti}$ (Sr-vacancy adjacent to interstitial Ti) [117], antisite defects such as $Ti_{Sr}$ and $Sr_{Ti}$ [118], or oxygen vacancies [119]. However, the detailed mechanism for the defect-driven ferroelectricity is not well understood. Scenarios such as defect-increased tetragonality [119,120], and the electric dipole of $V_{Sr}$-related defect centers, have been considered [116,117].

## 2.3  Electronic structure

SrTiO₃ is a band insulator, and can be *n*-type doped with Nb, La, or oxygen vacancies. It has an indirect bandgap 3.25 eV and a direct bandgap 3.75 eV, experimentally determined from valence electron-energy loss spectroscopy (VEELS) [121]. The valence band is mainly composed of the $2p$ orbitals of oxygen atoms, while the conduction band is mainly composed of the Ti $3d$ orbitals [122]. In the cubic phase, the $3d$ orbitals split into $t_{2g}$ and $e_g$, as shown in Figure 2.1 e [123]. Figure 2.1 f [124] shows a typical DFT calculation of the band structure. The $d_{xy}$, $d_{yz}$, and $d_{xz}$ orbitals of the $t_{2g}$ bands are degenerate at the conduction band minimum, the $\Gamma$ point, as shown in Figure 2.1 f [124]. The indirect band gap and the direct band gap observed in VEELS are assigned to $R \to \Gamma$ and $\Gamma \to \Gamma$, respectively [121]. In the tetragonal phase, the primitive cell is doubled, and the folding of the first Brillouin zone maps both the $R$ and $\Gamma$ of the cubic structure into $\Gamma$ [125]. Stress and strain can also modify the band structure. Refer to the ARPES work by Chang et al. [126], or theoretical calculations by Khaber et al. [127], for examples.

The $t_{2g}$ manifold is expected to further split due to the AFD transition and atomic spin-orbit coupling (SOC). The expected energy scale is from a few meV to tens of meV. However, since the early characterization by Uwe et al. [128,129] with Raman scattering and Shubnikov-de Haas (SdH) oscillations, to later by, for example, Chang et al. [130] with ARPES, and theoretical calculations such as Marques et al. [131], no consensus about how large the splitting should be has been reached.

The details of the band structure are still difficult to obtain experimentally. The critical features with a size around a few meV, such as the aforementioned spin-orbit splitting or tetragonal splitting, are too small for ARPES to resolve. Magnetotransport measurements yield widely varying effective masses [124,129,132] and do not agree well with DFT calculations [133-135] (We present a more comprehensive discussion on the SdH for LaAlO$_3$/SrTiO$_3$ and related 2DES in Sec. 5.6). Standard DFT tends to overestimate the $c/a$ ratio and the AFD rotation angle, while underestimating the bandgap [125]. These problems have been taken care of in the modern hybrid DFT calculations such as LDA+U, GW, etc., but which method is most reliable for the rest of the band structure is an open question. For a more extensive discussion see, for example, Evarestov [136].

## 2.4 Defects

SrTiO$_3$ defects greatly affect the structural, electronic, magnetic, and optical properties. For example, Sr-vacancies and related defect centers may be responsible for increased tetragonality [120] and emergent ferroelectricity in SrTiO$_3$ films [116-118,137]. Oxygen vacancies are the most abundant and important species of point defects in SrTiO$_3$. They are easy to introduce, remove and manipulate, and have wide influence on the properties. For instance, they can provide electron conduction, blue light emission, and dilute magnetism. In this section, we will mostly focus on oxygen vacancies, and will briefly discuss other defect complexes, e.g., terraces. Readers are referred to the recent review by Marshall et al. [138] for STM studies of the surface defects especially nanolines, the work by Szot et al. [139] for dislocations, and the tutorial by Sanchez et al. [140] for defects related to treated surfaces such as vacancy puddles and meandering terraces.

### 2.4.1 Oxygen vacancies

Oxygen vacancies (V$_O$, or $V_O^{\cdot\cdot}$ using Kröger–Vink notation [141] for doubly ionized oxygen vacancies) are the most abundant and investigated point defects in SrTiO$_3$ [138]. Oxygen vacancies can be introduced by annealing SrTiO$_3$ under low oxygen partial pressure, which is one of the primary ways of creating $n$-type SrTiO$_3$. Oxygen vacancies can also be introduced into PLD-grown thin films when growing under lower oxygen partial pressure [142]. The distribution and migration of the isolated oxygen vacancies is typically analyzed using a thermally-activated ionic diffusion model [143,144]. The diffusion constant, ranging from $\sim 10^{-14}$ cm$^2$s$^{-1}$ at $T = 295$ K to $\sim 10^{-4}$ cm$^2$s$^{-1}$ at $T = 1,600$ K [143], is a function of many variables including temperature, oxygen partial pressure, hopping paths (which are found to be dominated by the edges of the TiO$_6$ octahedra, see [145,146]), electric field, and light [147,148]. The distribution of oxygen vacancies is not always uniform [149,150]. Beyond the single vacancy diffusion model, oxygen vacancies may cluster [142], such as the apical-divacancy configuration predicted by Cuong et al.[145]. They may also cluster around dislocation cores [151], and have higher ionic mobility around dislocations

[152], twin walls (predicted for CaTiO$_3$ [153,154]). They may also perturb the crystal structure [155] and induce AFD rotation [156]. They have been implicated in the large thermoelectric power observed for SrTiO$_3$ [157].

Each oxygen vacancy nominally dopes the sample with two electrons, $2e^-$. Those electrons may remain localized and may not be necessarily conductive [158]. Some carriers freeze-out and become localized at low temperature, which was observed for the 2DES at the LaAlO$_3$/SrTiO$_3$ interface [150] as well as oxygen vacancy-doped SrTiO$_3$ [159]. The wavefunction spread for the vacancy-derived electrons are expected to depend on the vacancy configuration [160] and the overall structure (e.g., cubic vs. tetragonal) [155]. There may also be subtler differences between Nb-doped SrTiO$_3$ and oxygen-reduced SrTiO$_3$ (See Sec. 2.5).

Oxygen vacancies cause dramatic changes in the optical properties. Upon introduction of oxygen vacancies, the color of SrTiO$_3$ changes from transparent to blue [120,161]. Upon the application of an electric field at elevated temperature, the oxygen vacancies migrate, resulting in a gradient of color [161]. Photoluminescence peaks at 430 nm are observed for oxygen deficient SrTiO$_3$ samples, regardless of how the oxygen vacancies are introduced [162]. Upon cooling, the blue (430 nm) peak shifts toward green (550 nm). At $T = 160$ K, a new peak shows up at 380 nm. After the introduction of oxygen vacancies, SrTiO$_3$ can be highly photoconductive. By exciting with sub-bandgap light (2.9 eV), an increase in the conductivity by 2 orders of magnitude is observed that lasts for days [163]. The reversible switching controlled by UV light and water immersion for LaAlO$_3$/SrTiO$_3$ has also been reported [164,165].

### 2.4.2 Terraces

"Terraces" are unit-cell-sized (3.904 Å) steps that appear due to miscut angles of the SrTiO$_3$ crystal. The width of terraces can vary tremendously. Typical terraces widths of nominally (100) SrTiO$_3$ range from around 70 nm − 500 nm, corresponding to miscut angles α = 0.05° − 0.3°. Typically, terraces are not visible (under atomic force microscope) for samples with mixed termination. Surface treatments that produce TiO$_2$ termination (Ref. [4] and Sec. 3.2.2) yield clean and flat terraces; such methods are used nearly universally prior to the growth of epitaxial layers.

Currently, the impact of the terrace structure on 2D transport at SrTiO$_3$ interfaces is not clear. For the 2DES at the LaAlO$_3$/SrTiO$_3$ interface, two reports suggest that the terraces may enhance the mobility [166] or impede transport perpendicular to the terrace direction [167]. Fix et al. [166] compared samples with miscut angle α = 0.2° to those with α = 8° (denser terraces). Higher mobility with roughly a third of carrier concentration is observed for the α = 8° samples, at $T = 10$ K. Fix et al. indicate that the terraces may act as traps for low-mobility carriers [166]. On the other hand, Brinks et al. [167] reported anisotropic resistance for different transport directions. Brinks et al. suggested that the observed anisotropy is due to the difference in transport along or perpendicular to the terrace direction [167]. However, the impact of terraces has not been distinguished from other known sources of anisotropy, e.g., ferroelastic domains [168] (Sec. 5.2).

### 2.5 Superconductivity

In 1964, Schooley et al. [15] observed superconductivity in oxygen-reduced SrTiO$_3$ at $T_c \sim 250 -$ 280 mK. SrTiO$_3$ was the first complex oxide found to be superconducting, and the inspiration for Bednorz and Muller in the search for new superconducting compounds with high critical temperatures [16]. The critical temperature of the superconductivity as a function of carrier concentration traces out a "dome" peaked at $T_c \sim 450$ mK (Figure 2.1 g) [169,170]. This non-monotonic dependence on carrier concentration is familiar to cuprates [171], heavy-fermion systems [172], and iron-based superconductors [173].

SrTiO$_3$ is superconducting at very low carrier concentration, as low as $10^{15}$ cm$^{-3}$ [174-176]. This low carrier concentration implies that if it is indeed a Fermi liquid, it must have a Fermi energy that is low compared to phone energies. Superconductivity of SrTiO$_3$ therefore overlaps the nonadiabatic ($E_F < \hbar\omega_D$) and even anti-adiabatic regime ($E_F \ll \hbar\omega_D$), where the BCS-Midgal-Eliashberg theory is not applicable [177]. Lin et al. [178] argue that the insensitivity of $T_c$ and $H_{c2}$ to defect concentration implies s-wave pairing, as a consequence of Anderson's theorem [179].

There are many unanswered questions concerning the nature and pairing mechanism governing superconductivity in SrTiO$_3$. Here we discuss the most important outstanding questions (in no particular order).

*Is SrTiO$_3$ a multigap superconductor?* Early tunneling spectroscopy measurements by Binnig et al. [180] revealed two distinct superconducting gaps in the Nb-doped SrTiO$_3$, suggesting that the superconductivity is supported by at least two bands. The double gaps were not observed in the oxygen-deficient samples of the same work [180]. However, no follow-up experiments reported a second superconducting gap in either SrTiO$_3$ [181] or SrTiO$_3$-based 2DES [182], nor did a recent scanning tunneling microscopy experiment on Nb-doped SrTiO$_3$ by Ha et al. [183].

*The nature of the electron pairing* in SrTiO$_3$ has been pondered since the discovery of superconductivity in SrTiO$_3$ in 1964 and has remained "enigmatic for half a century" [177]. A paired, non-superconducting phase was first proposed by Eagles in 1969 [184] and subsequently used to explain [185] the two-step resistance drop of the superconducting transition reported by Tainsh et al. [186] in Zn-doped SrTiO$_3$. With the development of SrTiO$_3$-based heterostructures, questions about the nature of the superconducting gap and pairing have been revisited. A pseudogap signature was reported by Richter et al. by planar tunneling measurements on top-gated LaAlO$_3$/SrTiO$_3$ [182]. Cheng et al. used a single-electron transistor—fabricated within the LaAlO$_3$/SrTiO$_3$ interface, to show that electrons are paired far outside the superconducting regime [187] (See Sec. 6.5).

Conventional BCS pairing is built on attractive electron-electron interactions that are usually mediated by electron-phonon coupling. Experimentally, strong electron-photon coupling (described as polarons) has been observed in ARPES [188,189] and tunneling spectroscopy [181,190]. Those polarons are assigned to the LO4, LO3, and LO2 modes [181,190], with the largest energy scale being $\hbar\omega_{LO4} \sim 90 - 100\ meV$, depending on the techniques employed [181,188-190] and no difference between bulk or 2DES is found. The coupling strength decreases with increased carrier concentration [181,189]. The possibility of those LO modes giving rise to pairing has been considered [177,191], despite the questionable applicability of Eliashberg theory. (See [177] for the treatment and [192] for the follow up discussions.) Besides the aforementioned LO modes, other alternatives have been considered. Candidates range from the soft phonons of the AFD modes [193], soft phonons of the FE modes [194], plasmons with polar optical phonons [195], plasmons [192], and cooperative Jahn–Teller distortions [196].

A relatively new picture links superconductivity to the ferroelectric quantum critical point [197], analogous to the description of superconductivity in cuprates driven by magnetic quantum critical point for the cuprates. However, the link between superconductivity and ferroelectricity is not clear. It was reported by Itoh et al. [113] that the isotope $^{16}$O to $^{18}$O substitution can induce ferroelectricity in SrTiO$_3$. Recently, Stucky et al. [115] reported that the same isotope substitution raised $T_c$ by 50% as well as the upper critical field $H_{c2}$ by a factor of two. However, though Ca-substitution is also known to be able to induce ferroelectricity [109,110], de Lima et al. reported weakened superconductivity upon Ca-substitution [70].

# 3 SrTiO$_3$-based heterostructures and nanostructures

## 3.1 SrTiO$_3$-based two-dimensional electron systems

The first two-dimensional electron system (2DES) was formed by trapping electrons on the surface of liquid helium [198,199]. In solid state systems, the development of silicon MOSFETs, and modulation-doped III-V semiconductors have also lead to fundamental physics discoveries such as the integer [200] and fractional quantum Hall effect [201]. The profound discoveries and important device applications of 2DES formed from semiconductors have motivated research into the properties of SrTiO$_3$-based 2DES. They can accumulate at surfaces, or interfaces. While most research has focused on the LaAlO$_3$/SrTiO$_3$ interface [6], there are many other SrTiO$_3$-based heterostructures. Yang et al. have calculated band offsets for 42 candidate polar perovskite compounds at the interface of SrTiO$_3$ [202]. In this section, we summarize common features and highlight differences among the various predictions and measurements.

### 3.1.1 SrTiO$_3$ surface

A 2DES can be formed on the surface of SrTiO$_3$ by cleaving it in vacuum. Santander-Syro et al. [203] and Meevasana et al. [204] reported electronic structure of SrTiO3 crystals that were cleaved in vacuum at 20 K or below. By studying the surface using angle-resolved photoemission spectroscopy (ARPES), they found universal dispersive bands for undoped $n_{3D} < 10^{13}$ cm$^{-3}$ through the highly doped ($n_{3D} = 10^{20}$ cm$^{-3}$) substrates. Meevasana et al. only observed the formation of a 2DES after exposing the sample to intense ultraviolet radiation. The samples studied by Santander-Syro et al. and Meevasana et al. were mixed SrO- and TiO$_2$-terminated. Di Capua et al. [205] prepared 2DESs on the surface of TiO$_2$-terminated SrTiO$_3$ by initial etching in a hydrofluoric acid solution followed by oxygen annealing. That the surface is conductive is confirmed by scanning tunneling microscopy (STM).

Barquist et al. [206] also prepared TiO$_2$-terminated SrTiO$_3$ and measured metallic behavior down to 50 mK. Sarkar et al. prepared anatase and rutile TiO$_2$/SrTiO$_3$ [207] and found conductivity differences for rutile and anatase TiO$_2$ compared with LaAlO$_3$, which they attributed to differences in bond angles between Ti and O.

A buried 2DES in SrTiO$_3$ was produced by Kozuka et al. [208] using a 5.5 nm thick $\delta$-doped region of Nb-doped SrTiO$_3$. Low-temperature transport showed a superconducting transition at $T_C = 370$ mK. The thickness inferred from in-plane upper critical field $H_{C2}^{\parallel}$ is $d_{Tinkham} \sim 8.4$ nm. Ionic gel liquid gated SrTiO$_3$ [209,210] has also resulted in the formation and control of 2DESs in SrTiO$_3$.

### 3.1.2 LaAlO$_3$/SrTiO$_3$

The 2DES formed at the LaAlO$_3$/SrTiO$_3$ interface, first reported by Ohtomo et al. [6], is the most extensively studied conductive SrTiO3-based heterointerface. LaAlO$_3$/SrTiO$_3$ has a critical thickness of 4 u.c. of crystalline LaAlO$_3$ [17] and the SrTiO$_3$ substrate must be TiO$_2$-terminated [6,211]. This interface also has been found to exhibit "emergent" behavior such as an electric gate tunable metal insulator transition for critical thickness ($\sim$ 3 u.c.) [17,19] and magnetism [29,212,213] (Sec. 4.6). The interface is superconducting [30] and controllable with an electric field [31] (Sec. 4.7). Magnetism has also been reported at this interface [29,212,213] (Sec. 4.6).

The LaAlO$_3$ layer need not be (100) oriented or even crystalline. Chen et al. [214] found that amorphous-LaAlO$_3$/SrTiO$_3$ can produce a conducting interface. The conducting interface was attributed to oxygen vacancies [214]. A conductive interface on (110) and (111) SrTiO$_3$ was observed [215,216]; both interfaces are non-polar. Even "conventional" LaAlO$_3$/SrTiO$_3$ interfaces can be enhanced with a SrCuO$_3$ overlayer to promote oxygen exchange with the surface, leading to a reduced number of oxygen vacancies, which is correlated with higher-mobility transport [217].

### 3.1.3 SrTiO$_3$ based heterostructures and quantum wells

In addition to bare SrTiO$_3$ and LaAlO$_3$/SrTiO$_3$, a variety of other heterostructures have conductive interfaces [218-228], although some are insulating [219,223,229-234]. Trilayer SrTiO$_3$/LaTiO$_3$/SrTiO$_3$ heterostructures were found to be conductive even to one unit cell of LaTiO$_3$ [235] and believed to support a 2DES and 2DHG on the LaAlO$_3$/SrTiO$_3$ and SrTiO$_3$/LaAlO$_3$ interface, respectively [236]. The GdTiO$_3$/SrTiO$_3$ quantum wells have been explored as candidate Mott insulators [237-242]. Others result in ultrahigh carrier density like NdTiO$_3$/SrTiO$_3$ [243]. A spinel/perovskite interface $\gamma$-Al$_2$O$_3$/SrTiO$_3$ [244,245] with extreme mobility enhancement (as high as 140,000 cm$^2$/Vs) has also been reported.

## 3.2 Thin-film growth techniques

A variety of techniques are used to grow SrTiO$_3$ and related heterostructures. These include atomic layer deposition (ALD), pulsed laser deposition (PLD), molecular beam epitaxy (MBE), laser MBE, hybrid MBE, and sputtering. After a brief discussion of substrate preparation and surface treatment, we will then describe the growth techniques to highlight the differences for the reader. Interested readers are referred to the reviews of Chambers [246], Demkov et al. [247], and Koster et al. [248] for more details.

### 3.2.1 Substrates

There are several methods for preparing SrTiO$_3$ substrates and targets, including the solid-state reaction [249], flame fusion [250], and flux methods. In the flame fusion method, also known as the Verneuil process, a fine powder of the starting material passes down a tube along with oxygen. To grow SrTiO$_3$ a mixture of highly purified SrCO$_3$ and TiO$_2$ powders are used. The falling powder is heated by combustion with hydrogen gas. The powder forms a "sand pile," which is mostly polycrystalline, but on the very tip of the pile, near the flame, a small single crystal grain will form. The single crystalline material, known as a boule, will begin to grow on top of the pile from the single seed grain. A disadvantage of this method is that the crystal grows in a large thermal gradient that can lead to stresses and dislocations in the crystal. Another consideration when growing SrTiO$_3$ is that SrO will evaporate at the temperatures needed to melt the SrTiO$_3$ (1920°C) so excess SrCO$_3$ must be added to the feed powder [250].

While the solid-state reaction method can be used to prepare substrates, it is more often used to create PLD targets. SrTiO$_3$ in powder form is obtained from the solid-state reaction of high purity SrO and TiO$_2$ powders that are mixed in proper molar ratios. The mixture is then annealed at high temperatures (above 1200°C) for 12-24 hours. This thermal annealing step initiates the solid state reaction process and results in the formation of stoichiometric SrTiO$_3$ powder with crystalline phase. X-ray diffraction is generally performed to verify the crystalline cubic phase of SrTiO$_3$. Using mechanical pressure, the SrTiO$_3$ powder is then pressed into desired shapes and sintered to form a solid and dense SrTiO$_3$ target.

SrTiO$_3$ is used extensively as a substrate for fabrication of various oxide materials such as high $T_C$ superconductors (Ex:YBCO) and magnetic oxides (manganites),and oxide interfaces such as

LaAlO$_3$/SrTiO$_3$. The near-perfect cubic structure and lattice matching with most perovskites and layered perovskites makes it a suitable substrate [140]. The clean surface and abrupt interface formed with SrTiO$_3$ play an important role in the rich phenomena observed at SrTiO$_3$-based oxide interfaces. For example, Using PLD, Thiel et al. [17], have demonstrated that with precise growth control over film thickness, a sharp insulator-metal transition near the LaAlO$_3$/SrTiO$_3$ interface can be achieved for 3-4 u.c of LaAlO$_3$ on SrTiO$_3$.

There have been a lot of effort to integrate SrTiO$_3$ with Si. The initial motivations were to find a high-permittivity replacement for a long time. Epitaxial growth was first reported by McKee et al. [251] at Oak Ridge National Laboratory and Hallmark et al. [252,253] at Motorola. The large strain on the SrTiO$_3$ film was associated with ferroelectricity at room temperature [108]. Park et al. [254] showed that LaAlO$_3$/SrTiO$_3$ could be successfully grown on (scalable) silicon substrates, yielding 2DES with comparable properties, including the ability to create nanostructures using conducting AFM lithography.

A second desirable substrate alternative, lanthanum aluminate – strontium aluminum tantalite, ((LaAlO$_3$)$_{0.3}$(Sr$_2$TaAlO$_6$)$_{0.7}$), also known as LSAT, was first developed to grow high $T_c$ superconductors. It has the advantage of being able to be grown by the Czochralski method, is twin-free and is available in much larger wafer sizes. LaAlO$_3$/SrTiO$_3$ 2DESs have been demonstrated on LSAT by Bark et al. [255].

The substrate may also be used to strain-engineer the heterostructures grown on top of it. Besides aforementioned SrTiO$_3$/Si, SrTiO$_3$/DyScO$_3$ also showed ferroelectricity at room temperature [97]. The effects of strain on the 2DES formed at the interface of LaAlO$_3$ and SrTiO$_3$ [255] were studied using NdGaO$_3$ (NGO), LSAT, DyScO$_3$ (DSO), and GdScO$_3$ substrates. These substrates were used to biaxially strain epitaxial SrTiO$_3$ between -1.21% to +1.59%. It was found that SrTiO$_3$ compressively strained more 1% resulted in insulating films. Compressively strained films were less conducting but had an increase of the critical thickness of LaAlO$_3$, 15 u.c. for -1.21% strain.

### 3.2.2 SrTiO$_3$ surface treatment

Under suitable conditions, techniques such as pulsed laser deposition (discussed later in section 3.2.3) allow materials to grow layer-by-layer with atomic precision, resulting in atomically smooth surfaces. However, in order to achieve this layer-by-layer growth, an atomically flat substrate with controlled surface termination (either TiO$_2$ or SrO) is essential. As we will see later on, the specific surface termination also plays a critical role in the functionality of the 2DES [211]. For a topical review of perovskite surface treatments please see Sanchez et al. [140].

A TiO$_2$ termination can be achieved by selective surface etching treatments such as buffered HF solution developed by Kawasaki et al. [4] and Koster et al. [256]. First, the SrTiO$_3$ substrate is immersed in an HF solution for 20-30 seconds under ultrasonication, followed by rinsing with distilled water. The substrates are then subjected to a high temperature annealing at 950°C in an oxygen atmosphere for 1-2 hours. This single termination is desirable for the heteroepitaxy of abrupt interfaces, preventing the intermixing of ion species at the interface during the deposition. Later in 2008 Kareev et al. [257] developed the "Arkansas etch", consisting of initially soaking the substrate in DI water followed by soaking in (3:1) mixture of HCl:HNO$_3$, which claim to achieve TiO$_2$-termination with lower surface roughness compared with HF etching.

### 3.2.3 High-pressure RHEED

Reflection high-energy electron diffraction (RHEED) is a commonly used technique for monitoring the growth of the thin film [258]. The techniques relies on diffraction pattern of the electron beam of high energy at low glancing angle to the sample surface. While one can learn the crystal structure, stress, texture of the film from the diffraction pattern, the layer-by-layer growth is monitored by the oscillations of the intensity of the specular spot of RHEED. The maximal (minimal) intensity happens when the roughness of the surface is at its minimum (maximum), corresponding to a pristine or fully- covered (half-covered) surface. The growth mode can also be told from the shape of the oscillation (See, for example, [258]).

High-pressure RHEED was developed by Guus J. H. M. Rijnders [5] to operate at higher base pressure (from oxygen partial pressure), which extends the explorable parameter space of the growth of oxide heterostructures, and is a major advance led to high quality $LaAlO_3$/$SrTiO_3$. In high-pressure RHEED, the electron beam is enclosed in a differentially pumped tube, which extends almost to the sample, to reduce scattering of the electron beam to the molecules in the chamber and ensures the visibility of the RHEED pattern.

### 3.2.4   Pulsed Laser Deposition

Pulsed laser deposition (PLD) is a thin film fabrication technique to prepare various combinations of oxide thin films, interfaces, and heterostructures [259]. Though vapor deposition of thin films using various types of lasers had been known for years, its utility as a method to fabricate high-quality thin films began in the early 1980s. Since its first application in the growth of YBCO thin films [260], PLD has played an instrumental role in the early development of thin film high-$T_c$ superconductor compounds [260,261].

Figure 3.1 a is a simplified drawing for a typical PLD setup, consisting of a laser source and optics, a target material holder, a sample holder with a heater attachment, and a high-pressure RHEED system to monitor the oxide thin film growth process. Generally, the chamber is connected to high-vacuum pumps that control the vacuum level in the chamber in the pressure range of ultra-high vacuum (UHV) to atmospheric pressure. The chamber is also fitted with various gas lines with fine control (inert, oxygen, and nitrogen) to create the desired gas environment inside the chamber during the deposition.

In the PLD process, a laser pulse is directed at a target; the target material is ablated due to the localized thermal heating and a resultant plasma forms. The plasma expands into the surrounding vacuum in the form of a *plume* of energetic species containing atoms and ions of the target material. This plume reaches the substrate surface with an average energy per particle of 0.1 to > 10 eV, depending on the pressure of the background gas environment. As a result of the short, high-energy laser pulses, the evaporated material is not in thermodynamic equilibrium. Crucially, the relative amount of different compounds in the plume corresponds to the target composition, even for constituents of very different melting points. In contrast to most other deposition techniques, a desired thin-film stoichiometry can be easily achieved with PLD by using an appropriate target. Adjusting the laser power can trigger cation off-stoichiometry [262,263]. The ablated species are collected on a substrate that is mounted on a heater. The separation between the target and the sample is an important parameter to be controlled, along with background pressure that can thermalize the ablated species. The substrate material serves as a template that acts like an *epitaxial* or seed layer for the incoming species to grow.

The laser typically used is a pulsed high-energy excimer laser (KrF 248 nm) with variable repetition rate (usually 1-5 Hz). The laser beam is focused to a spot size of a few mm and energy on the order of 100 mJ. The laser power is a very crucial parameter and should be large enough to create local heating that is larger than the vaporization of the target material in order to get ablation of the target material. It is also important to position the target material that is to be ablated. Usually, the laser beam is incident at an angle of 45°, which allows a vertical ablation of the target. The target position is defined in such a way that the laser hits the target near the focal point of the focusing lens, where basically the beam has a more uniform energy within the spot size. The target specimen is rotated and rastered to get a uniform ablation and to avoid formation of ablated rings or lines at the same area during the ablation.

The next important parameter in PLD is the separation between the target source and substrate. Usually this spacing is about 6-10 cm. Since the ablated plasma contains a high-energy species, to avoid the bombardment of these high energy species (when very near to target) at the substrate surface and also to avoid a possible non equilibrium deposition (very far from target), it is important to locate the substrate at the desired position from target. The substrate should also be located at the center area of the incoming plasma, where the most uniform deposition takes place.

In order to achieve crystalline cubic phase for $SrTiO_3$ in thin film form, a high temperature deposition is required. Generally, $SrTiO_3$ grows epitaxially and in single cubic phase at elevated temperatures ranging from 550°C and above. Below this temperature $SrTiO_3$ thin films often forms polycrystalline or amorphous phase. $SrTiO_3$ grows epitaxially above 450°C in the (001) direction of substrate; below this temperature, there is no diffraction peak from $SrTiO_3$ film indicating an amorphous phase for $SrTiO_3$. Oxygen partial pressure during growth controls the concentration of oxygen vacancies [142], which, in turn, have extensive effects on various properties (See Sec. 2.4.1).

### 3.2.5   Atomic-Layer Deposition

Atomic layer deposition (ALD) [264] is a layer-by-layer growth mode where the growth of each layer is self-limiting. This self-limiting nature allows for the growth of films at monolayer level, with uniform coverage on arbitrarily large substrates. ALD also has the ability to conformally coat three dimensional and high aspect ratio objects as well as allowing for precise composition control. The first ALD system was developed for the growth of ZnS films for flat-panel display applications [265]. ALD has since been used to grow a wide range of materials from elemental, oxides, nitrides, and sulfides, and in many contemporary applications [266]. ALD is an important industrial tool for depositing a variety of oxide-based materials such as high-k dielectrics [267], photovoltaics [268], and solid-oxide fuel cells [269].

With ALD, the substrate is exposed to a gaseous species known as a precursor (Figure 3.1 b). A precursor gas A is reacted with the surface and terminates, with monolayer coverage, after all of the reactive sites are used. Gas A is then purged from the system, after which the next precursor gas B is then introduced and reacted with the new surface formed from gas A. After gas B is finished it is purged from the system and gas A is introduced again. An ALD cycle is a sequence of (1) dose A, (2) purge A, (3) dose B, (4) purge B, after which the sequence repeats. To grow an oxide material, an oxidizer such as $H_2O$ or $O_3$ is inserted into each dose/purge step.

One advantage of ALD is its inherently self-limiting growth where single-species layers can be produced. Also high aspect ratios. However, an atomically flat substrate is required to grow high quality films. Also, each precursor must be flushed from the growth chamber before the next precursor can be introduced. With ALD and suitable precursors, various $SrTiO_3$-based heterostructures have been reported, from $SrTiO_3$ film on Ru and $TiO_2$-colated Ru [270], amorphous $LaAlO_3/SrTiO_3$ [223,271], $LaAlO_3/SrTiO_3$ [272], (partially amorphous) $LaAlO_3/SrTiO_3$ on $SrTiO_3$-buffered Si (001) [273].

### 3.2.6   Molecular Beam Epitaxy

Molecular beam epitaxy (MBE) [274-276] is an epitaxial growth technique allowing for the highest quality films, in particular in GaAs/AlGaAs heterostructures [277] but also for the growth of oxides. For a recent review of oxide MBE see Schlom [278]. Figure 3 c is a simplified drawing. An effusion cell is used to sublimate a solid source material in order to create a "molecular beam" that travels in a line of sight toward the substrate. The sublimation takes place in a high or ultra-high vacuum (as low as $10^{-12}$ mbar), resulting in very small levels of impurities. Generally, epitaxial growth is achieved using very low growth rates. The growth of oxides by MBE is challenging because the presence of oxygen has the potential to disrupt the molecular beams. This problem is dealt with by pumping to deal with the oxygen. In some instances, atomic oxygen or ozone may be used to increase reactivity.

Several variations of MBE are used for the growth of oxides. For $SrTiO_3$-based 2DES, the quantum Hall effect with lowest filling factors were observed in δ-doped $SrTiO_3$ grown in oxide MBE by Matsubara et al.[279], with mobility as high as 18,000 cm$^2$/Vs. The first reported growth of $LaAlO_3$/$SrTiO_3$ grown by MBE was by Segal et al. [280]. They use elemental cells with Sr, La, and Al reacted with molecular oxygen at $3 \times 10^{-7}$ Torr to grow films on commercially available $TiO_2$-terminated $SrTiO_3$. SrO termination was also produced by growing a monolayer of Sr and reacted with $O_2$.

With Laser MBE [281,282], a pulsed UV laser ablates a target in order to create the molecular beam. Kanai et al. first used laser MBE to grow the oxide thin film $(Ca,Sr)CuO_2$ [281]. Laser MBE is very similar to PLD in that all elements are ablated simultaneously. Each also provides unit cell by unit cell growth. The main distinguishing feature of laser MBE is that the pressure is much lower (typically < $10^{-9}$ Torr) and therefore the mean free path is such that the elements travel in "molecular beams". A technique used by Herklotz et al. [283], which has been alternatively called sequential PLD and laser MBE, involves sequentially ablating binary-oxide targets (SrO and $TiO_2$) to grow $SrTiO_3$ and the Ruddelson-Popper series $Sr_{n-1}Ti_nO_{3n+1}$. Lei et al. use a method they term ALL-Laser MBE to grow high quality $LaAlO_3$/$SrTiO_3$ [284] The saw no evidence of $O_2$ vacancies because the growth occurred at high $O_2$ pressure, a regime not accessible by other growth techniques.

### 3.2.7 Sputtering

Sputtering involves bombarding a target with high-energy particles. It is a widely used industrial technique that can produce large, uniform epitaxial films [285,286]. Epitaxial $LaAlO_3$ was first deposited on $SrTiO_3$ (100) by sputtering by Lee et al [287], although it was found that the layer was insulating. In that early work, it was not specified whether the $SrTiO_3$ termination was controlled. Dildar et al [288] used sputtering to grow $LaAlO_3$ on $TiO_2$-terminated $SrTiO_3$ but did not find a conductive interface. Podkaminer et al [289] were able to grow conductive $LaAlO_3$/$SrTiO_3$ (001). The key ingredients of their recipe were sputtering in pure Ar and their single crystal $LaAlO_3$ target to sustain the large required background pressure of $O_2$ and O [290].

### 3.3 Device Fabrication

Several methods for fabricating $LaAlO_3$/$SrTiO_3$-based devices have been reported. The most widely used techniques, and those that will be discussed in more detail here, include conventional photo- or electron beam-lithography using hard masks or thickness modulation of the $LaAlO_3$ layer, bombardment with Ar or O ions to locally create insulating regions, and conductive AFM lithography.

### 3.3.1 Lithography - Thickness Modulation, Hard Masks, and Angle Deposition

The first lithographically patterned $LaAlO_3$/$SrTiO_3$ devices were demonstrated by Schneider et al. [291]. They used a hard mask of $LaAlO_3$ to define regions in which there is greater than 4 u.c. of crystalline $LaAlO_3$ (Fig. 3.2 (a)). Regions of amorphous $LaAlO_3$ are used to define regions with greater than 4 u.c. critical thickness of crystalline $LaAlO_3$. 2 u.c. of epitaxial $LaAlO_3$ is grown on $SrTiO_3$. Conventional photolithography is then used to define liftoff regions, on which amorphous $LaAlO_3$ is grown. Following liftoff of resist and amorphous $LaAlO_3$, 3 u.c. of crystalline $LaAlO_3$ is deposited. Regions that previously had resist now have greater than the critical thickness of crystalline $LaAlO_3$ on $SrTiO_3$. The smallest conducting structures realized by this method were ~200 nm by patterning the resist with e-beam lithography. Amorphous $AlO_x$ has also been used as a hard mask [292].

Similar to Schneider, e-beam lithography [293] defines regions that will be crystalline $LaAlO_3$/ $SrTiO_3$ while other regions have amorphous $LaAlO_3$/ amorphous $SrTiO_3$/ $SrTiO_3$. Cu/PMMA deposited before e-beam to avoid charging of surface. Unlike Schneider, there is no "pre deposition" of crystalline $LaAlO_3$ before resist deposition and patterning. Devices can also be patterned in a single lithographic step according to Maniv et al. [294]. They deposited hydrogen silsesquioxane (HSQ) onto $TiO_2$-terminated $SrTiO_3$. Exposure to e-beam results in a $SiO_2$ hard mask, after which $LaAlO_3$ is deposited. The lithography-defined features are around 100 nm while the resulting quantum dot is about 10 nm, inferred from the level spacing.

One problem with a hard mask is that it remains on the device following processing. Banerjee et al. demonstrated a "direct patterning" [295] method for $LaAlO_3/SrTiO_3$ that avoids this issue. Amorphous $AlO_x$ is deposited onto a $SrTiO_3$ substrate and subsequently patterned using UV photolithography. The photoresist developer reacts with the exposed $AlO_x$ to form water-soluble alkali-metal aluminates, resulting in $TiO_2^-$ terminated $SrTiO_3$. $LaAlO_3$ was deposited through the $AlO_x$ mask and the remaining $AlO_x$ is removed with NaOH.

Quasi-1D nanowires with width about 50 nm have been created without e-beam lithography by the use of angle deposition [296], as shown in Figure 3.2 (b). An initial hard mask of amorphous oxide with trenches is deposited onto $TiO_2$-terminated $SrTiO_3$. The sample is then tilted by an angle θ, and a second layer of amorphous oxide is deposited. The first layer shades a region of $SrTiO_3$, resulting in a bare region on which to deposit LaAlO3.

### 3.3.2 Ion beam irradiation

Ion beam irradiation has been used to make microstructures and nanostructures at the $LaAlO_3/SrTiO_3$ interface (Fig. 3.2 (c)). A beam of ions will create local defects in $LaAlO_3$, resulting in a conducting 2DES becoming insulating in regions irradiated by the ions. Aurino et al. [297] used focused Ar+ ion beams to switch conducting 4-10 u.c. $LaAlO_3/SrTiO_3$ to insulating.

Several other groups have utilized ion beams to make devices in $LaAlO_3/SrTiO_3$. Mathew et al. [298] used a hard mask of brass to define conducting regions using Ar+ ions. They also demonstrated pattering using He ions. Azimi et al. uses silicon stencil masks to define structures as small as 150 nm using He ions [299]. Three terminal devices, top-gated field-effect devices with a $Si_3N_4$ gate dielectric, were made using $O_2$ ion-irradiation by Hurand et al. [300].

One advantage ion beam lithography has over the lithography techniques covered in Section 3.3.1 is that ion irradiation-defined insulating regions may become metallic using high temperature oxygen annealing [301]. Retention of nanostructures using a $SrCuO_2$ capping later [302] was shown to improve the quality of sub-micron devices. Intermodulation electrostatic force microscopy (ImEFM) was used to image the surface potential. Capped nanostructures have a more homogeneous surface potential than uncapped nanostructures, resulting in size-independent carrier density and mobility below 1 μm.

### 3.3.3 Conductive-AFM lithography

Nanoscale devices can be created using conducting atomic force microscopy (c-AFM) [19,303-305]. This approach was based upon a report by Thiel et al. [17] showing that $LaAlO_3/SrTiO_3$ heterostructures with a thickness of 3 unit cells can be metastably switched between insulating and conducting phases using +/- 100 V applied to the back of the $SrTiO_3$ substrate. Cen et al.[303] demonstrated that nanoscale control of the metal-insulator transition can be achieved using a voltage-biased c-AFM tip that is scanned over the top LaAlO$_3$ surface, as shown in Figure 3.3. Devices with characteristic features as small as 2 nm (at room temperature) have been created. The source of the metastability was revealed by Bi et al [164] to be related to a "water cycle" that exchanges protons between the c-AFM tip and the LaAlO$_3$ surface via a water meniscus. Subsequent experiments showed that the hysteretic conduction reported by Thiel et al. [17] is mediated by atmospheric ions [306], therefore can be viewed as the global version of the c-AFM lithography. The role of surface protonation in regulating conductivity at the LaAlO$_3$/SrTiO$_3$ interface was investigated more systematically by Brown et al. [165] (see section 4.2.5). Using conductive AFM lithography, a variety of device concepts have been demonstrated, including photodetectors [307], diodes [307], transistors [303,308], quasi-1D superconducting channels [309,310], (superconducting) single electron transistors [187,311], and electron waveguides [312] (See Sec. 6).

# 4 Properties and phases of LaAlO$_3$/SrTiO$_3$

In this section, we review the reports on the basic properties of the LaAlO$_3$/SrTiO$_3$ two-dimensional electron system (2DES). We summarize the current state of understanding of many facets that have been investigated intensively, including conductivity mechanism, the origin of a critical thickness for conduction, the confinement profile, band structure, superconductivity, and emergence of magnetism. We also briefly discuss the reports and theories on the possible coexistence of the superconductivity and magnetism. The quantum transport of this 2DES will be discussed in more depth in Sec. 5 and Sec. 6.

## 4.1 Conductivity with a threshold thickness

The discovery of conducting behavior at the LaAlO$_3$/SrTiO$_3$ interface initiated a broad search to elucidate the origin of the 2DES. It was found that in order for the interface to conduct, several empirical conditions must be satisfied. First, the LaAlO$_3$ layers must exceed a critical thickness [17]. Second, the SrTiO$_3$ must terminate with TiO2. [6]. Third, La/Al stoichiometry ratio should be off (less than 0.97) [313]. Several other factors are noteworthy: the growth of LaAlO$_3$ on SrTiO$_3$ is not equivalent to the growth of SrTiO$_3$ on LaAlO$_3$ [314]. Also, substrate strain can significantly shift the critical thickness to be as high as 12 unit cells with a strained LSAT substrate [255]. Surface termination (protonation) of the LaAlO$_3$ can switch the interface conductivity of LaAlO$_3$/SrTiO$_3$ heterostructures with up to 7 unit cells of LaAlO$_3$ [165]. We discuss these empirical observations and explanations below.

### 4.1.1 Confinement thickness (the depth profile of the 2DES)

The confinement profile of the 2DES and how far it extends into the SrTiO$_3$ is closely tied to the properties of the system. Basletic et al. used conductive atomic force microscopy (c-AFM) to map the resistance profile of the interface by probing the cleaved edge of a LaAlO$_3$/SrTiO$_3$ heterostructure (Figure 1.2) [18]. Two samples grown under different oxygen partial pressures $P(O_2)$ were compared. They found that annealed samples had a 2DES that was highly confined (~7 nm) at room temperature, while the 2DES of unannealed samples extended far into the bulk SrTiO$_3$ [18]. Copie et al. [315] performed a similar c-AFM experiment at cryogenic temperatures, and found that the confinement depth increases from ~8 nm at room temperature to ~12 nm at $T = 8$ K [315].

It may seem surprising that electron confinement persists to low temperature, given that the permittivity in SrTiO$_3$ can be as large as 30,000 at 10 K. While the permittivity is large, it is also highly nonlinear. By including this nonlinearity into a self-consistent tight-binding model, computational models have shown that the 2DES can remain strongly confined so long as the electron density is high enough to sufficiently reduce the permittivity [316,317]. The enhancement of the permittivity at cryogenic temperatures is the result of a softening of an optical phonon mode where the $Sr^{2+}$ and $Ti^{4+}$ move opposite to the $O^{2-}$ octahedra (Figure 2.1 b). This phonon mode is sensitive to the local electric field strength, and thus the permittivity experienced by the 2DES is affected by the local electron density. Because the electron penetration into the $SrTiO_3$, and thus the confining potential, is influenced by the permittivity, local density perturbations can alter the band structure of the material, providing an intrinsic mechanism for phase separation of the 2DES [318].

Several groups have used optical methods to probe the spatial extent of the 2DES. Using infrared ellipsometry to look at Berreman modes—dynamic charge fluctuations within the 2DES that alter the reflectivity in a polarization-dependent manner [319]—Dubroka et al. [320] found that the 2DES is composed of two parts: a sharp peak within 2 nm of the surface and a long tail with a characteristic penetration depth of 11 nm into the $SrTiO_3$. These features are supported by self-consistent tight-binding calculations that indicate that the sharp density peak is associated with the Ti $t_{2g}$ $d_{xy}$ band. The long tail stems from the Ti $t_{2g}$ $d_{xz}$ and $d_{yz}$ bands, which have a much lighter band mass perpendicular to the interface [316,317]. Photoluminescence experiments have examined the Auger recombination of triplet electron-electron-hole excitations, and also been used to measure the penetration depth by varying the energy of the exciting photons [321]. Additionally, hard x-ray photoemission spectroscopy (HAXPES) has been used to map the electron density at the interface [322]. By varying the angle of incidence on the sample, depth-dependent information can be obtained. Using this method it was determined that the 2DES is strongly confined (at room temperature) with the first u.c. of the $SrTiO_3$.

## 4.2 Metal-insulator transition and critical thickness

### 4.2.1 "Polar catastrophe"

The most popular and widely cited mechanism used to explain the origin of the 2DES is the "polar catastrophe" [211]. When a polar material is attached to a non-polar material, there is a built-in electric field that results in a large potential difference between the two surfaces. As the thickness of the polar material increases, the size of the potential difference can lead to electronic and structural instabilities or reconstruction. For example, Ge is a non-polar material, while GaAs is polar; Ge/GaAs heterostructures result in an atomic reconstruction in which surface roughening and intermixing of the Ge and Ga at the interface compensate the buildup of this field [323,324].

The polar catastrophe mechanism has been proposed as the primary mechanism for conductivity at the $LaAlO_3/SrTiO_3$ [211]. Due to their perovskite crystalline structures, in the [001] direction, $LaAlO_3$ may be regarded as a series of $La^{3+}O^{2-}$ and $Al^{3+}O_2^{2-}$ planes while $SrTiO_3$ a series of $Sr^{2+}O^{2-}$ and $Ti^{4+}O_2^{2-}$ planes. Thus, $LaAlO_3$ makes a series of +1,-1,+1,-1,... planes and the $SrTiO_3$ planes are uncharged (Figure 4.1 b,c). When $LaAlO_3$ is then grown on $SrTiO_3$, there is an uncompensated, charged plane at the interface which results in the buildup of an electric field that can be negated if 0.5 electrons per u.c. are transferred to the interface. In order to prevent a polar catastrophe, the system undergoes an electronic reconstruction in the case of $n$-type ($TiO_2$ terminated $SrTiO_3$) interfaces (Figure 4.1 a,d). The charge carriers are transferred from the $LaAlO_3$ surface valence bands (Oxygen $2p$ bands) to the $SrTiO_3$ conduction bands (Ti $3d$ $t_{2g}$ bands) and metallic behavior is observed at the interface [6].

While the full validity or applicability of the polar catastrophe mechanism has been extensively debated in the oxide community, it can account for some striking experimental findings. For instance, the observation of a critical thickness for conduction [17] can be viewed as a crossover of the $LaAlO_3$ valence band maximum and $SrTiO_3$ conduction band minimum [325,326]. At a critical point, the valence band at the LaAlO$_3$ surface crosses the SrTiO$_3$ conduction band at the interface and charge is transferred to the interface. In this case, altering the polarization of the LaAlO$_3$ overlayers would then alter the critical thickness. This approach has been taken by Reinle-Schmitt et al. [224] by intermixing Sr and La, creating La$_x$Sr$_{1-x}$AlO$_3$, with the polarization of the material depending linearly upon $x$ (Figure 4.1 d,e,f). Thus, as Sr was increased, the polarization decreased and the critical thickness for the onset of a metallic interface subsequently increased.

The built-in electric field in the LaAlO$_3$, which is responsible for the charge transfer to the interface, has been detected using a few different approaches. Huang et al. [327] directly measured this field using scanning-tunneling microscopy and spectroscopy on cross sectioned samples to probe the band minima and maxima of the conduction and valence bands, respectively, as the tip was moved across the interface (Figure 4.1 g). As the LaAlO$_3$ is a dielectric, the built in electric field should also give rise to electrostriction, deformation of a dielectric due to an electric field, in the LaAlO$_3$. This deformation has been observed by measuring the $c$-axis of the LaAlO$_3$ as the number of layers increases by Cancellieri et al. [328], where electrostriction effects vanish after the critical thickness is reached, indicating an electronic reconstruction had occurred. In addition to this, Singh-Bhalla et al. [329] used Pt contacts on the LaAlO$_3$ to perform tunneling measurements through the LaAlO$_3$ to a metallic interface. The built-in electric field was estimated by analyzing the tunneling rate with respect to applied voltage.

Inconsistencies arise when the electric field is mapped in other ways. Using HAXPES and XPS to map shifts of the core levels of La, Sr, Al, and Ti at different depths, only a small shift was observed by multiple groups [280,330-332] in the LaAlO$_3$ core levels that is roughly an order of magnitude smaller than that expected for the polar catastrophe to lead to an electronic reconstruction. Furthermore, the size of the shift seems largely independent of the LaAlO$_3$ thickness which is also unexpected in a polar catastrophe scenario [331]. These differences, however, may be accounted for if oxygen vacancies form at the LaAlO$_3$ surface or if the measurement process excites particle-hole pairs which then act to screen the built in electric field.

In order to avoid the polar catastrophe, 0.5 of an electron must be transferred to the interface. This process should lead to a carrier concentration of $n = 3 \times 10^{14}$ cm$^{-2}$. Experimentally measured carrier concentrations (from the Hall effect), however, are generally an order of magnitude lower [7,8], begging the question: where are the missing electrons? One explanation postulates that there are two populations of electrons at the interface, one that is tightly bound to the interface and strongly localized through impurities and defects, while the other is more loosely bound and mobile [52]. The existence of low-mobility or fully localized carriers has been put forth to explain the even lower carrier concentrations inferred from Shubnikov-de Haas oscillations (see Sec. 5.6). Still, some heterostructures (e.g., GdTiO$_3$/SrTiO$_3$) exhibit a full charge transfer [238].

Another issue arises with the numerical value of the critical thickness. Some calculations predicted that 4 u.c. of LaAlO$_3$ are necessary for a conductive interface, consistent with most experiments. Other treatments [326], however, predicted a critical thickness of 5 u.c. before the surface LaAlO$_3$ valence bands cross the SrTiO$_3$ conduction bands. This discrepancy has not yet been resolved.

Even more problematic for the polar catastrophe model are multiple observations of conductive interfaces for which there is no polar discontinuity. These include amorphous LaAlO$_3$ on SrTiO$_3$ [214,223] and heterostructures in the [110] direction [333]. In both cases, there are no charged layers in the LaAlO$_3$ to give a diverging electric potential and thus there is no polar catastrophe to avoid by charge transfer. In both cases, however, the interface is found to be conductive above a critical thickness with properties very similar to those found for crystalline $n$-type [001] LaAlO$_3$/SrTiO$_3$ heterostructures. In a similar vein, the polar catastrophe model predicts a $p$-type interface for SrO terminated SrTiO$_3$, but this interface is always found to be insulating, regardless of the LaAlO$_3$ thickness.

### 4.2.2 Oxygen Vacancies

A second possible source of electrons for the 2DES are oxygen vacancies. As the O sites in both SrTiO$_3$ and LaAlO$_3$ have a formal valence of $-2$, oxygen vacancies results in $n$-type doping. Oxygen vacancies are able to dope bulk SrTiO$_3$ at carrier concentrations exceeding about $n \sim 10^{17}$ cm$^{-3}$ (See Sec. 2.4.1). In the case of interfaces such as LaAlO$_3$/SrTiO$_3$, the concentration of the oxygen vacancies is a function of the oxygen partial pressure, $P(O_2)$, and the temperature during growth [120,334-336]. Conduction in bulk SrTiO$_3$ typically appears when $P(O_2) \leq 10^{-4}$ mbar [212]. The concentration of the oxygen vacancies can be significantly reduced by post annealing the sample under higher $P(O_2)$. Pressures higher than $10^{-2}$ mbar lead to 3D, island-like growth [212].

If the 2DES stems from oxygen vacancies in bulk SrTiO$_3$, however, there are several problems. First, EELS (electron energy loss spectroscopy) suggests that there are not enough oxygen vacancies in the SrTiO$_3$ bulk to account for the observed charge densities, and that the compensating charge is located towards the LaAlO$_3$ overlayers [337]. Furthermore, oxygen vacancies in bulk SrTiO$_3$ leads to electron localization at F-centers in the bulk [338,339] and on nearby Ti $e_g$ orbitals [160], potentially giving rise to the observed magnetic moments at the interface [340] (see Sec. 4.6).

Another potential source of electron doping comes from oxygen vacancies in the LaAlO$_3$ overlayers [341-343]. In this case, first principles [341], tight binding [342], and DFT [343] calculations show that oxygen vacancies at the LaAlO$_3$ surface result in states within the LaAlO$_3$ band gap at the surface. These in-gap states can transfer electrons to the SrTiO$_3$ conduction band at the interface, negating the built-in electric field caused by the polar discontinuity at the LaAlO$_3$/SrTiO$_3$ interface. This process helps to explain why HAXPES and XPS studies do not observe the shifts in core orbitals expected from the polar catastrophe model [331,344].

A basic question arises of how a critical thickness arises if oxygen vacancies are responsible for the transition to a metallic interface. First, for crystalline LaAlO$_3$, the formation of oxygen vacancies at the surface becomes energetically favorable as the thickness of the LaAlO$_3$ overlayer grows [341,343]. As the number of oxygen vacancies reaches a critical point, percolation leads to extended states at the interface and the appearance of metallic behavior [342]. For amorphous LaAlO$_3$/SrTiO$_3$ heterostructures, however, this mechanism of oxygen vacancy formation does not appear. Liu et al. [345] used Ar$^+$ milling to reduce the LaAlO$_3$ thickness below the critical thickness for both amorphous and crystalline LaAlO$_3$ layers. The crystalline layers exhibited a metal-to-insulator transition, while the amorphous LaAlO$_3$ layer did not. The interpretation is that the amorphous LaAlO$_3$ forms more mobile oxygen vacancies that migrate to the SrTiO$_3$ bulk and were thus unaltered by the milling process, unlike for the crystalline LaAlO$_3$ heterostructures. This picture then explains how resistive switching can take place as a result of a strong back voltage when the overlayer thickness is near the critical thickness in amorphous LaAlO$_3$ [346].

### 4.2.3 Interdiffusion

A third mechanism for generating carriers at the LaAlO$_3$/SrTiO$_3$ 2DES interface involves cation intermixing. From surface x-ray diffraction (SXRD) [347-349], medium-energy ion spectroscopy (MEIS) [350], and electron energy loss spectroscopy (EELS) [211], intermixing of the La and Sr within the first few unit cells of the substrate after growth has been observed. This intermixing can act as a source of doping in the substrate as the La$^{3+}$ replaces the Sr$^{2+}$ at the interface, requiring some compensation in nearby Ti layers, transforming Ti$^{4+}$ to Ti$^{3+}$. Intentional delta doping with Cd and Mn, however, did not give rise to metallic behavior, as reported by Fix et al. [351,352], nor did significant mixing of Sr and La in the overlayers during growth, as reported by Reinle-Schmitt et al. [224]. Furthermore, LaAlO$_3$/SrTiO$_3$ alloys are insulating, presenting significant problems for this mechanism of charge transfer [353].

### 4.2.4   Polar Interdiffusion and Antisite Pairs

The previous explanations of producing the 2DES focus on a single aspect of the LaAlO$_3$/SrTiO$_3$ interface. It is possible to consider simultaneously multiple possible doping schemes. In these models, the built-in field introduced by the polar discontinuity at the interface drives certain crystal defects to be favored [354-356]. In the n-type interfaces, this leads to favoring the formation of oxygen vacancies at the surface which then act as in-gap states which the built in field depopulates, forming the 2DES in the SrTiO$_3$ conduction band. This would then produce an insulating LaAlO$_3$ surface as well as cancel the built-in polar field, explaining various HAXPES studies that did not find a built-in field [280,330-332].

This mechanism alone, however, is insufficient to explain the absence of the built-in field below the critical thickness as the oxygen vacancy formation is only favored spontaneously for $\geq 4$ u.c. of LaAlO$_3$. Instead, Al-Ti mixing occurs, forming a compensating dipole field, which consists of localized states [354]. When spontaneous oxygen vacancies are then favored, a portion of the transferred electrons compensates the Al-Ti intermixed pairs, reducing the eventual carrier density from the ideal value of 0.5 electrons per u.c.

Another potential mechanism is related to a LaAlO$_3$ buckling transition that occurs because of the polarization field [355]. This buckling would result in the smooth formation of a 2DES for >5 u.c. of LaAlO$_3$; however, the introduction of oxygen vacancies at the surface leads to a sharp transition at 4 u.c. This would then explain the reduction of the built in field as well as the sudden transition to a metallic interface at 4u.c.

In *p*-type interfaces, two mechanisms can possibly account for an insulating interface. In the first case, the formation of oxygen vacancies is not favored and instead the polarization field is compensated by cation intermixing which results in the formation of either charge-neutral or otherwise localized states [354]. In the latter case, oxygen vacancies form and the resulting electrons in the surface in-gap states are transferred to the interface, compensating the holes introduced by the polarization field electronic reconstruction [356].

### 4.2.5   Role of surface adsorbates

In traditional semiconductor devices, the conduction layer is often located several hundred nanometers away from the donor layer and buried deep within the structure. By contrast, LaAlO$_3$/SrTiO$_3$ interfaces are only a few unit cells from the structures surface. As a result, the surface chemistry, e.g., adsorbates and surface reconstructions, can profoundly influence the behavior of the 2DES. The importance of the surface can be seen from the ability to switch locally between an insulating state and a conducting state using voltages applied from a local probe [303] (See Sec. 3.3.3). By performing conductive AFM lithography in controlled atmospheric environments, it was found that adsorbed water was necessary to switch the interface conductivity [164]. Further studies on this effect found that surface protonation plays a significant role in mediating the metallic nature of the interface (Figure 4.2).  Remarkably, a deprotonated surface becomes insulating for LaAlO$_3$ thicknesses up to 7 u.c. [165]. LaAlO$_3$/SrTiO$_3$ interfaces, after being exposed to low-energy oxygen plasma, become strongly insulating, for thickness up to 8 u.c. of LaAlO$_3$ [357].

### 4.2.6 Hidden FE-like distortion - Strain induced instability

A final mechanism for influencing the 2DES formation at the LaAlO$_3$/SrTiO$_3$ interface involves the lattice mismatch between LaAlO$_3$ and SrTiO$_3$ [358,359]. Due to this mismatch, the oxygen octahedra in the LaAO$_3$ may undergo a rotation, straining the SrTiO$_3$ and lead to a strain-induced charge transfer to the near-interface layers of the SrTiO$_3$ [358]. The lattice mismatch also leads to $c$-axis elongation in the SrTiO$_3$ and subsequent rotation of oxygen octahedra, leading to charge transfer associated with a pseudo-Jahn-Teller distortion [359].

## 4.3 Ferroelectric and Piezoelectric Properties

SrTiO$_3$-based heterostructures can have piezoelectric and ferroelectric-like responses due to the presence of the polar layer. Using piezoforce microscopy, the electromechanical response of the LaAlO$_3$ can be switched locally and hysteretically [360-362]. This switchable response is indicative of a ferroelectric transition in the LaAlO$_3$ thin film which is absent in bulk LaAlO$_3$. By varying $P(O_2)$ during growth and the humidity/temperature at the time of measurement, two mechanisms of generating this ferroelectric-like behavior were identified by Li et al. [361]. At low $P(O_2)$, there are a large number of oxygen vacancies in the heterostructure; from electrostatic considerations, there are two minima in the oxygen vacancy formation energy, one at the interface and the other at the LaAlO$_3$ surface. Electric fields applied to the sample then drive the oxygen vacancies to one of these minima or the other, giving rise to an effective switchable polarization, which is responsible for the electromechanical response. In the case that the oxygen vacancies are not abundant enough (high $P(O_2)$ growth conditions), no switchable electromechanical response is observed unless the humidity is high in the measurement chamber. In this case, surface adsorbates can serve a similar purpose, as also observed by Huang et al.[306]; positive voltage attracts positive ions to the surface, while a negative voltage removes them. Hysteresis is only observed in the presence of ions generated from a vacuum ion gauge.

A characteristic of ferroelectric materials is that mechanical strain may also alter the polarization and thus switch the piezoresponse of the material. To test this, Sharma et al. [363] used a grounded AFM tip to apply mechanical strain to a LaAlO$_3$/SrTiO$_3$ heterostructure. With increasing force, it was found that the resistance of the interface could be switched from a high to a low value (Figure 4.3 c). Furthermore, PFM measurements also showed that the piezoresponse of the heterostructure changed abruptly when 500 nN of force was applied. As the mechanical strain is not sensitive to the environmental conditions, the mechanism of action was attributed to the movement of oxygen vacancies from the surface of the LaAlO$_3$ to the interface.

In addition to mechanical strain, the effects of epitaxial strain can be considered. By growing LaAlO$_3$/SrTiO$_3$ heterostructures on different substrates with different lattice mismatches, the epitaxial strain can be varied from compressive to tensile, with marked effects on the critical thickness for interface conduction. Bark et al. [255] found that all samples with tensile strain were insulating. In the case of compressive strain, however, the interface could be made conducting, but the critical LaAlO$_3$ thickness increased with increasing strain. Compressive strain in the SrTiO$_3$ causes the Ti atoms to shift away from the interface, inducing a polarization in the SrTiO$_3$ which partially cancels the polar field in the LaAlO$_3$ and subsequently increases the critical thickness.

## 4.4 Electronic band structure

### 4.4.1 Theory

The conduction band of SrTiO$_3$ is derived from Ti $3d$ orbitals, with the band minimum at the Γ point. The $e_g$ orbitals are split from the $t_{2g}$ orbitals by the crystal field (Figure 2.1 e). Within LaAlO$_3$/SrTiO$_3$ heterostructures, interface confinement and spin-orbit coupling (SOC) further split the $t_{2g}$ orbitals, pushing up the $d_{xz}, d_{yz}$ orbitals, making the $d_{xy}$ orbital the lowest energy conduction band. These effects have been reproduced by first-principles calculations which have successfully produced a full set of parameters for producing Hubbard model calculations of the LaAlO$_3$/SrTiO$_3$ system [364].

These properties, however, are subject to change from multiple effects. The electrostatic confinement at low-temperature is heavily weakened by the very large dielectric constant. This is caused by a soft optical phonon mode where the Sr$^{2+}$ and Ti$^{4+}$ move out of phase with the oxygen octahedral (Figure 2.1 b), giving rise to a strong lattice screening of electric fields. At large electric fields, this phonon mode stiffens and the dielectric constant decreases. The degree of electron confinement becomes strongly dependent upon the electric field of the electron liquid and thus the electron density [365]. As the electron density changes, the nature of the conduction band may change not just by the addition of more bands. As shown by Khalsa and MacDonald [316] using a tight binding model, the orbital character of the lowest band may change from $d_{xz}, d_{yz}$ at low density to $d_{xy}$ at high density. Other work using DFT calculations find the $d_{xy}$ band to be lower in energy [325,366,367].

Another effect which can be accounted for are lattice deformations. Such deformations arise from the tetragonal transition in bulk SrTiO$_3$ and the epitaxial strain arising from the slight lattice mismatch in the LaAlO$_3$ and SrTiO$_3$ lattice constants. By carrying out a DFT calculation where the lattice was allowed to relax, Zhong and Kelly [368] were able to show that certain deformations, such as a GdFeO$_3$-like deformation (tilt of TiO$_6$ octahedra along the b-axis followed by a rotation about the c-axis) at the interface, can give rise to hybridization of the $t_{2g}$ orbitals at the interface. Additional lattice deformations were argued by Lee and Demkov [369] to be a pseudo-Jahn-Teller effect which further splits $d_{xy}$ orbital from the other $t_{2g}$ orbitals.

Orbital hybridization also arises from the inclusion of SOC. In this case, the broken inversion symmetry along with the atomic SOC gives rise to Rashba SOC (see Figure 4.4 a for a three-band tight-binding band structure with SOC). As the confinement and the distortion of the octahedra determine the energy splitting of the $t_{2g}$ bands, it was found that the Rashba spin-orbit splitting as a function of momentum is highly sensitive to this confinement strength [370]. SOC is discussed in greater detail in Sec. 5.4. Spintronic effects attributed to SOC are discussed in Sec. 5.8.

### 4.4.2 Experiment
#### 4.4.2.1 ARPES
Angle-resolved photoemission spectroscopy (ARPES) gives an experimental means to directly measure the band structure of a sample (see Figure 4.4 b for an example of the band structure extracted via ARPES). From these measurements, the band masses may be extracted in addition to the signatures of other effects, such as SOC. Due to the anisotropy of the bands, each of the $t_{2g}$ bands is described by a light ($m_l^*$) and a heavy ($m_h^*$) effective mass, with $m_l^* \sim 0.5 - 0.7 m_e$ and $m_h^* \sim 9 - 20 m_e$ [203,204,371,372]. King et al. [371] extracted a Rashba SOC parameter for the $d_{xy}$ $n = 0$ ($\alpha_0 = $ 1.4meVÅ) and the $n = 1 (\alpha_1 = 3$ meVÅ) subbands.

ARPES measurements of SOC in SrTiO$_3$ have been controversial. Of particular interest is the possible appearance of massive SOC in SrTiO$_3$, reported by Santander-Syro et al. [373] The ARPES measurements revealed two rings at the Fermi surface with a spin energy splitting of $\Delta \sim 100$ meV. Mapping these onto the difference of the minima of the two spin bands, $k_R$, the expected result is $k_R \sim 6 \times 10^{-3}$Å$^{-1}$, but instead they found $k_R \sim 0.1$Å$^{-1}$. This giant splitting was also accompanied with a lifting of the degeneracy at the $k = 0$ point, which is not expected for Rashba SOC. When another, similarly prepared SrTiO$_3$ sample was examined using similar techniques, however, no massive splitting was observed [374]. It has been suggested [375] that the discrepancy can be traced to oxygen vacancies in the surface layer which were present in the case of giant splitting and absent in the case where no splitting was observed.

Another result which stands out was a potential polaron signature (Figure 4.4 c) in the LaAlO$_3$/SrTiO$_3$ band structure reported by Cancellieri et al. [376]. In ARPES measurements, polarons appear with a peak-dip-hump structure, where the different parts of the polaron are probed. The peak is then the electron band which contributes to the polaron while the hump is the signature of the phonon modes which contribute to the polaron and the dip the energy difference between the two. In the case of LaAlO$_3$/SrTiO$_3$, the peak arises from hybridization of the $d_{xy}$ and $d_{yz}$ bands. The hump was then composed of a hard LO3 phonon mode (oxygen cage breathing mode) and a soft TO1 phonon mode (associated with the polar instability of SrTiO$_3$). This result could explain the temperature dependence of the electron mobility in LaAlO$_3$/SrTiO$_3$ heterostructures as the LO3 mode sets the low temperature limit of the mobility, while the TO1 mode changes its character after the tetragonal distortion, leading to a significant temperature dependence of the mobility.

### 4.4.2.2 Ellipsometry

Using infrared ellipsometry, Dubroka et al. [320] were able to probe the band structure of the LaAlO$_3$/SrTiO$_3$ heterostructure. The electron gas was found to have two components: a highly concentrated portion within 2 nm of the interface and a more diffuse tail, spreading over ~11nm. For the band properties, an effective charge carrier mass was found of $m^* \sim 3.2 m_e$. Additionally, the mobility of the 2DES was found to be very sensitive to the frequency, which is characteristic of polaron quasiparticles (possibly Frohlich polarons [188]) acting as the charge carriers, in line with the ARPES results of Canciellieri et al. [376].

### 4.4.2.3 Photoluminescence

Photoluminesescence (PL) experiments in bare SrTiO$_3$ and LaAlO$_3$/SrTiO$_3$ heterostructures show a temperature-dependent shift in the decay of the PL signal [377-379]. This shift is linked to a change in dominant electron-hole recombination method. The switch is from Auger recombination (where an electron-hole pair recombine and the energy is dissipated by phonons) to single-carrier trapping. These results, in combination with persistent photocurrent measurements, suggest that there exist oxygen vacancies which act as electron traps. These SrTiO$_3$ in-gap states may then provide the electrons for photocurrent measurements and the single-carrier traps in the photoluminescence experiments.

### 4.4.2.4 Other techniques

Several other techniques have also been used to probe the electronic band structure. Using X-ray photoemission spectroscopy (XPS) [332,380] and cross-sectional scanning-tunneling microscopy (STM) [327], the band bending of the conduction band could be observed in LaAlO$_3$/SrTiO$_3$ either by the shifts in the core levels of the Ti (XPS) or by the electron tunneling energy as the interface is approached (STM). Additionally, x-ray magnetic circular dichroism (XMCD) measurements [51,233,381] reveal that the conduction band is dominantly composed of Ti $d_{xy}$ electrons.

## 4.4.3 Lifshitz transition

A Lifshitz transition appears as a change in the symmetry of the Fermi surface as the Fermi energy is tuned. In the case of LaAlO$_3$/SrTiO$_3$, as the other measurements of the band structure reveals, the Ti $d_{xy}$ band splits from the other $t_{2g}$ bands and has a light electron mass in the directions parallel to the interface while the other $t_{2g}$ bands, the $d_{xz}, d_{yz}$ bands, have a light and a heavy mass depending on the direction of propagation. As the Fermi energy increases, these bands begin to fill and the 2DES undergoes a Lifshitz.

Theoretically, this transition has been probed using a three-band tight-binding model which includes both SOC and an on-site interaction through Hartree [11] and Fock [382] corrections. The results show that as the Lifshitz transition is crossed, the heavier bands are preferentially filled, producing a dome structure where two bands are being simultaneously filled.

Experimentally, the Lifshitz transition has been identified using magnetotransport measurements. Joshua et al. showed that there is a critical density for LaAlO3/SrTiO3 heterostructures at which the magnetotransport properties abruptly change [383,384]. Similar experiments were conducted by Smink et al. [385], and the data were fit to a two-band tight-binding model. The data was also shown to be consistent with a more complex 3-band model in which there is a Lifshitz transition. The anisotropic magnetoresistance measurements mentioned here are described in more detail in Section 5.3.

## 4.5  Defects, doping, and compensation

Perovskites are susceptible to numerous defects which can subtly or profoundly influence the properties of the system. Thiel et al. [386] used bicrystal substrates to study the effects of grain boundaries on transport. They found that the conductivity of the interface across the grain boundary is very sensitive to the grain boundary angle, with even small angles between the lattice vectors causing the conductivity of the sample to drop by several orders of magnitude. This is because dislocations along the boundary act as nucleation centers for charge-depleted regions in the 2DES. As the angle increases, these dislocation centers become more closely spaced, leading to insulating regions. Based on the angle dependence of the conductivity, these charge depleted regions were found affect a ~5 nm diameter region.

Other defects, such as vacancies, may be treated by manipulating the growth conditions. Ariando et al. [212] showed that by varying the oxygen partial pressure, the conductivity of the interface could either be enhanced or suppressed, with a transition to 3D conduction as the oxygen vacancies migrate into the SrTiO$_3$ bulk. La vacancies and La/Al cation intermixing were investigated by Breckenfeld et al. [387], who found that that La-deficient and Al-excessive samples were highly conductive.

Theoretically, a large number of potential intrinsic defects were treated by Lou et al. [388] using DFT. They found that under reducing conditions, paired charged oxygen vacancies are likely to form. These vacancies present in-gap states that may serve to create leakage currents in these devices and were thus found to be the most problematic defect.

## 4.6  Magnetism

SrTiO$_3$ and LaAlO$_3$ are both non-magnetic materials, so the first report of magnetism at the LaAlO$_3$/SrTiO$_3$ interface in 2007 [29] was quite surprising. There are several prominent reported features: (i) In the temperature dependence of the resistance there is a characteristic Kondo-like minimum (Figure 4.5 a), followed by a saturation at lower temperature. (ii) The magnetoresistance shows a distinct hysteresis. Although the hysteresis was later ascribed to a magnetocaloric effect [389] – there were many follow-up reports of magnetism. To date, the consistency of the results is still lacking, as is a comprehensive picture for the origin and mechanism of the magnetism. In this section, we first discuss some of the most notable experimental studies. We then argue that it is necessary to classify the observed evidence into two categories (i) a proper ferromagnetic phase and (ii) a metamagnetic phase that is a result of the multiband nature of LaAlO$_3$/SrTiO$_3$ and many body effects. This topic has also been reviewed recently by Pai et al. [54] and interested readers are encouraged to look there for more details.

### 4.6.1 Experimental evidence

The evidence for magnetism at the LaAlO$_3$/SrTiO$_3$ interface arise from a number of different measurements, including SQUID magnetometry [212], cantilever-based torque magnetometry [390], scanning SQUID magnetometry [391] [340,392], magnetic force microscopy (MFM) [213], X-ray magnetic circular dichroism (XMCD) [51,233,381], and β-NMR [393]. Indirect signatures of magnetism arise in transport studies, and include magnetoresistance anisotropy (Sec. 5.3), giant (negative) magnetoresistance (Sec. 5.3) and anomalous Hall effect (Sec. 5.5).

*Kondo effect*

The interaction between a dilute magnetic moment and itinerant carriers is known as the Kondo effect [394]. A key characteristic is a resistance minimum at finite temperature followed by an upturn and a saturation at lower temperatures. The resistance vs. temperature curve ($R(T)$) follows a universal scaling function $R_K(T/T_K)$ with a single parameter: the Kondo temperature $T_K$. The Kondo effect was one of the main signatures of magnetism in LaAlO$_3$/SrTiO$_3$ first reported by Brinkman et al. [29]. Subsequently, the Kondo effect was widely reported, for LaAlO$_3$/SrTiO$_3$ [295,395,396], electrolytically-gated SrTiO$_3$ [209,397], LaAlO$_3$/EuTiO$_3$/SrTiO$_3$ trilayer structures [398], and in δ-doped LaTiO$_3$/SrTiO$_3$ [399].

The Kondo temperature is found to increase with carrier concentration [209,395]. The observed negative magnetoresistance, measured with a magnetic field that is in-plane and perpendicular to the transport direction, has been explained within the context of Kondo scattering [209], but this negative magnetoresistance may have a different origin, as argued by Diez et al. [400]. Han et al. [401] reported weaker Kondo signatures in samples that were expected to have a higher concentration of oxygen vacancies (growth at lower oxygen partial pressure without post annealing). The explanation was that in the samples with more oxygen vacancies, the conduction networks extended into the bulk and were able to evade the dilute local moments, resulting in the weaker coupling between the iterant carriers and localized moments [401].

The Kondo upturn can be suppressed by light [402,403], even with sub-bandgap excitation of 650 nm [403]. The suppression of the Kondo effect has been attributed to light-induced charge redistribution [402] and destruction of Kondo coherence [403,404].

*SQUID*

Using a SQUID magnetometer, Ariando et al. [212] reported magnetic hysteresis loops of LaAlO$_3$/SrTiO$_3$ (Figure 4.5 b). The hysteresis loop persisted up to room temperature. There was also a monotonic increase of the measured magnetization as the oxygen partial pressure $P(O_2)$ increases from $P(O_2) = 10^{-6}$ until it peaks at $P(O_2) = 10^{-2}$.

*Scanning SQUID*

Bert el al. [391] imaged dipole-like ferromagnetic patches (Figure 4.5 d) by using a scanning SQUID microscope to measure LaAlO$_3$/SrTiO$_3$. The spatial distribution was found to be highly non-uniform. The magnitude of the magnetization and size also varies from patch to patch. Kalisky et al reported a 3 u.c. critical thickness for the magnetic patches [340], which is the same as the insulator-to-metal transition in LaAlO$_3$/SrTiO$_3$ [17]. The patches can also be manipulated by using the scanning SQUID probe to push on the sample surface, and they found the orientation of the patches to align preferentially along the terraces of the sample [392].

The same scanning SQUID magnetometer is capable of measuring diamagnetic susceptibility as well, which can provide a measure of superfluid density [391,405]. The reported superfluid density was found to be tunable via gating from the back of the sample, while the ferromagnetic patches do not shift within the same range of back gate voltages (-70V to 390V). There was also no correlation found between the ferromagnetic patches and the superfluid density [391,405,406]. Curiously, follow-up scanning SQUID experiments by Wijnands et al. [407] found no magnetic patches on samples grown under nominally identical conditions.

*Magnetic torque magnetometry*

Li et al [390] reported in-plane, superparamagnetic-like order of $LaAlO_3/SrTiO_3$ using cantilever-based magnetometry (Figure 4.5 c). The reported magnetization was ~0.3 $\mu_B$ per interfacial unit cell. The magnetization was independent of temperature up to 40K, and found to persist beyond 200K.

*β-detected nuclear magnetic resonance (β-NMR) and neutron scattering*

Salman et al. [393] used β-NMR to study the magnetism of $LaAlO_3/SrTiO_3$. The magnetization of the samples was derived from the relaxation rate of the spin-polarized radioactive $^8Li$ atoms. The spins of $^8Li$ are inferred from the spins of the electron as a product of the β-decay. They reported a magnetization corresponding to $\sim 1.8 \times 10^{-3} \mu_B/\text{u.c.}$, if uniform magnetization is assumed, which is significantly smaller than that reported by Li et al. [390]. However, if spatial variation is taken into consideration, a local density of $\sim 10^{12} \mu_B/cm^2$, which is on the same order of magnitude reported by Bert et al. [391].

*Polarized neutron reflectometry*

Fitzsimmons et al. [408] reported a null result for the magnetic order of $LaAlO_3/SrTiO_3$ using polarized neutron reflectometry. In this method the magnetization is obtained from the asymmetry in specular reflectivity, it being a function of wave vector transfer and neutron beam polarization. The observed magnetization was found to be close to the signal-to-noise limit of the measurement, in fields as large as 11T and temperatures as low as 1.7K.

*Magnetic force microscopy (MFM)*

Using magnetic force microscopy (MFM), Bi et al. [213] reported electronically controlled ferromagnetism up to room temperature at the $LaAlO_3/SrTiO_3$ interface (Figure 4.5 e). The interaction between the magnetized tip and the sample surface alters the amplitude, phase, and frequency of the tip oscillation. A gold top gate enables the interface to be depleted of electrons. Magnetization was found only when the sample was gated into the insulating state, and when the MFM probe is magnetized in-plane. The thickness window for magnetic response was found to be 8-25 u.c [409].

*X-ray magnetic circular dichroism (XMCD)*

The magnetic moment at the interface can be probed locally using X-ray magnetic circular dichroism (XMCD). The XMCD signal comes from the asymmetry in the polarization dependent X-ray absorption spectra caused by the magnetic moment. XMCD signal of the Ti $L_2$, $L_3$ edges was reported by Lee et al. [381] and Salluzzo et al. [51,233]. The corresponding magnetic moment is very weak, from $\sim 0.01 - 0.1$ $\mu_B$ per Ti atom in Ref. [381], or $< 0.01$ $\mu_B$ in Ref. [233]. The above numbers were obtained at $T = 3 - 10$ K. The XMCD signal is not observed at room temperature [381]. While Lee et al. [381] assigned the moment to the $d_{xy}$ orbital of the $Ti^{3+}$ site, the magnetic moment may also be related to the oxygen vacancies, as stronger XMCD signal is observed for the sample with higher concentration of oxygen vacancies [51].

### 4.6.2 Two types of magnetism

The observed magnetic signatures appear to have two distinct origins: (i) proper ferromagnetism due to localized moments antiferromagnetically coupled to the iterant electrons (similar to the picture proposed by Salluzzo et al. [51,233] and (ii) metamagnetism resulting from a combination of the multiband effect, SOC, and many body effects. There are several reasons for treating the metamagnetism separately from the aforementioned ferromagnetism. For example, the proper ferromagnetic long-range order can extend up to room temperature, but the metamagnetism has a reported critical temperature of approximately 35K. Additionally, the reported characteristic field $B_p$ of the metamagnetism is inversely dependent on the carrier concentration. $B_p$ can be as large as 14 T (which would be unphysically large as a coercive field for ferromagnetism), for carrier concentrations $n_e$ approaching the Lifshitz transition $n_L$.

*Ferromagnetism*

Depending on the concentration of iterant carriers $n_e$ and the concentration of the localized moments $n_m$, the phase of the system can be in several states: (1) a nonmagnetic insulator, if $n_m \gg n_e$, i.e., $n_m$ is too high, (2) a ferromagnetic insulator, if $n_m \simeq n_e$ (roughly of the same order of magnitude) where the concentration of the itinerant carriers $n_e$ is enough to support long range ferromagnetic order, which survives to room temperature [51,233], but not enough to form a conducting state, or (3) a Kondo-like conductor without ferromagnetic long range order if $n_m \ll n_e$. In the later state the long-range ferromagnetic order will be destroyed by the flood of iterant carriers. There can also be an accompanying decrease of the concentration of the localized moment $n_m$ due to the global offsets of the chemical potential – if the localized moments are intrinsic, as is the case with Ti3+. The remaining magnetic moments will still be present, as manifested in the characteristic temperature dependent resistance minimum.

The previously discussed ferromagnetism is responsible for most of the direct measurement of magnetization such as scanning SQUID [340,391,392], SQUID [212,396], cantilever magnetometry [390], MFM and MeFM [213,409], and MCD [410]. This picture does not provide any constraint on local variations of the order parameter, and it does not preclude a non-uniform carrier concentration leading to a phase separation as reported by Ariando et al. [212] or a non-uniform distribution of magnetic patches in the scanning SQUID measurements reported by Bert et al. [391]. However, it does account for the null result reported by Fitzsimmons et al. [408], as in that case the carrier concentration was not actively controlled.

This still does not answer what exactly the local moments are. The local moments responsible for the ferromagnetism could be magnetic impurities, or they may have a more intrinsic origin such state localized by oxygen vacancies. The impurity level of substrates varies across vendors but are mostly in the ppm level [411]. Indeed, as in the case of dilute magnetism in semiconductors, a ppm level of magnetic impurities is enough to induce the magnetization that has been reported [411-413]. However, a strong connection between the magnetism and oxygen vacancies was reported by Rice et al. [410]. The magnetic circular dichroism signal was observed only for the samples annealed under low oxygen partial pressures (to introduce oxygen vacancies), and disappears once the samples were re-oxygenated [410]. The oxygen vacancies also strongly affect the magnetism of Fe-substitute $SrTiO_3$ ($Sr(Ti_{1-x}Fe_x)O_{3-\delta}$) [414]. Several theoretical studies have linked oxygen vacancies (or vacancy clusters) with magnetism. Vacancies produce in-gap states [415] with various charge states, vacancy clustering configurations [145], and associated magnetic moments of Ti atoms as a function of their relative position with respect to the oxygen vacancies or vacancy clusters [160]. In addition to the Ti-3d orbital possessing a moment, the F-center (single electrons at the vacancy sites) may also possess a magnetic moment [338,339]. See Sec.

2.4.1 for discussions about the spatial distribution of the oxygen vacancies. Other defects such as Sr-vacancies have also been considered to be responsible for the observed magnetism [416].

*Metamagnetism*

The second type of magnetism, or "metamagnetism," is due to a combination of the multiband effect, SOC, and electron pairing. The observed signatures are the anisotropy of the magnetoresistance (See Sec. 5.3), giant negative magnetoresistance (Sec. 5.3), and the anomalous Hall effect (See Sec. 5.5). The local moments can nonetheless influence this metamagnetism and all of the relevant transport properties, but only via the proximity effect. The reports on magnetotransport are discussed in more detail in Sec. 5.3 and Sec. 5.5, and electronic pairing without superconductivity is discussed in Sec. 6.5.

## 4.7 Superconductivity

Superconductivity in LaAlO$_3$/SrTiO$_3$ heterostructures was first reported by Reyren et al. [30]. The superconducting transition temperature $T_c \cong 200$ mK is close to that of bulk SrTiO$_3$. The in-plane upper critical field $H_{c2}^{\parallel}$ is expected to be greater than out-of-plane upper critical field $H_{c2}^{\perp}$. This result is due to the absence of the flux penetration effect when the field is in-plane [417]. The characteristic thickness, $d_{Tinkham} \sim 10$ nm [418,419] is an order of magnitude smaller than the Ginzburg – Landau correlation length [420] $\xi_{GL} \sim 100$ nm [30,419,421].

From here, Caviglia et al. [31] used back-gated samples to map out a phase diagram (Figure 4.6 a) of the system as a function of temperature and carrier density. The results show an insulating phase at low carrier densities which transitions into a superconducting state at intermediary carrier densities. This phase diagram has been confirmed by other groups [292,422]. As all of these transitions appeared to be Berezinskii-Kosterlitz-Thouless (BKT) transitions, Schneider et al. [423] performed a series of experiments to find the order parameters at the critical point of the transition, fitting the superconductor-insulator transition into the three-dimensional *XY* universality class. This transition is of particular experimental interest as it is one of few experimentally accessible quantum phase transitions. For more details please consult the recent review article by Lin et al. [424]. The dome shape of the $T_c$ as a function of the superfluid density (from scanning SQUID) was also reported by Bert et al. [405] (Figure 4.6 c).

The peak of this superconducting dome is in the vicinity of the Lifshitz transition [383,384]. Using a three-band model with repulsive interactions included through Hartree [11] and Fock [382] corrections, Maniv et al. have shown that for some chemical potential range after the Lifshitz transition, two bands are simultaneously occupied. They then argue that this may indicate that instead of the dome being centered on the Lifshitz transition, superconductivity appears only after the transition and persists until the second band is depopulated by interactions and the system enters a conducting state.

In addition to this superconducting dome, Richter et al. [182] showed that the interface also has a pseudo-gap in the insulating region (Figure 4.6 c). As a result, the temperature-carrier density phase diagram of LaAlO$_3$/SrTiO$_3$ interfaces looks very similar to that of high-$T_c$ superconductors, but at much lower temperatures. By measuring the critical magnetic field of this pseudo-gap, Fillis-Tsirakis et al. [425] argue that a phase coherent paired state persists into the insulating region.

Herranz et al [426] showed that the behavior of the superconducting region is sensitive to the crystal structure by comparing superconducting (001) and (110) LaAlO$_3$/SrTiO$_3$ samples. Due to the anisotropic band masses of the $t_{2g}$ bands, these two orientation have different band structures, with (001) interface having the $d_{xy}$ band as the lowest band, while, for the (110) interface, the $d_{xz}, d_{yz}$ bands are the lowest. In the critical magnetic field measurements performed, this difference in confinement led to substantial changes in the anisotropy of the critical magnetic field, both in-plane and perpendicular to the interface. As a result, the superconducting thickness for the (110) interfaces was found to be $d\sim$24-40 nm, 2-3 times as thick as for the (001) interface.

The electron pairing may be derived from that of $SrTiO_3$ (See Sec. 2.5 for superconductivity for the bulk $SrTiO_3$), such as the proposed BCS-pairing owning to the LO phonons [427,428] of $SrTiO_3$. However, this picture needs to treat the non-adiabatic pairing properly if the Midgal-Eliashberg theory is to be applied (See Sec 2.5). On the other hand, the confinement and reduced dimensionality may give rise to more exotic pairing, such as d-wave pairing from the electron-electron interactions between itinerant and localized electrons across the interface [429], topological superconductivity in the $d_{xz}/d_{yz}$ band from particle-hole fluctuation [430], and surface resonance (oscillatory $T_c$ as a function of the confinement thickness $d$)[428]. With reports on the coexistence emergent magnetism and superconductivity, other scenarios were also proposed. We describe both the experimental reports and proposed theories in the next section.

### 4.7.1 Coexistence of superconductivity and magnetism

The coexistence of superconductivity and ferromagnetism at the sample level has been reported using three different techniques. (1) Dikin et al. [431] reported a hysteretic magnetoresistance at, and outside of the superconducting transition. (2) Li et al. [390] reported in plane magnetization of the sample using cantilever-based torque magnetometry. Superconducting transition of the same sample was observed separately via transport measurement. (3) Bert et al. [391], using scanning SQUID magnetometry, mapped the diamagnetic susceptibility (due to superconductivity) as well as the magnetization. The ferromagnetic patches were not correlated with the superconducting regions.

The question is whether the superconductivity and ferromagnetism truly coexist or phase separated? The latter is suggested by, for example, Mohanta et al. [432]. Locally ferromagnetic puddles could form at the cluster of oxygen vacancies [432]. On the other hand, since the s-wave pairing is not compatible with ferromagnetic order, more exotic scenarios that support true coexistence have also been proposed [433-435]. They are mostly a version of either p-wave pairing, or Fulde–Ferrell–Larkin–Ovchinnikov (FFLO) (finite-momentum pairing) [436,437], and with the help of the Rashba SOC at the interface.

Michaeli et al. [433] suggest *helical* FFLO pairing at the $LaAlO_3$/$SrTiO_3$ interface, with the direction of magnetization perpendicular to the transport direction. The Rashba interaction helps the pairing strength to surpass the Pauli depairing limit. The pairing strength is strongest in the low-disorder limit, with an additional local maximum at intermediate impurity concentration. Band specialization occurs and pairing is assigned to the $d_{xz}/d_{yz}$ band. Maximally superconducting transition temperature is expected to occur at nonzero magnetic field. With Rashba SOC, Banerjee et al. [435] proposed a real-space spiral ordering of spins with pitches of tens of nanometers. This model explains the non-hysteric superparamagnetism reported by Li et al in the torque magnetometry. The patches reported by Bert et al. were assigned to broken spirals due to defects. The ferromagnetic order is thought to be mediated by the $d_{xz}/d_{yz}$ band, while superconductivity is assigned to the $d_{xy}$ band and assumed to be conventional. Notably, polar distortion of (Ti-O-Ti) bond may help stabilizing ferromagnetic domains over spirals. Fidkowski et al. [434] considered a spin-orbit mixed s/p-wave pairing of $d_{xz}/d_{yz}$ at the interface, with $d_{xy}$ mostly localized. The s/p-wave pairing is hence compatible with finite magnetic field, and further mediate superconducting grain in the bulk of $SrTiO_3$ close to interface. The SC order is topological, and a 1D wire with p-wave nature would be expected to have Majorana zero modes [438].

## 4.8 Optical properties

### 4.8.1 Photoluminesce experiments

The photoluminesence properties of LaAlO$_3$/SrTiO$_3$ heterostructures appear to be dominated by the SrTiO$_3$ bulk [377,378]. Thus, at high temperature (>170K) a broad blue peak is observed in the photoluminesence spectrum. At lower temperatures, other suppressed effects become apparent, such as single-particle trapping. This leads to a two-peaked structure in the photoluminesence spectrum in addition to a change in the photoluminescence lifetime, as observed by Yamada et al. [377].

### 4.8.2 Second Harmonic Generation

Second-harmonic generation (SHG) is a nonlinear optical process by which the frequency of reflected light is double that of the incident light [439]. In centrosymmetric materials with broken inversion symmetry, the SHG signal a weighted average of the polar asymmetry felt by the electrons. Thus providing a measure of the nonlinear susceptibility of the electron gas. Savoia et al. [440] used SHG to probe LaAlO$_3$/SrTiO$_3$ interfaces of differing thicknesses. An abrupt transition in the SHG signal was found at 3u.c. thick LaAlO$_3$ and another transition over 4-6u.c. which result from a precursor effect of the polar catastrophe transferring electrons to localized states at 3u.c. followed by the formation of the conducting layer over 4-6 u.c. Using selection rules [441] to pick out specific features of the nonlinear susceptibility in the SHG response by manipulating the polarization of the incident light, Rubano et al. [442] were able to show that the SHG response can also contain information about structural properties, showing that LaAlO$_3$ induces a rotation of the oxygen octahedra as well as a relationship between the SHG signal and the lattice mismatch between SrTiO$_3$, and the LaAlO$_3$, LaGaO$_3$, and NdGaO$_3$ overlayers.

# 5 Quantum transport in LaAlO$_3$/SrTiO$_3$ heterostructures and microstructures

## 5.1 Two-dimensional transport

The physics is drastically different in lower dimensions. In some heterostructures, electrons (or holes) can be confined at the interface. Because the confinement can be very narrow, subbands can form due to quantum confinement of the electron (hole) wavefunction perpendicular to the heterostructures planar dimension (conventionally referred to as the *z*-direction). This quantization leads to steps in the 2D density of states and many other notable and interesting effects. A proper 2DES strictly refers to conditions in which only the lowest subband is occupied [28].

Electrons in 2D or 3D are usually well described by Fermi liquid (FL) theory (reviewed, e.g., by Schultz [443]). However, there are notable departures from FL behavior that are often ascribed to strong electron correlations. There have been many recent reviews on strongly correlated electron systems, see, for examples, [444-447]. In the oxide literature, 2D electrons are often referred to as 2D *electron liquids* (2DELs) [36]. As we will see below, the degree to which electron correlations manifest themselves can depend on many parameters, e.g., magnetic field, carrier density, and temperature. Therefore, in this review we will use the more agnostic term "2D electron system" (2DES) to encompass all behavior that is 2D, regardless of the degree of electron correlation. Systems that are not clearly 2D but confined will be described as *quasi*-2D or quasi-2DES.

## 5.2 Inhomogeneous Transport

Many experiments performed on the LaAlO$_3$/SrTiO$_3$ system indicate that electron transport is spatially inhomogeneous, percolative, or phase-separated [212,448]. Spatially inhomogeneous transport (i.e., channels) were postulated to explain the unusual temperature-independent metal-insulator behavior with a threshold near the conductance quantum $\sigma_0 \sim e^2/h$ (Figure 5.1 a and b) [395,449]. ($\sigma_0 \sim 4e^2/h$ corresponding to the conductance from a single Cooper pair was reported by Caviglia et al. [31,423].) Low-temperature scanning probe measurements identified a candidate mechanism: ferroelastic domains. Two back-to-back reports [450,451], using different local-probe techniques, showed that ferroelastic domains play a major role not only in transport [168,450,452] but in many other aspects. (See Merz et al. [453] and Gray et al. [454] for other characterizations.) Ferroelastic domains have been shown, using scanning SQUID microscopy, to modulate current transport [450]. Ferroelastic domains can form along different crystallographic axes, separated by sharp domain walls (See Sec. 2.1). The domains affect the transport significantly. They not only modulate the current density, they also introduce strong anisotropy (Figure 5.1 c, d, and e) [168], and shift the critical temperature for superconductivity (Figure 5.1 f and g) [452].

Ferroelastic domains can be controlled by gating [455,456]. The domain walls may be charged or polarized [457], as seen under scanning SET [451]. (Readers are referred to the recent review by Salje et al. [60] for studies on the details of the twin walls.) Domain walls may attract defects such as oxygen vacancies (predicted for CaTiO$_3$ [154]). A form of magnetism (different from the ferromagnetic patches under SQUID in, say Bert et al. [391]) also appears to be coupled to the ferroelastic domains, as observed by Kalisky et al. [458]. Ngo et al. [459] performed hysteretic tunneling magnetoresistance measurements with Co/Ti/LaAlO$_3$/SrTiO$_3$ heterostructures. The tunneling magnetoresistance has a two-fold symmetry while the coercive field exhibits four-fold symmetry, and changes configuration only when the sample is warmed through $T_{AFD} = 105$ K.

The existence of ferroelastic domains—or any other mechanism that can lead to highly inhomogeneous electron transport—has the potential to significantly alter the nature of electron transport. The most obvious implication is that the transport can (under certain conditions) be dominated by 1D behavior. The nature of 1D transport is fundamentally different from higher dimensions (see Giamarchi [460] for a modern and approachable overview of the subject). The most notable departure is the breakdown of Fermi-liquid (FL) behavior. The apparent non- FL behavior in SrTiO$_3$-based 2DES may in fact be related to the breakdown of 2D transport and the emergence of 1D (or quasi-1D) transport.

## 5.3 Anisotropic Magnetoresistance

Anisotropic magnetoresistance (anisotropic MR or AMR) is widely reported for LaAlO$_3$/SrTiO$_3$ [212,396,461-467], and related [209,468]. The main findings can be categorized as follows:

(1) *Transport with out-of-plane magnetic field:* most commonly positive MR [7,209,212,396,463,468]. In the small-to-intermediate field regime, the MR arises from weak localization (at lower carrier concentration) or antilocalization (at higher concentration). At larger fields, quantum oscillations are sometimes observed (See Sec. 5.6).
(2) *Transport with in-plane magnetic field, perpendicular to current direction:* the magnetoresistance, though sometimes positive ($\Delta R > 0$) at smaller field, eventually changes sign at large magnetic field [209,212,400,421,463,464]. A negative magnetoresistance as large as $\Delta R/R = -70\%$ was reported [400]. The negative magnetoresistance persists up to $T = 20$ K [400].

(3) *Transport with in-plane magnetic field, parallel to current direction:* MR is usually negative [212].

The in-plane anisotropy of the magnetoresistance is a combination of (2) and (3). However, there is a marked dichotomy of the in-plane anisotropy for carrier concentration $n_e < n_L$ and $n_e > n_L$, where $n_L$ is the Lifshitz point [383,384,469]. As the magnetic field is swept in plane, a sinusoidal magnetoresistance is observed when carrier concentration $n_e < n_L$, as shown in Figure 5.2 b. The anisotropy becomes irregular or of higher harmonics when $n_e > n_L$ (Figure 5.2 c and e). The anisotropy vanishes at $T \sim 35$ K [461]. A similar temperature-dependence of MR anisotropy of SrTiO$_3$ (111) has been reported [466], showing six-fold symmetry [466,467].

## 5.4 Spin-orbit coupling

Gate tunable spin-orbit coupling (SOC) of SrTiO$_3$-based 2DES has been widely reported [421,426,470-475]. Rashba SOC is generally expected due to the inversion-symmetry breaking at the interface, however both k-linear and k-cubic Rashba SOC may be possible. The k-linear Rashba interaction Hamiltonian takes the form $H_{R1} = \alpha_{R1} i(k_-\sigma_+ - k_+\sigma_-)$, where $\sigma_\pm = (\sigma_x \pm i\sigma_y)/2$ and $k_\pm = k_x \pm ik_y$ while the k-cubic Rashba interaction $H_{R3} = \alpha_{R3} i(k_-^3\sigma_+ - k_+^3\sigma_-)$. Theoretical calculations suggest that the Rashba SOC would be dominantly k-cubic in $d_{yz}/d_{xz}$ [370,476,477] with energy scale few meVs to ~20 meV. On the other hand, for the $d_{xy}$ band, the Rashba interaction is predicted to be either k-linear [476], k-cubic or negligible [477].

Spin-orbit interactions have been associated with a variety of experimental signatures, including: (1) weak localization (WL) and/or weak anti-localization (WAL), (2) Violation of the Pauli (Chandrasekhar–Clogston) paramagnetic limit (3) Spin injection or spin-charge conversion such as the inverse Edelstein effect, and (4) Spin- and angle-resolved photoemission spectroscopy (SARPES). The strength of the SOC is generally found to be tunable (by applying gate voltage) in all cases (except (4), in which the tunability is not explored [373]. However, there are some inconsistencies among reports. For example, the dependence of the SOC strength on the gate voltage inferred from WAL [470] and those deduced from the violation of the Chandrasekhar—Clogston limit are opposite to each other [421]. Furthermore, a change of sign of the Rashba coefficient is reported for the case of spin-injection [478] (Sec. Sec. 5.8). The reports with SARPES are also controversial. Santander-Syro et al. [373] reported an extremely large spin-splitting of 100 meV. But the spin splitting is not observed in the follow up work by McKeown Walker et al. [374] (See Sec. 4.4.2.1). Here we briefly discuss the results from the first two methods (using magnetotransport, Table 5.1). The spintronic-related works will be discussed in Sec. 5.8.

**Table 5.1 Reports on the spin-orbit coupling for SrTiO$_3$-based 2DES**

| Report | Sample | Experiment | Carrier concentration $n$ (cm$^{-2}$) | Analysis | Type | Spin-splitting energy $\Delta$ (meV) | Note |
|---|---|---|---|---|---|---|---|
| Caviglia et al. [470] | LaAlO$_3$/SrTiO$_3$ | WAL in out-of-plane magnetoconductance |  | MF | DP | 1 – 10 meV |  |
| Ben Shalom et al. [421] | LaAlO$_3$/SrTiO$_3$ | In plane critical field for superconductivity | $3.0 - 9.5 \times 10^{13}$ | Violation of Clogston-Chandrasekhar limit |  | 0.0 to > 2.0 meV | $H^*$ is used instead of $H_{c2}^\parallel$ |
| Lee et al. [209] | Electrolyte gated SrTiO$_3$ | WAL in out-of-plane magnetoconductance | $6.97 \times 10^{13} - 1.13 \times 10^{14}$ | HLM |  | $H_{SO} = 1.9 - 2.4$ T |  |
| Kim et al. [419] | 1 at.% doped Nb:SrTiO$_3$ d = 3.9 nm – 457 nm | Angular dependence critical field for superconductivity |  | WHH | Neither DP nor EY | 2 meV | Since the δ-doped layer is sandwitched by the buffer layers on the both side, no apparent symmetry breaking presents. The violation of Pauli limit may be intrinsic to SrTiO$_3$ |
| Nakamura et al. | Bulk SrTiO$_3$ | WAL in out-of-plane magnetoconductance | $2 - 8 \times 10^{12}$ | ILP | k-cubic Rashba Not EY | 0.1 – 0.3 meV |  |

| Reference | System | Method | Carrier density (cm$^{-2}$) | Model | Mechanism | $\Delta$ or $\alpha$ | Notes |
|---|---|---|---|---|---|---|---|
| Chang et al. [479] | LaAlO$_3$/SrTiO$_3$ | WAL in out-of-plane magnetoconductance | $2.4 - 6.3 \times 10^{13}$ | MF | | 7.0 meV | |
| Stornaiuolo et al. [472] | LaAlO$_3$/SrTiO$_3$ | WAL in out-of-plane magnetoconductance | | MF | | | |
| Hurand et al. [473] | LaAlO$_3$/SrTiO$_3$ | WAL in out-of-plane magnetoconductance | $1.2 - 2.2 \times 10^{13}$ (offset by $1.69 \times 10^{13}$) | MF | DP | $6.5 - 9.0$ meV | Linear dependence of $\Delta$ on gate voltage |
| Liang et al. [474] | LaAlO$_3$/SrTiO$_3$ LaVO$_3$/SrTiO$_3$ | WAL in out-of-plane magnetoconductance | $0.8 - 4.6 \times 10^{13}$ (LaAlO$_3$/SrTiO$_3$) $1.2 - 5.0 \times 10^{13}$ (LaVO$_3$/SrTiO$_3$) | ILP and HLN | $k$-cubic Rashba (need to set the contribution to from the linear Rashba to 0 to fit the model) | $0.5 - 2.8$ meV (LaAlO$_3$/SrTiO$_3$) $0.5 - 3.0$ meV (LaVO$_3$/SrTiO$_3$) | Non-monotonic gate dependence. The peak is assigned to the avoided crossing. |
| Herranz et al. [426] | LaAlO$_3$/SrTiO$_3$ (001) and (110) | WAL in out-of-plane magnetoconductance | $2 - 8 \times 10^{13}$ (001) $4 - 16 \times 10^{13}$ (110) | MF | | | (110) exhibits weak gate dependence |
| Singh et al. [475] | LaAl$_{1-x}$Cr$_x$O$_3$/SrTiO$_3$; x = 0, 0.1, 0.2 | WAL in out-of-plane magnetoconductance | $8.5 - 12 \times 10^{13}$ | MF | DP | $\alpha \cong 2.0 - 6.5$ meV $\cdot$ nm | No dependence on Cr-substitution. Linear dependence of $\Delta$ on gate voltage |

### 5.4.1 WL and WAL in the magnetoconductance

In order to get a good quantitative match between experimental and theoretical results for the magnetoresistance, spin-orbit effects must be included. These effects fall into two major categories with competing mechanisms of action. The more simplistic mechanism is known as the Elliot-Yafet (EY) mechanism of spin relaxation [480,481]. In this case, the electron spins relax as the spin-orbit interaction causes them to not be in a pure spin state and thus their spin may change with each scattering event, leading to a case where the spin-orbit relaxation time is proportional to the elastic scattering lifetime, $\tau_{SO} \propto \tau_{el}$. The other mechanism, the Dyakonov-Perel (DP) mechanism, is the result of spin precession in an inhomogeneous magnetic field between scattering events and thus the spin-orbit relaxation time scales inversely with the elastic scattering lifetime, $\tau_{SO} \propto 1/\tau_{el}$ [482]. Utilizing these relaxation mechanisms to understand the magnetoresistance of two-dimensional systems has led to two classes of models. The Maekawa-Fukuyama (MF) equation ($k$-independent corrections) [483] and the Hikami-Larkin-Nagaoka (HLN) equation ($k$-cubic corrections) [484] use the EY mechanism to describe the magnetoresistance, while the results of Iordanskii, Lyanda-Geller, and Pikus (ILP) assume the DP relaxation mechanism through a spin-dependent vector potential (has both $k$-linear and $k$-cubic corrections) [485]. Below we describe the experimental results for LaAlO$_3$/SrTiO$_3$ and the applications of these models to understand the dominant relaxation method.

In the first report of Rashba SOC in LaAlO$_3$/SrTiO$_3$ using WAL, Caviglia et al. [470] used MF and obtained gate tunable $\tau_{SO}$ that decreases with gate voltage. $\tau_{SO}$ decreases sharply by 3 orders of magnitude. $\tau_i$ has a power-law dependence and decreases with increased temperature ($\tau_i \propto T^{-p}$ for some power $p$). A WL to WAL crossover can be seen in Figure 5.3 a. The onset of sharp increase of $\Delta$ (decrease in $\tau_{SO}$) roughly matches the gate voltage value for insulator – superconductor transition at lower temperatures (Figure 5.3 b). $\tau_{SO} \propto 1/\tau_{el}$ is observed therefore DP is suggested.

On the other hand, Nakamura et al. [471] used ILP and found the WAL can be fitted only when the k-linear dependence is rejected (leaving only k-cubic term). No $\tau_{SO} \propto \tau_{el}$ is observed therefore EY is excluded. Later, Liang et al. [474] reported a non-monotonic evolution of the coupling strength as a function of gate voltage. Also with ILP, k-cubic Rashba interaction is identified. The Rashba interaction is attributed to the $d_{xz}/d_{yz}$ band, which has spin splitting peaked at the avoided crossing. This assignment also explains the non-monotonic gate dependence of the coupling strength. The opposite dependence on gate reported by Caviglia et al. [470] and Ben Shalom et al. [421] may also be explained if the range covered by them are on two different sides of the maximum.

### 5.4.2 Violation of the Chandrasekhar-Clogston limit in the upper critical field

SOC is one of the possible explanation for violation of Chandrasekhar-Clogston limit. As the in-plane magnetic field is applied, Ben Shalom et al. [421] observed a three-stage evolution of the magnetoresistance as a function of the field, as shown in Figure 5.3 c: (i) superconducting state for $0 < B < \mu_0 H_{c2}^{\parallel}$ (ii) resistive state with nearly constant resistance for $\mu_0 H_{c2}^{\parallel} < B < \mu_0 H^*$ and (iii) a sudden drop of resistance when $\mu_0 H^* < B$. Both $H_{c2}^{\parallel}$ and $H^*$ increase with decreasing carrier concentration. The Clogston–Chandrasekhar limit $H_{c2,BCS}^{\parallel} < 1.76 k_B T_c/(\sqrt{2}\mu_B)$ ($k_B$: the Boltzmann constant; $\mu_B$: the Bohr magneton) is violated at lower carrier concentration. Ben Shalom et al. assigned $H^*$ to SOC, and obtain the energy scale from $g\mu_B H^* = \varepsilon_{SO}$, and spin-orbit scattering time from $\varepsilon_{SO} = h/\tau_{SO}$. $\varepsilon_{SO}$ increases with decreasing carrier concentration. $H^*$ becomes large and inaccessible ($H^* > 18T$) for the experimental setup at carrier concentration $n_{Hall} = 3.0 \times 10^{13}\ cm^{-2}$. The effect of $H^*$ is observed up to $T = 100K$ [421]. Later, in δ-doped SrTiO3, Kim et al. [419] fitted the observed angular dependence of $H_{c2}$ to Werthamer-Helfand-Hohenberg (WHH) theory [486] (which extended the result of Gor'kov on $H_{c2}$ of Type-II superconductors by the including Pauli spin paramagnetism and spin-orbit scattering off impurities) and obtained an energy scale ~2 meV for SOC (Figure 5.3 d). In the thickness dependence of $\tau_{SO}$ and $\tau_{el}$, no traits of DP or EY are observed. Furthermore, since there is no apparent surface inversion symmetry breaking of the δ-doped SrTiO3 in [419] because both sides of the δ layer is covered with undoped SrTiO3 buffer layer, Kim et al. suggest that the observed violation of the Pauli limit is intrinsic to SrTiO3.

## 5.5 Anomalous Hall Effect

For SrTiO3-based heterostructures, a change of the Hall coefficient $R_H$ (slope of the Hall resistance $R_{xy}$) with applied magnetic field (Figure 5.4 a) is commonly observed [7,292,383,384,416,421,462,465,487-491]}[468]. The origin of this nonlinearity is still unclear. It was thought to be due to multiband transport [7,292,465,474]. However, recently Gunkel et al. [416] pointed out that the multiband conduction model was able to capture the field dependence of the Hall coefficient $R_H$ except the small upturn near the zero field, as shown in Figure 5.4 b. Gunkel et al. introduced an additional term $R_0^{AHE}$, with Langevin-type polarizable spin-1/2. The $R_0^{AHE}$ fits the upturn well, but the observed (1) temperature independent saturation field (2) temperature dependent $R_0^{AHE}$ are not expected for the Langevin-type model. Joshua et al. probed this nonlinearity of the Hall resistance $R_{xy}$ with a large in plane plus a small out of plane magnetic field [384] Figure 5.4 c. The critical field $B_p$ for the onset of the change of $R_H$ increases with decreasing carrier concentration when $n_e > n_L$ and diverges as $n_e$ approaches $n_L$.

## 5.6 Shubnikov–de Haas (SdH) Oscillation

Shubnikov–de Haas (SdH) oscillation is very commonly reported (Figure 5.5 a). So far, there have been 15 reports of SdH for SrTiO3-based 2DESs (See Table 5.2). From the archetypical LaAlO3/SrTiO3[7-12], ionic liquid gated LaAlO3/SrTiO3 [13], SrCuO2 capped LaAlO3/SrTiO3 [14], with insertion of 1 u.c. LSMO [245,492]. δ-doped SrTiO3 [208,279,493,494], Mott Insulator interface GdTiO3/SrTiO3 [238].

SdH is a powerful tool to probe the Fermi surface, as the frequency $F$ of the SdH is related to the Fermi surface area via the Onsager relation $F = (\Phi_0/2\pi^2)A)$. Dimensionality and the confinement of the 2DES can also be learned from the angular dependence of the SdH, as no SdH should be observed with an in-plane magnetic field. The 2D-3D crossover at about thickness $d = 64\ nm$ for δ-doped SrTiO3 was calculated by Kim et al. [494]. The 3D conduction nature is indeed observed in two earlier reports [6,336] whose samples were grown at low oxygen partial pressure $P(O_2) = 10^{-4} - 10^{-6}$ mbar without post annealing and carrier concentration far beyond typical range $10^{12} - 10^{14}$ $cm^{-2}$.

Although SdH is commonly observed for this system, a comprehensive picture built upon all of the aforementioned reports has proven difficult. Experimentally determined parameters such as the effective masses vary from $0.62$ $m_e$ to $2.7$ $m_e$. As many as five distinct SdH frequencies have been reported for a single device [14]. Properly assigning the $t_{2g}$ subbands convolved by Zeeman energy, Landau levels, tetragonal energy, atomic and Rashba SOC is challenging.

**Table 5.2 Reports on SdH of SrTiO₃-based 2DESs and related**.

| | Sample | $n_{Hall}$ $(cm^{-2})$ | $n_{SdH}$ $(cm^{-2}), F(T), m_{SdH}$ $(m_e)$ | Mobility $(cm^2/V.s)$ | Note | Band assignment of the observed SdH/QHE |
|---|---|---|---|---|---|---|
| Ohtomo et al. [6] | unannealed LaAlO₃/SrTiO₃ grown under low oxygen partial pressure P(O₂) =10⁻⁴ to 10⁻⁶ torr. | $3 \times 10^{14} - 2 \times 10^{18}$ | | $\mu_{Hall} = 10{,}000$ | Bulk conduction Fano linshape | |
| Herranz et al. [336] | 10⁻⁶ mbar LaAlO₃/SrTiO₃ | $1.6 \times 10^{16}$ | $F = 11.1$ | $\mu_{Hall} = 18{,}000$ | Bulk conduction | |
| Kozuka et al. [208] | 1% δ Nb-SrTiO₃ d = 5.5 nm | $4.7 \times 10^{13}$ | $F = 37.5$ $1.8 \times 10^{12}$ $1.24 - 1.26$ $m_e$  $F = 27.1$ $3.6 \times 10^{12}$  $F = 99.4$ $4.8 \times 10^{13}$ | $\mu_{Hall} = 1{,}100$ | Thickness of superconductivity 8.4 nm (inferred from in plane upper critical field) | |
| Ben Shalom et al. [7] | LaAlO₃/SrTiO₃ | $5 \times 10^{13}$ | $F = 57.5 - 76$ $\sim 10^{12}$ $2.1$ $m_e$ | $\mu_{Hall} = 5{,}500$ and 500 (non-oscillating) | Saw nonlinear Hall | Proposed valley = 3 and spin = 2 degeneracy. |
| Caviglia et al. [8] | LaAlO₃/SrTiO₃ | $1.05 \times 10^{13}$ | $F = 35$ $1.69 \times 10^{12}$ $1.45$ $m_e$  $F = 50$ $2.4 \times 10^{12}$ | $\mu_{Hall} = 2{,}860$[1] | Gate dependence for different sample Shift towards higher frequency | |
| Jalan et al. [493] | δ La-SrTiO₃ d = 3 nm | $3 \times 10^{13}$ | $F = 27, 54$ $1.3 \times 10^{12}$ $1.56$ $m_e$ | $\mu_{Hall} = 1{,}500$ $\mu_Q = 2{,}600$ | Saw nonlinear Hall Shift of 1/B | |
| Lee et al. [209] | Electrolyte gated SrTiO₃ | $2.6 \times 10^{13}$ | $F = 62.5$ $3 \times 10^{12}$ | | | |
| Kim et al. [494] | 0.2% δ Nb-SrTiO₃ d = 11nm - 292 nm | | $F = 100, 1.37 m_e$ $F = 90, 1.36 m_e$ $F = 80, 1.32 m_e$ 18% of $n_{Hall}$  124 nm sample: $F = 110, 1.12 m_e$ 28% of $n_{Hall}$ | | Bulk like in 124 nm sample (SdH crossover at d = 64 nm, c.f. SC at 124 nm) | |
| Moetakef et al. [238] | GdTiO₃/SrTiO₃ | $3 \times 10^{14}$ | $F = 637, 1.01 m_e$ (only low field) $F = 1170, 1.08 m_e$  Two bands model: $3.1 \times 10^{13}, 5.7 \times 10^{13}$  Single band model: $2.8 \times 10^{13}$ | $\mu_{Hall} = 322$ $\mu_Q = 295$ | No nonlinear Hall | Assigned to spin-splitting of $d_{xy}$ |
| Allen et al. [134] | La-SrTiO₃ d = 1280 nm and 800 nm | $d = 1280$ $nm$ $3.6 \times 10^{17} cm^{-3}$  $d = 800$ $nm$ $1.2 \times 10^{18} cm^{-3}$ | 4.2e17 17.5T [111]  1.82e18 55.2 and 9.55 T [001] 1.41 me | $\mu_{Hall} = 37{,}000$  $\mu_{Hall} = 33{,}000$ | | Dispersion and angular analysis |
| Fête et al. [10] | LaAlO₃/SrTiO₃ | $2.5 \times 10^{12} - 4.8 \times 10^{12}$ | 18T 0.87e12 1.25me  55.9T 2.7e12 2.7me  Or 2.2 Rashba band | $\mu_{Hall} = 3{,}900$ $- 6{,}900$ | $g^* = 5.2, -3.4$ | |
| Lin et al. [135] | Bulk, Nb-SrTiO₃ and reduced SrTiO₃ | $10^{17} - 10^{20} cm^{-3}$ | $F_1 = 10 - 200, 1.5 - 4.0 m_e$ $F_2 = 8 - 200, \sim 1.8 m_e$ $F_3 = 7 - 200, \sim 1.5 m_e$ | | | |

---

[1] The 250 mK data. 6,600 is presented in different dataset.

| | | | $F_2$ starts at $1.2 \times 10^{18} cm^{-3}$ $F_3$ starts at $1.6 \times 10^{19} cm^{-3}$ | | | |
|---|---|---|---|---|---|---|
| McCollam et al. [14] | SrTiO$_3$ (2 u.c.)/SrCuO$_2$ (1 u.c.)/LaAlO$_3$/SrTiO$_3$ | $3 \times 10^{13}$ | $F = 8, 0.9 m_e$ $F = 19, 0.9 m_e$ $F = 36, 0.9 m_e$ $F = 83, 2.0 m_e$ | $\mu_Q = 1{,}800$ $\mu_Q = 2{,}400$ $\mu_Q = 1{,}700$ $\mu_Q = 800$ | No apparent harmonics | |
| Xie et al. [9] | LaAlO$_3$/SrTiO$_3$ | $4.7 - 7.5 \times 10^{12}$ | Low Field $F = 20$ $0.48 - 0.58 \times 10^{12}$ $0.72 - 0.9\, m_e$ High field $F = 60$ $1.44 - 1.59 \times 10^{12}$ $1.19 - 1.32\, m_e$ | $\mu_{Hall} = 5{,}500 - 7{,}600$ | QHE $\Delta \nu = 4$ | Magnetic breakdown bands × 4 |
| Chen et al. [245] | a LaAlO$_3$/LSMO(1 u.c.)/ SrTiO$_3$ | $5.5 \times 10^{12}$ | $F = 21$ | $\mu_{Hall} = 8{,}500^2$ | See Trier et al. [492] | |
| Maniv et al. [11] | LaAlO$_3$/SrTiO$_3$ | $1.7 - 2.0 \times 10^{13}$ | $F = 33 - 49$ $1.7 - 2.3 \times 10^{12}$ | $\mu_{Hall} = 6{,}000$ and 500 (non-oscillating) | Nonmonotonous filling of band 2 | |
| Matsubara et al. [279]. | La-SrTiO$_3$ d = 10 nm | $0.8 - 1.2 \times 10^{12}$ | $B < 3.3T$: $F = 6.4$ $3 \times 10^{11}$ $0.62\, m_e$ $B > 3.3T$: $F = 12.9$ $6 \times 10^{11}$ $1.17\, m_e$ | $\mu_{Hall} = 13{,}000 - 18{,}000$ | QHE $\nu = 4, 6$ | 2 bands |
| Trier et al. [492] | a LaAlO$_3$/LSMO(1 u.c.)/ SrTiO$_3$ | $3.0 - 7.8 \times 10^{12}$ | $B < 6T$: $F = 12.8$ $B > 6T$: $F = 25$ $3.8 - 8.5 \times 10^{11}$ $1.03 - 1.40\, m_e$ | $\mu_{Hall} = 5{,}760 - 11{,}416$ $\mu_Q = 2{,}350$ | QHE $\Delta \nu = 10, 20$ | 10 bands |
| Yang et al. [12] | LaAlO$_3$/SrTiO$_3$ | $1.48 \times 10^{13}$ | $1.65 \times 10^{12}, 2.40 \times 10^{12}$ (two samples, both $1.9\, m_e$ and $g^* = 5$) | $\mu_{Hall} = 1{,}872$ $\mu_Q = 203$ | Upturn of parallel at 50T | $d_{xz}/d_{yz}$ |
| Zeng et al. [13] | Ionic liquid gated LaAlO$_3$/SrTiO$_3$ | $5 \times 10^{12}$ | 0.7 me | $\mu_{Hall} = 6{,}600^3$ | | |

*Missing carriers.* In all of the aforementioned reports for SdH at SrTiO$_3$-based 2DESs, the carrier concentration determined from SdH ($n_{SdH}$) is consistently smaller than the carrier concentration determined from the Hall effect ($n_{Hall}$). The ratio $n_{SdH}/n_{Hall}$ ranges from 4% [493] to 90% [279]. There are at least three possible (non-mutually exclusive) explanations. The first explanation is that nontrivial degeneracies may present, from the 3-fold valley degeneracy [7], 2-fold spin degeneracy, to 4-fold magnetic breakdown orbits [9]. The second explanation is that there may exist unknown carriers who suffer from extensive scattering and have a mobility too low to show SdH. This scenario seems to be supported by increased $n_{SdH}/n_{Hall}$ coincides with higher mobility of the samples, larger $\Delta R/R$ of SdH. Also, in the dimensionality crossover comparison by Kim et al. [494], the inconsistency is smaller for the $d = 124$ nm sample, compared to $d = 37$ nm sample. This inconsistency also disappears in the 3D sample ($d = 800$ nm, 1280 nm) in Allen et al. [134]. These low mobility carriers may also be the reason for imperfect QHE reported so far [9,279,492]. A third possibility is that the transport may be dominated by quasi-1D channels (where 2D Fermi surface no longer exists), which may be supported by, for example, the twin walls of the ferroelastic domains. With the transport dominated by quasi-1D character, the observed SdH oscillations may be the steady magnetic depopulation of quasi-1D modes [495,496].

*Subband assignment.* What are the subbands responsible for the observed SdH frequencies? The observed frequencies were often assigned by their effective mass, $d_{xy}$ if $m^*_{SdH}$ is lighter (typically $0.7 - 1\, m_e$)

---

[2] 70,000 was reported for other sample.
[3] Sample A. Other sample D was tuned to 19380.

[238], and $d_{xz}/d_{yz}$ if $m^*_{SdH}$ is heavier ($m^*_{SdH} \gtrsim 2\, m_e$) [12]. Detailed assignment is especially difficult when five frequencies with corresponding energy levels differ by only few tenths of an meV [14].

As the gate voltage increases, the oscillations shift toward larger Fermi surface (due to the population of the carriers [8,10] (Figure 5.5 b). However, at higher carrier concentration, Maniv et al. [11], reported a non-monotonic behavior of the SdH frequency as a function of the gate voltage $V_G$. The SdH frequency decreases after $V_G = 3V$. Maniv et al. suggested that the SdH is from $d_{xz}/d_{yz}$, and they are repelled by $d_{xy}$ at higher carrier concentration. For follow up, see Smink et al. [385] and Maniv et al. [382]. Notably, the carrier concentration supported in GdTiO3/SrTiO3 [238] is an order of magnitude larger than all the other reports.

As the field increases, various types of change in $F$ has been reported. For example, Jalan et al. reported a shift of the frequency [493]. The two Rashba branches $F^+$ and $F^-$ shift in opposite directions with the increased field in Fête et al. [10]. For harmonically related frequencies, the lower harmonic disappears or changes into higher harmonic after some field strength, most commonly $F \to 2F$ [238,279,492], except $F \to 3F$ in [9]. The $F \to 2F$ is usually explained with spin-splitting, except for Ref. [279]. We discuss the work of Xie et al. [9] and Matsubara et al. [279] in the next section.

## 5.7 Quantum Hall Effect

Quantum Hall effect (QHE) is still elusive for SrTiO3-based 2DESs. Instead of perfectly quantized at $G_{xy} = \nu e^2/h$ at integer filling factor ν, the observed plateaus of the Hall conductance $G_{xy}$ is often not perfect, limited to higher filling factors, and with nontrivial degeneracies, such as $\Delta \nu = 2$ in Ref. [279], 4 in Ref. [9], and 10 or 20 in Ref. [492].

To explain the observed $\Delta\nu \sim 10, 20$ (raw value $\nu = 30.6, 41.0, 53.5, 68.5, 88.5$, etc), Trier et al. [492] proposed a picture consisting of 10 simultaneously populated bands with $s = \pm 1/2$ degeneracy. At $B < 6T$, $\Delta\nu \sim 20$ and SdH frequency $F = 12.8T$ is observed. At $B > 6T$, on the other hand, the plateaus become $\Delta\nu \sim 10$, and SdH frequency becomes twice as large, at $F = 25T$. This is attributed to the splitting of the spin degeneracy. The 10-band picture is also used to explain $n_{Hall} \sim 10\, n_{SdH}$. As the gate voltage increases, the second band (of the 10 bands) has largest response.

Xie et al. [9] reported $\Delta\nu \sim 4$ with raw values $\nu = 23, 26.7, 31.5, 35, 38.4, 42$ (one of the four sets of presented data) for field $B > 7\,T$ (Figure 5.5 c). A change of SdH frequency from $F = 20T$ at $B < 4.3T$ to $F = 60T$ (three times as large) at $B > 4.3T$ is observed (Figure 5.5 e). Xie et al. proposed a magnetic breakdown orbit between the two lowest lying band (the assignment/nomenclature of $d_{xy}$ or $d_{xz}/d_{yz}$ was avoided), as shown in Figure 5.5 d. The orbit has 4-fold degeneracy, to account for the $\Delta\nu \sim 4$. The orbit also has to have Fermi surface area $A_{MB} = 3A$, where $A$ is the area of before the magnetic breakdown for $B < 4.3T$, to explain the three fold change of the SdH frequency.

Matsubara et al. [279] observed $\nu = 4, 6$, which are currently the lowest filling factor observed in SrTiO3-based system so far. The $\Delta\nu = 2$ is attributed to spin degeneracy. There are two relevant subbands in the picture (called EB1 and EB2, EB1 has lower energy). Compared to $d_{xy} = EB1$ and $d_{xz}/d_{yz} = EB2$ assignment for interface 2DESs, Matsubara et al. pictured that the lowest lying EB1 would have more $d_{xz}/d_{yz}$ character. Both of the bands are dominated by $d_{xz}/d_{yz}$ at the Fermi level. A change of the SdH frequency from $F = 6.4T$ at $B < 3.3T$ to $F = 12.9T$ at $B > 3.3$ is observed. Instead of explaining this by the lifting of the spin-degeneracy, Matsubara et al. argued that the SdH is dominated by EB2 for $B < 3.3T$. As the Landau index would be relatively high for EB1 therefore the amplitude would be smaller.

On the other hand, EB1 dominates $B > 3.3T$, as the quantum limit for EB2 may have been reached. This interpretation requires the Fermi surface to satisfy $2A_{EB1} = A_{EB2}$.

## 5.8 Spintronic Effects

Oxide heterostructures show increasing promise for spintronic [497,498] applications. The concept of semiconductor spintronics emerged from the successful application of the electron spin degree of freedom for data storage and sensing applications. However, as a full-blown replacement for electron charge, many effects are still missing. In this section we review efforts to measure effects related to spin injection, polarization, diffusion and detection. Many of the experiments performed so far were inspired by experiments first performed with semiconductors like silicon and GaAs.

Large, gate controllable Rashba SOC (Sec. 5.4) makes SrTiO$_3$-based 2DES attractive for possible spintronic applications such as spin field effect transistors. Possible integration of gate controllable phases of LaAlO$_3$/SrTiO$_3$ may enable more functionality. There have been several attempts to measure spin polarization via the Hanle effect [499-505]; spin-charge conversion by spin pumping followed inverse Edelstein effect (IEE) [478,506].

Hanle measurements are often carried out with a three-terminal geometry (Figure 5.6 a) with a ferromagnetic–insulator–2DES junction. Current is sourced between the ferromagnetic top electrode and one of the interface contacts, and the nonlocal voltage is measured from the third terminal (Figure 5.6 a and b). The tunneling magnetoresistance is also measured as a function of out-of-plane magnetic field. A Lorentzian lineshape is expected from the Hanle effect: when magnetic field $B^{\perp}$ is applied, spins of the carriers undergo Larmor precession at $\omega_L = g\mu_B B^{\perp}/\hbar$, and creates spin-imbalance voltage drop. The spin lifetime can be calculated from the width of the Lorentzian lineshape. The obtained spin lifetime $\tau_{sf}$ ranges from a few ps to few tens of ps. For example, $\tau_{sf} = 50$ ps at $T = 2$ K and length scale $l_{sf} = \sqrt{D\tau_{sf}} \approx 1$ µm in Reyren et al. [500]. Gate-dependence of $\tau_{sf}$ is reported by Kamerbeek et al. [504] and Han et al. [501] but not Inoue et al. [503]. Both the thickness of the insulating layer and the tunneling resistance are reported to be critical parameters [503]. While the Hanle signal is observed, it may not necessarily arise from spin accumulation. However, whether there exist spin accumulation with lifttime long enough to account for the Lorentzian lineshape is a question, see Ref. [503] for discussions.

Efficient spin–charge conversion is reported by Lesne et al. [478] (Figure 5.6 c and d) and later by Song et al. [506]. Both of the reports used spin-pumping to create spin imbalance at the interface. The spin imbalance is then converted to charge by Rashba SOC (the IEE effect). The efficiency is not only gate-controllable, but also unprecedentedly high, both as a 2DES and as a 3D equivalent. Lesne et al. apply spin pumping to NiFe/LaAlO$_3$(2 u.c.)/SrTiO$_3$ at $T = 7$ K. The spin-to-charge conversion efficiency, described by $\lambda_{IEE}$ (with peak value 6.4 nm) for 2D materials. To compare with 3D materials, $\lambda_{IEE}$ can be calculated from the spin Hall angle ($\vartheta_{SHE}$) for 3D materials via $\lambda_{IEE} = \vartheta_{SHE} l_{sf}$, where $l_{sf}$ is the spin diffusion length. $\lambda_{IEE} = 6.4$ nm for LaAlO$_3$/SrTiO$_3$ is more than ten times larger than the $\lambda_{IEE} = 0.43$ nm for W, which is known for large spin Hall angle. Furthermore, $\lambda_{IEE}$ and $\alpha_R$ (Rashba coupling strength) are tunable by gate voltage. The evolution of $\alpha_R$ is non-monotonic and changes sign after a critical gate voltage (Figure 5.6 d). Lesne et al. suggested that the sign and the magnitude of $\alpha_R$ may be different for $d_{xy}$ and $d_{xz}/d_{yz}$. Then if the critical voltage corresponds to the Lifshitz point $n_L$, the $\alpha_R$ may be dominated by $d_{xy}$ at lower gate voltages and by $d_{xz}/d_{yz}$ at higher gate voltage. The non-monotonic evolution is also reported by Song et al. [506] at $T = 300$ K, but the change of sign was not observed [478]. The IEE signal vanishes as the 2DES is depleted [506].

# 6 Quantum transport in LaAlO$_3$/SrTiO$_3$ nanostructures

Having discussed 2D quantum transport in the previous section, here we survey the recent observations that have been made possible by devices with reduced dimensionality, where the physics can be profoundly different. The fact that the boundary can have fundamental effects in semiconductors (or metals) is first recognized by David J. Thouless [507]. For LaAlO$_3$/SrTiO$_3$ or related 2DES, Goswami et al. [508] reported the first superconducting quantum interference devices (SQUID) in this system (Sec. 6.4). Superconducting single-electron transistors (SSET) formed at the LaAlO$_3$/SrTiO$_3$ interface have provided new insights into the physical origins of electronic phases. For example, a paired, non-superconducting phase is observed in the SSETs, with a pairing field more than an order of magnitude larger than the superconducting upper critical field (Sec. 6.5). At higher carrier density, a crossover from attractive to repulsive electron-electron interactions is observed (Sec. 6.6). Narrow conducting channels can exhibit universal conductance fluctuations (Sec. 6.2), Fabry-Perot interference effects, and conductance quantization in units of $e^2/h$ (Sec. 6.3).

## 6.1 Quasi-1D Superconductivity

The quasi 1-D superconductivity of c-AFM-defined nanowires is reported by Veazey et al. [310]. The width of the nanowires is $w \sim 10$ nm at room temperature, an order of magnitude smaller than the superconductivity coherent length $\xi_{SC} \sim 100$ nm [30]. The reported transition temperature is slightly lower than that of the "2D" counterpart, at $T_c \sim 200$ mK. The transition is wider and with a finite residual resistance ($\sim 1.2 - 5.0$ k$\Omega$). Disorder, local hotspots, or thermally-activated phase slips may contribute to the residual resistance. It is worth noting that, even though the samples are all atomically-flat under AFM, the properties for the superconducting nanowires created with c-AFM lithography created at different locations of the sample are different [310]. This may be related to the ferroelastic domains or other sources of inhomogeneity (e.g., oxygen vacancies). Wider structures created by e-beam with width $w \sim 500\ nm$ are reported by Stornaiuolo [293], for which UCF is observed when the superconductivity is suppressed by gating.

With a c-AFM defined Hall-bar structure, nonlocal transport is reported by Veazey et al. [509] and Cheng et al. [309]. A change of sign of the nonlocal resistance is observed, but the mechanism is not understood. Recently, nonlocal response attributed to Rashba-induced charge-to-spin conversion is reported [505].

## 6.2 Universal Conductance Fluctuations

Universal conductance fluctuations (UCF) [510] are reproducible, seemingly random, device specific oscillations in the conductance [511]. UCF is a manifestation of coherent scattering of electrons and is closely related to weak localization (WL). The major distinction is that only the averaged interference between the electrons of time-reversed paths is concerned in WL. UCF, on the other hand, is a result of interference between all possible electron paths, which depends on magnetic field, carrier concentration, impurity configuration, etc. UCF is therefore device specific, and the magnetotransport signature is often known as a *magnetofingerprint*. UCF is characterized by two length scales: the dephasing length $L_\varphi$ and the thermal length $L_T$.

UCF is observed in SrTiO$_3$-based devices with reduced lateral dimensions (typically with width on the order of a micron) [293,512-514]. The reported dephasing length are around $L_\varphi \sim 100 - 300$ nm at low temperature [512,513]. Note that the extrapolated dephasing rate $\tau_\varphi^{-1}$ does not vanish at $T = 0$ K for both reports [512,513]. Stanwyck et al. [513] suggested that Kondo scattering may occur, though a Kondo resistance minimum is not reported. UCF is also reported for the superconducting critical current of a Josephson junction created from LaAlO$_3$/SrTiO$_3$ [514].

## 6.3 Ballistic transport and dissipationless electronic waveguides

Over the past few years, various quasi-1D channels with micron-long coherence lengths have been reported [32,33]. The clean limit of a 1D quantum wire is particularly important for investigating Luttinger liquid physics and the nature of electron-electron interactions. Ron et al. [32] created parallel channels with alternating 1 u.c. and 3 u.c. of LaAlO$_3$ top layers. Conductance in steps of both $e^2/h$ and $2e^2/h$ were observed for the device with 28 parallel 4 μm-long channels, as shown in Figure 6.1 a and b. Ron et al. [32] suggested that the transport happens at the interfaces between those 3 u.c. capped sections and and 1 u.c. capped sections. Tomczyk et al. [33] saw Fabry–Pérot-like quantum interference in c-AFM defined electron cavity with length $0.25 - 4.0$ μm (Figure 6.1 c and d). Both reports suggest that the electron coherence length may be much longer than the value for superconducting order parameter $\xi_{SC} \sim 100$ nm obtained in early 2D report [30].

Electron waveguides with fully quantized lateral and spin modes were reported by Annadi et al. [312] recently. Figure 6.2 a is the typical design of an electron waveguide, created with c-AFM lithography. Figure 6.2 b shows quantized conductance plateaus in $e^2/h$, as a function of a sidegate voltage or chemical potential, at $T = 50$ mK. The subband structure is revealed from the transconductance $dG/dV_{sg}$ spectra (Figure 6.2 d). The subband structure can be described by lateral and vertical spatial quantum numbers as well as the spin degree of freedom. Figure 6.2 c shows the calculated wave functions for the lowest six subbands. Below a pairing field $B_p \sim 1$ T, the lowest two subbands are paired (but not superconducting), as reported by Cheng et al. [187] (Sec. 6.5). Reentrant pairing of different subbands at finite magnetic field is also observed, such as the $|0, 1, \downarrow\rangle$ and $|0, 1, \uparrow\rangle$ shown in Figure 6.2 d. States with the same spin exhibit avoided crossing, e.g., between $|0, 1, \downarrow\rangle$ and $|1, 0, \downarrow\rangle$.

## 6.4 Superconducting Quantum Interference Devices (SQUID)

Goswami et al. [508] reported the first SQUID devices at the LaAlO$_3$/SrTiO$_3$ interface. Two types of SQUIDs were fabricated- one with constrictions defined by lithography, the other with weak links from the nearly depleted 2DES right underneath a biased topgate (Figure 6.3 a). The oscillation frequency scales as expected with the area of the SQUIDs. The kinetic inductance is few orders of magnitude greater than the geometric inductance, and can be tuned by backgate. SQUID-like modulation of the critical supercurrent is also observed in SCO-capped rings by Aurino et al. [515]. The oscillations have minima at zero magnetic field, possibly an indication of ferromagnetic Josephson junction or p-wave pairing [515]. Josephson junctions are reported by Monteiro et al. [514], but the Fraunhofer patterns for the Josephson junctions were not reported.

## 6.5 Electron pairing without superconductivity

For the 2DES in LaAlO$_3$/SrTiO$_3$, strong electron pairing outside of the superconducting phase has been observed [187]. This phase was proposed by Eagles et al. [184] for SrTiO$_3$ (See Sec. 2.5). To probe electron pairing, Cheng et al. [187] used the single electron transistor (SET) geometry, as shown in Figure 6.4 a. Figure 6.4 b is the resulting differential conductance ($dI/dV$) as a function of side-gate voltage ($V_{sg}$) and out-of-plane magnetic field $B$. The features in the $dI/dV$ are due to paired or single electron tunneling, as described below.

As the field is increased from $B = 0$ T, several phases are observed for this representative device: (i) $B = 0$ T ~ 0.2 T: superconductivity. The SET is in superconducting state, and a peak due to Josephson current marked by black triangles. (ii) $B = 0.2$ T ~ 2.0 T: electron pairing without superconductivity. The transport is still dominated by paired tunneling, even though the superconductivity is destroyed by the out-of-plane magnetic field $B > \mu_0 H_{c2}^\perp$. Electrons remained paired until $B > B_p$. (iii) $B = 2.0$ T ~ 5.0 T: single electron tunneling. The electron pairs are destroyed by $B > B_p$, and the single electrons states are Zeeman-split by the applied magnetic field. (iv) $B = 5.0$ T ~ 6.0 T: reentrant pairing. The range for the reentrant pairing varies from subbands to subbands and device to device. The single-electron tunneling peaks, split from two different neighboring conductance diamonds, lock together and remain paired for about $\Delta B \sim 1.0$ T. The reentrant pairing is also observed in electron waveguides [312], see Sec. 6.3. Notably, the pairing field $B_p$ is very strong, compared to the upper critical field $\mu_0 H_{c2}^\perp$, and increases with decreasing carrier concentration. $B_p$ as large as ~ 11 T has been observed in an electron waveguide device [312] (Sec. 6.3).

The pair splitting is also observed by Prawiroatmodjo et al. [516], while Maniv et al. [294] reported otherwise. One possible explanation for the null result of Maniv et al. [294] is that the pairing field may exceed the maximum magnetic field applied (6 T) in the report.

## 6.6 Tunable Electron-Electron Interactions

In Section 6.5, the magnitude of the electron-electron interaction that gives rise to pairing was shown to be tunable as the pairing field changed with backgate voltage. But the sign of this interaction can also change. Cheng et al. [517] observed a distinct crossover from conductance diamonds to Andreev bound states in the differential conductance map of SSETs. Figure 6.5 b is an example of the crossover from conductance diamonds to loop-like features. At lower chemical potential, the electron-electron interaction is attractive, characterized by Hubbard $U > 0$. The pairing is local and BEC-like. As the chemical potential is raised by the increasing sidegate voltage, the interaction becomes repulsive with $U < 0$. The pairing becomes nonlocal, BCS-like, and the Andreev bound states therefore show up.

To explain this crossover, Cheng et al. [517] proposed that the effective interaction may have different signs in the $t_{2g}$ manifold: attractive in $d_{xy}$, and repulsive in $d_{yz}/d_{xz}$. Thus, at lower chemical potential when only $d_{xy}$ is populated, the effective interaction is attractive. As the chemical potential increased by sidegate, the $d_{yz}/d_{xz}$ start to be populated after the Lifshitz point, the effective interaction becomes repulsive. A tunable electron-electron is a useful feature that could potentially be used for quantum simulation.

# 7 Outlook

## 7.1 Outstanding physics questions

### 7.1.1 Polar catastrophe (or not)

A significant collective effort has taken place in which the existence and nature of the "polar catastrophe" in LaAlO$_3$/SrTiO$_3$ has been analyzed and questioned. The simplest picture of an ideal polar discontinuity is clearly more complex. However, it is worth emphasizing that much, if not all, of the interesting 2D electronic behavior depends only on the existence of a conducting interface, not on its origin. There are many remaining issues related to the conductive interface, in particular the low electron density as compared with the prediction of the polar catastrophe.

### 7.1.2 Coexistence of phases

One of the outstanding issues relates to the coexistence of various phases and understanding whether the coexistence takes place only in an average sense or whether the coexistence is intrinsic. One of the most striking examples is the coexistence of superconductivity and magnetism. Barring an unconventional pairing mechanism, the superconducting state should be incompatible with the existence of magnetism, leading to the conclusion that these two phases are separated in space. Local probes with high spatial resolution may be able to answer this question definitively.

### 7.1.3 Novel superconducting states (e.g., FFLO, other pairing symmetries)

The mechanism of electron pairing in $SrTiO_3$ remains unsolved, even more than a half century after its discovery. Recent experiments have shown that electron pairing is strong enough to survive in the absence of long-range superconducting order, which implies that it is either near unitary or in the BEC regime. Understanding the physics underlying electron pairing is central to many important questions and future applications.

### 7.1.4 Magnetism mechanism

The origin of magnetism is also an important outstanding question. As argued recently by Pai et al [438], there are likely two separate forms of magnetism that have been treated without distinction in the literature: one involving the ordering of magnetic moments that are related most likely to defects (i.e., oxygen vacancies), and the other related to electron-electron interactions. The former leads to a proper ferromagnetic phase, and has been reported at room temperature, while the latter appears only below ~35 K in magnetotransport measurements. A key question is to determine definitively the source of the magnetic moments. Future experiments in which oxygen vacancies are controlled in a given sample, and in which magnetism is probed locally, will help provide more convincing evidence, paving the way for future applications involving electronic control over magnetic properties.

### 7.1.5 Exotic phases (eg. Majorana physics)

The ability to modulate electronic properties, combined with intrinsic properties including magnetism, superconductivity, SOC, and low dimensionality, represents a powerful collection of properties with the potential for engineering new and potentially useful phases of quantum matter. The theoretical framework that motivated the first signatures of Majorana fermions in semiconductor nanowire-based devices [518] requires control over properties that appear to also be present in the $LaAlO_3/SrTIO_3$ system. However, no claims about Majorana physics have yet been made.

### 7.1.6 Luttinger liquids

The physics of one-dimensional electron systems will undoubtedly benefit from a new physical system that can create Luttinger liquids on demand. Luttinger liquids are strongly interacting phases in which the relevant excitations are derived from collective excitations of the strongly interacting system. Charge and spin degrees of freedom behave independently, with their own dispersion relations, and the charge itself can fractionalize. Ideal spinless (or spin-polarized) Luttinger liquids are routinely created in $LaAlO_3/SrTiO_3$ using conductive AFM lithography. The ability to sketch 1D quantum wires will enable new probes of this fascinating class of electronic materials, one which differs profoundly from transport in two or higher dimensions.

## 7.2 Future applications

### 7.2.1 Spintronics

Complex oxide heterostructures offer new possibilities for applications in spintronics. Recent reports of long-lived spin diffusion [504] and transport [519] in $LaAlO_3/SrTiO_3$ at room temperature suggest that spin-based electronic devices may indeed serve as building blocks for storing or transmitting spin-based information. It will be important to understand much better the electronically controlled ferromagnetic phases that exist at this interface.

### 7.2.2 Quantum simulation

One exciting prospect associated with the reconfigurable $LaAlO_3/SrTiO_3$ interface relates to its application in quantum simulation in 1 or 2 spatial dimensions. The precision with which the potential landscape $V(x, y)$ can be controlled can approach the mean spacing between electrons [303], meaning that in principle new lattice structures can be created for the purposes of constructing a Hubbard-model Hamiltonian. The Hamiltonians may describe the physics of high-temperature superconductors or other strongly interacting fermionic matter (e.g., lattice QCD). At the moment, the exact form of this Hamiltonian is not known, which means that structures can be created but it is not known what they are simulating. As foundational knowledge of the physics of the $LaAlO_3/SrTiO_3$ interface increases, it will be possible to "close the loop" between theory, models, and experiments, enabling simulation of Hamiltonians that are currently intractable using known analytical or numerical methods. Being able to perform this kind of simulation can potentially lead to important conceptual advances in the study of quantum matter.

### 7.2.3 Qubits/quantum computing

At the moment, it is not clear how oxide heterostructures can be used to store qubits or perform gate operations for a quantum computer. One of the reasons is that the requirements for a spin-based quantum computer are well-defined two-level systems. Such physical qubits have not yet been identified, in part due to the high electron density. It may be possible in the future, but there are also issues related to independent gating of these spin qubits. Topologically-protected qubits (e.g., Majorana fermions) may be possible, but they too have not been demonstrated. Systems with flying qubits along quantum channels may be possible, or it may be possible to construct hybrid systems in which storage qubits come from, e.g., N-V centers, and are coupled by quantum wires. Or it may be possible to create a superconducting qubit. A lot of work is left to do along this front, but it definitely seems worth pursuing.

### 7.2.4 Sensing

Quantum sensing seems to be a possible application for quantum nanostructures. Electron waveguides are sensitive to disorder, and it may be possible to sense single spins that are nearby (although the dipole-dipole coupling is going to be weak). They can be used as nanoscale THz sensor (and sources).

# 8 Acknowledgements

We thank ONR N00014-15-1-2847 for financial support.

# 9 Figures

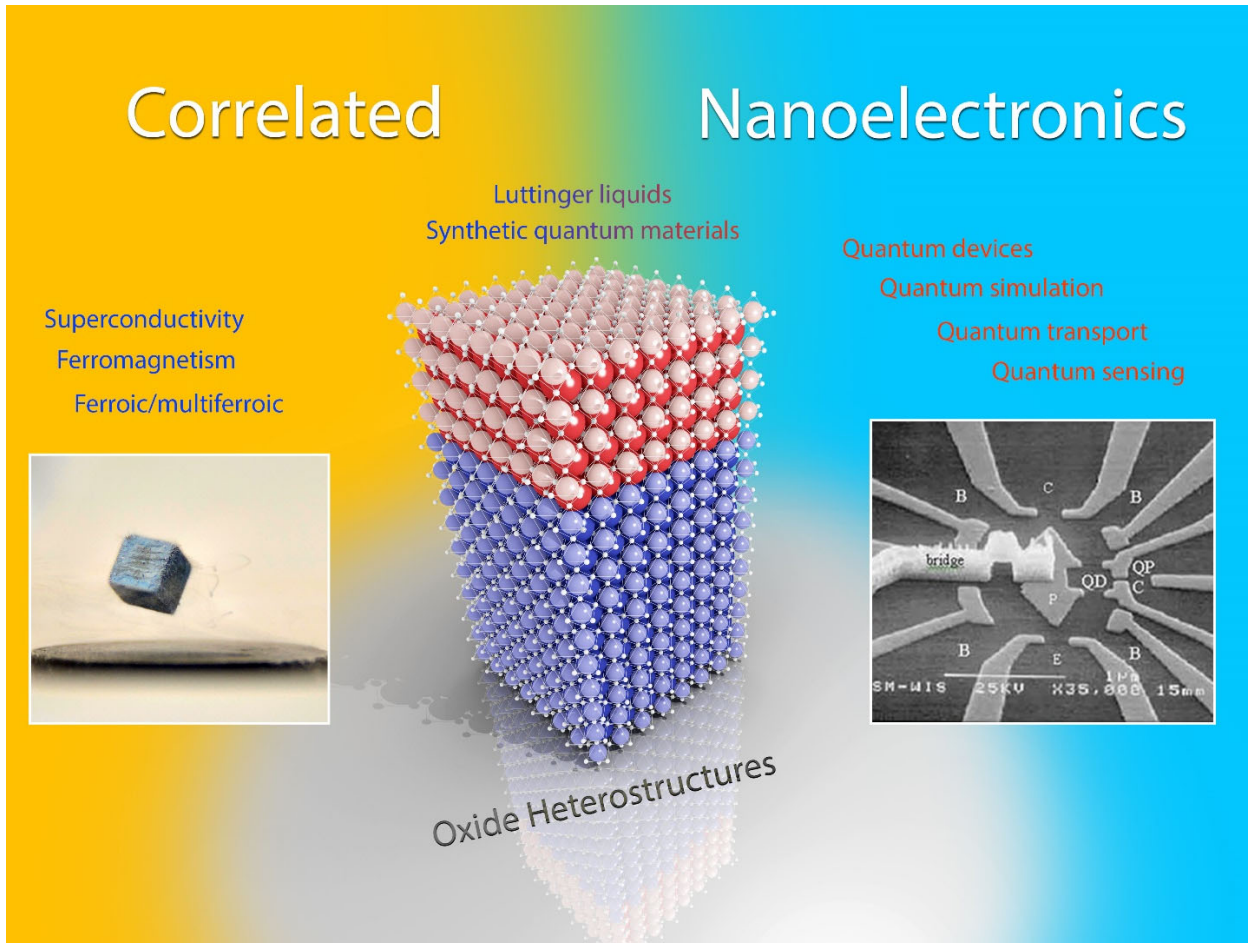

**Figure 1.1 Correlated Nanoelectronics**: a real interface at the metaphorical interface between complex-oxide heterostructures and nanoscale quantum devices. *Left Inset*: levitating magnetic on a high $T_c$ superconductor. *Right Inset*: A which-path device created out of 2DES at GaAs-AlGaAs. See [520]. (Reprinted by permission from Macmillan Publishers Ltd: *Nature* [520], copyright 1998.)

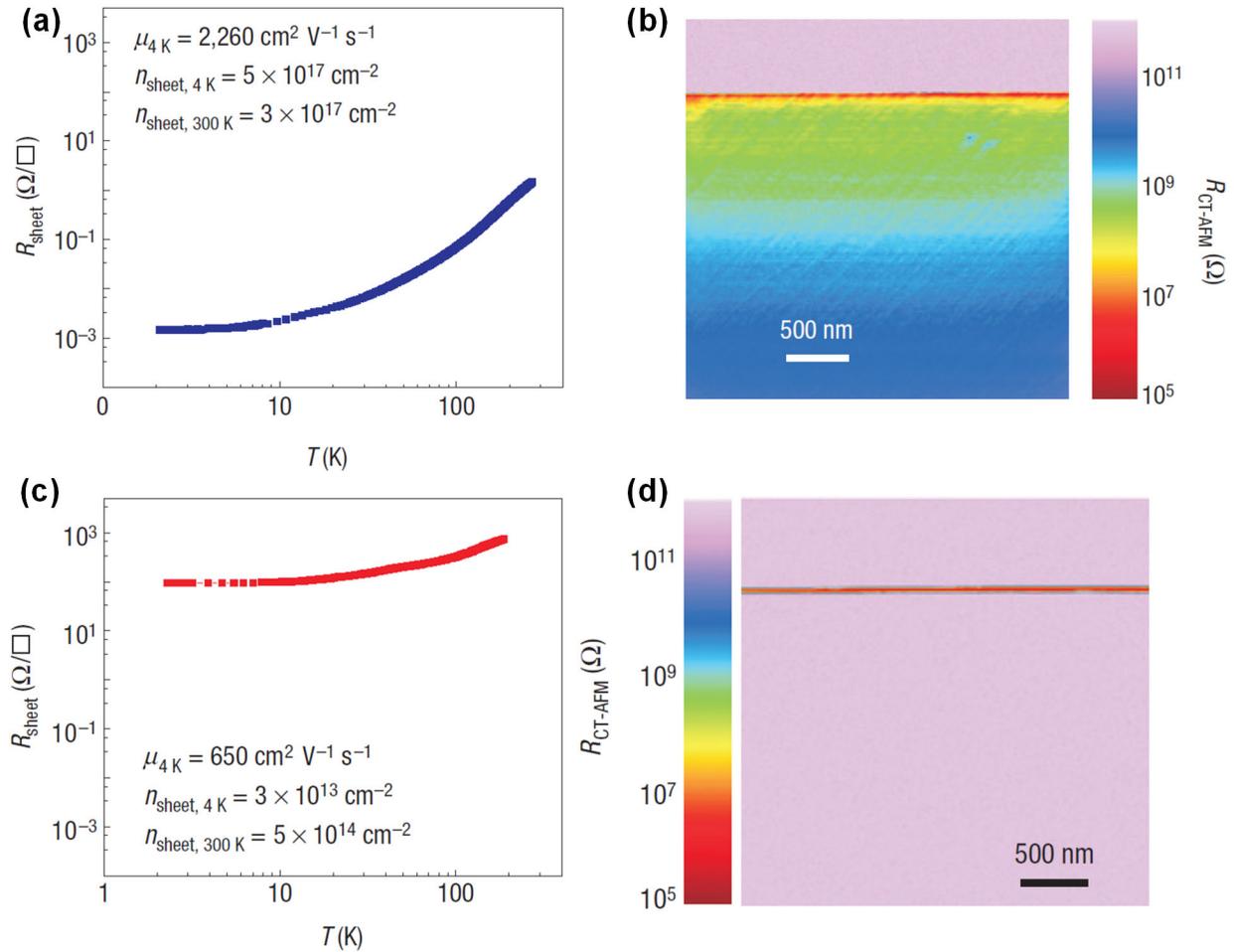

**Figure 1.2 Electron profile probed by Conductive-tip AFM (CT-AFM), for the unannealed and annealed samples**. (a) The sheet resistance as a function of the temperature for the unannealed sample. The 2D carrier concentration $5 \times 10^{17}$ cm$^{-2}$ is unphysically high [6]. (b) The electron profile for the unannealed sample seen under CT-AFM. The electrons are not well confined and spread into the SrTiO$_3$ bulk. (c) The sheet resistance for the sample annealed at high oxygen partial pressure ($P(O_2) =$ 400 mabr) after the growth. (d) The CT-AFM image for the annealed sample. The electrons are confined at the interface with width $\sim$ 7 nm. ((a)-(d): Reprinted by permission from Macmillan Publishers Ltd: *Nature Materials* [18], copyright 2008)

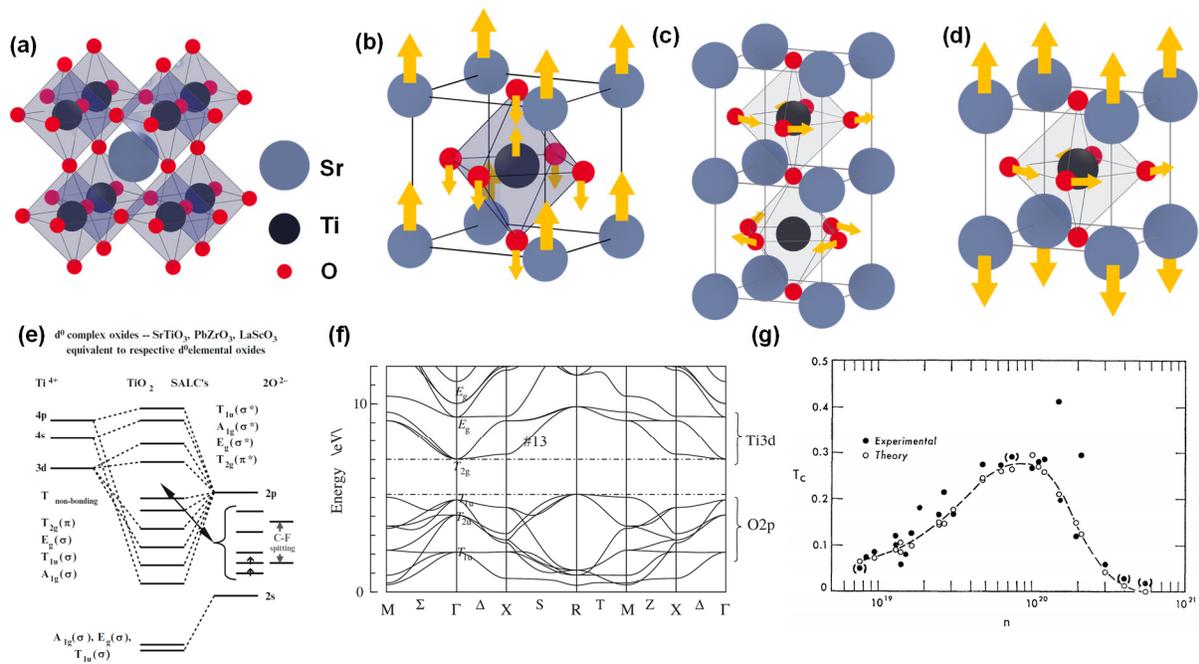

**Figure 2.1 Properties of SrTiO₃.** (a) The crystal structure of SrTiO$_3$. (b) Ferroelectric displacement. (c) Antiferrodistortive rotation. The rotation of neighboring oxygen octahedral are antiphase to each other. Rotation about the c-axis is shown. (d) The tetragonal transition due to AFD rotation. (e) The orbital hybridization of SrTiO$_3$. The conduction band is from the Ti 3d orbitals, and most importantly, the t$_{2g}$ manifold. The primary contribution of the valence band is the $2p$ orbitals of oxygen atoms. [123] (Reprint from *Long Range Cooperative and Local Jahn-Teller Effects in Nanocrystalline Transition Metal Thin Films*, Volume 97 of the series Springer Series in Chemical Physics pp 767-808, with permission of Springer.) (f) An DFT calculation of the electronic structure for cubic phase SrTiO$_3$. The $t_{2g}$ are the lowest lying conduction bands. The #13 is the $d_{xy}$ band. [124] (Reprinted from *J. Appl. Phys.* **113**, 053705 (2013), with the permission of AIP Publishing.) (g) The superconducting transition temperature $T_c$ of SrTiO$_3$ traces out a dome shape [169] (Reprinted figure with permission from C. S. Koonce, Marvin L. Cohen, J. F. Schooley, W. R. Hosler, and E. R. Pfeiffer, Physical Reviews, **163**, 380. Copyright 1967 by the American Physical Society.)

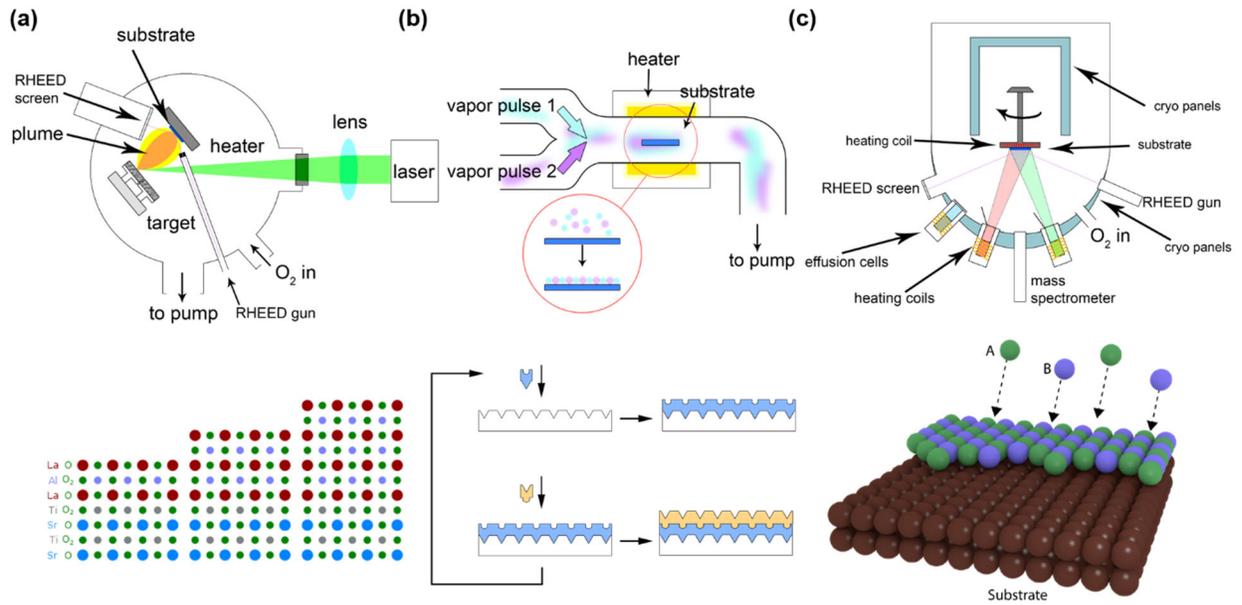

**Figure 3.1 Simplified schematics for common oxide thin film growth techniques.** (a) Pulsed laser deposition (PLD). (b) Atomic layer deposition (ALD). (c) Molecular beam epitaxy (MBE).

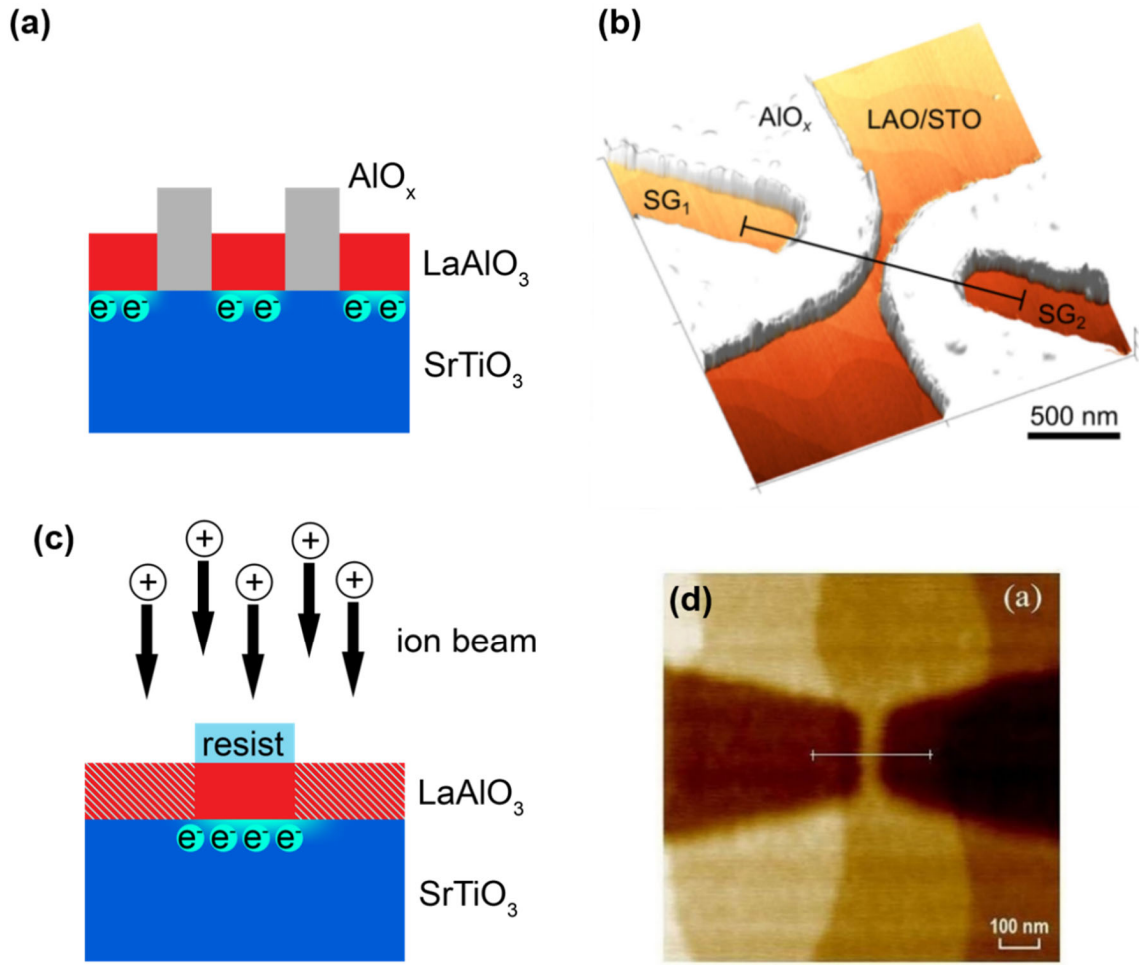

**Figure 3.2 Fabrication of devices at oxide interfaces.** (a) Amorphous $AlO_x$ is used as a hard mask to create insulating regions for photolithography. (b) AFM image of a sidegated junction using single-step photolithography [514]. Reprinted with permission from Monteiro et al., *Nano Lett.*, **2017**, 17 (2), pp 715–720. Copyright 2017 American Chemical Society. (c) Patterning with low energy (Ar) ion irradiation. The ion irradiation renders the exposed area insulating. (d) AFM image of the bridge structure created at $LaAlO_3/SrTiO_3$ using ion irradiation. Reprinted from Aurino et al., *Appl. Phys. Lett.* **102**, 201610 (2013), with the permission of AIP Publishing.

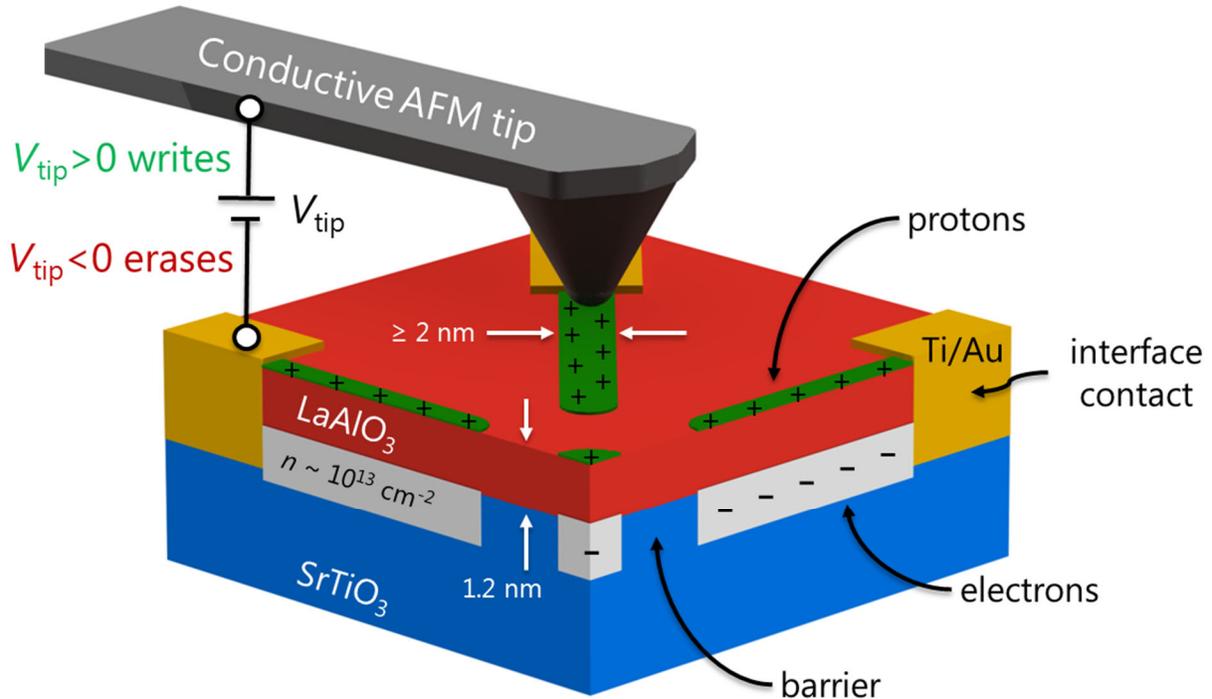

**Figure 3.3 Conductive-AFM lithography.** With positively biased AFM tip, conducting regions can be created because of the dissociation of the water molecules followed by surface protonation. The deposited protons act as local topgates and tune the region at the interface and underneath them conductive. The c-AFM lithography is reversible: written structure can be erased with negatively biased tip. The features as small as 2 nm have been achieved.

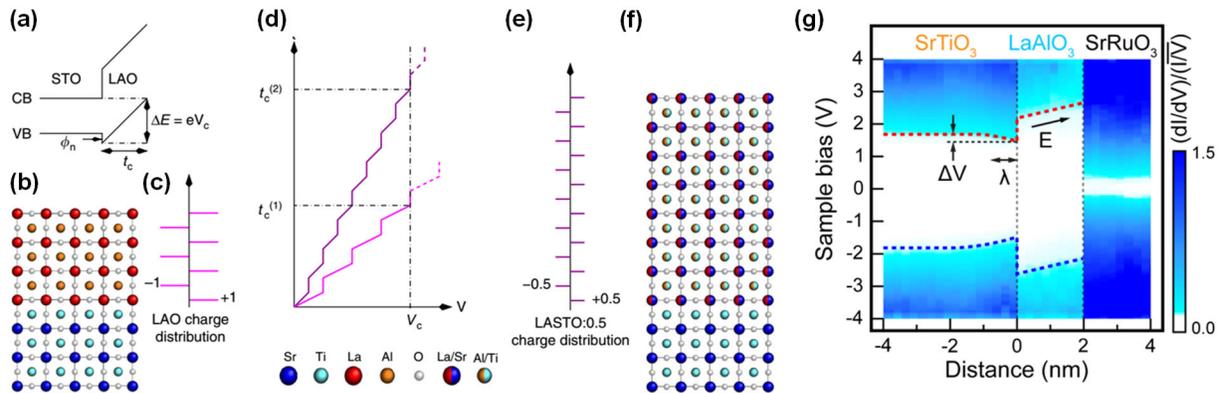

**Figure 4.1** The polar catastrophe is the result of the polar discontinuity caused by growing polar $LaAlO_3$ on non-polar $SrTiO_3$ (b,c). As the layers a deposited, an electric potential develops across the $LaAlO_3$ (d) which, after a critical thickness of ~4 u.c. $LaAlO_3$, raises the $LaAlO_3$ valence band at the surface above the $SrTiO_3$ conduction band at the interface (a), which was directly measured using scanning tunneling microscopy by Huang et al. on cross sectioned samples (g). This band shifting leads to an electronic reconstruction that transfers half an electron per unit cell to the interface. By substituting La with Sr in the overlayers (f), Reinle-Schmitt et al. [224] showed that the polarity (e) of the overlayers was changed and so was the potential (d), and reduce the density of charge carriers. ((a)-(f): Reprinted by permission from Macmillan Publishers Ltd: *Nature Communications* [224], copyright 2012) ((g): Reprinted figure with permission from Bo-Chao Huang et al., *Phys. Rev. Lett.* **109**, 246807 (2012). Copyright 2012 by the American Physical Society.)

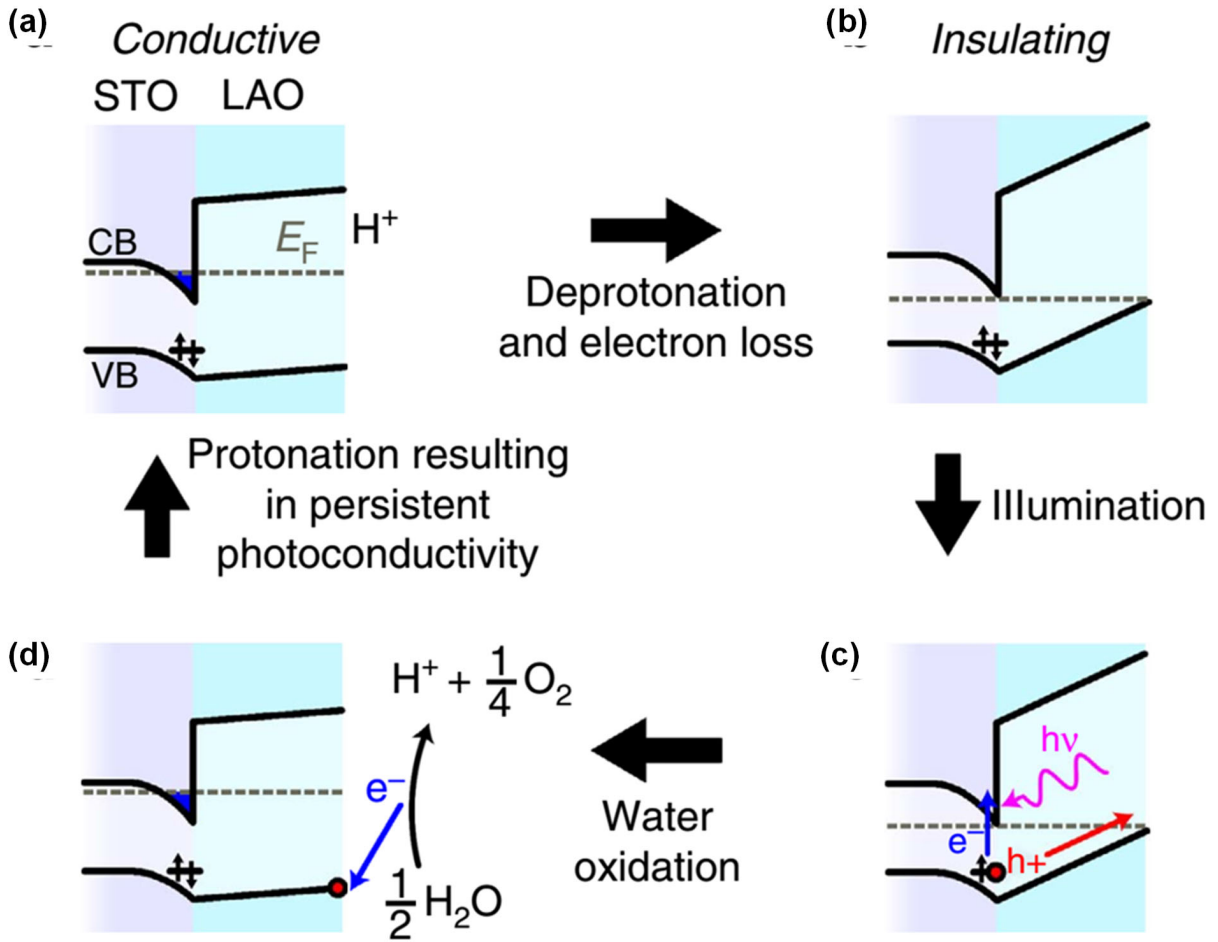

**Figure 4.2** Surface protonation has been found to contribute significantly to the conductivity of $LaAlO_3/SrTiO_3$ heterointerfaces, as illustrated in the conductivity hysteresis observed by Brown et al. In this process, a protonated surface (a) has an electron transferred to the surface by the disassociation of water. By deprotonating the surface (b) using solvents or oxygen plasma, the electron returns to the surface oxygen vacancy. When exposed to UV light (c), the oxygen vacancy then produces an electron-hole pair which are separated by the polar field of the $LaAlO_3$. The hole at the interface then oxidizes through water disassociation, again protonating the surface and stabilizing the conducting interface (a). (Reprint from [165] by permission from the authors.)

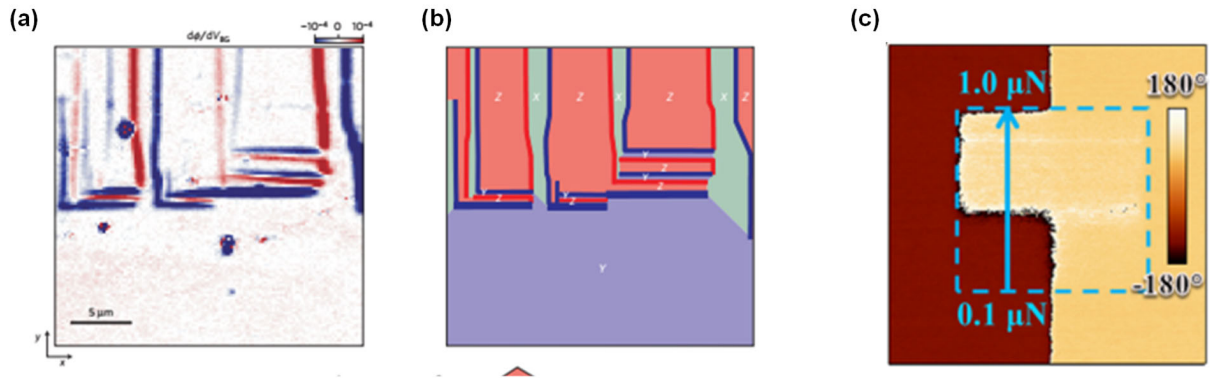

**Figure 4.3** (a) It is known that below 105K that SrTiO$_3$ undergoes a ferroelastic transition due to the rotation of the oxygen octahedra. This results in different domains determined by the direction of the rotation. In LaAlO$_3$/SrTiO$_3$ heterostructures, the effects of the transition are apparent in the transport properties at the interface. As imaged using a scanning SQUID by Honig et al., the current is enhanced at the domain walls (a), allowing the different domains to be identified by the angles of the walls between different domains (b). ((a) and (b): Reprinted by permission from Macmillan Publishers Ltd: *Nature Materials* [451], copyright 2013) Additionally, Sharma et al. used an AFM tip to mechanically drive a ferroelectric transition in the material, switching the piezoresponse through the migration of oxygen vacancies (c). (Reprinted with permission from *Nano Lett.*, **2015**, 15 (5), pp 3547–3551. Copyright 2015 American Chemical Society.)

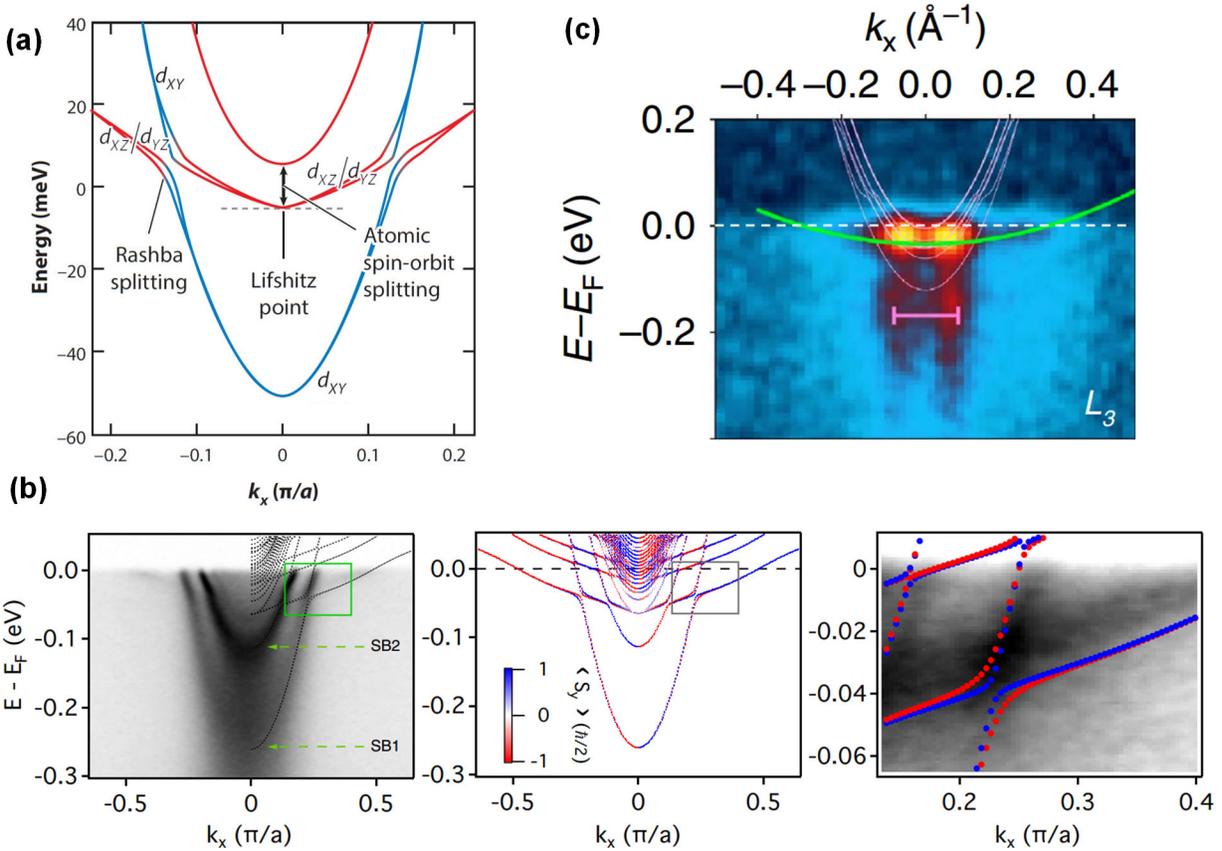

**Figure 4.4 The electronic band structure of LaAlO$_3$/SrTiO$_3$.** (a) The electronic band structure is dominated by the Ti $t_{2g}$ orbitals. In the (001) oriented heterostructures, the confinement potential at the interface causes the $d_{xy}$ band, which has a heavier band mass perpendicular to the interface, to be split from the $d_{xz}/d_{yz}$ bands which have heavy band masses in a direction parallel to the interface. As a result, the $d_{xy}$ bands have a highly symmetric Fermi surface, but the $d_{xz}/d_{yz}$ bands are highly asymmetric. Furthermore, SOC leads to hybridization of these bands after the Lifshitz transition, as shown by the tight-binding model of Kim et al. [476]. (Adapted and annotated from [42] by permissions from the authors.) (b) The band structure obtained by ARPES. (Reprinted figure with permission from S. McKeown Walker et al., *Phys. Rev. B* **93**, 245143 (2016). Copyright 2016 by the American Physical Society.) (c) The peak-dip-hump structure of polarons observed in APRES (Adapted from [376] under Creative Commons.)

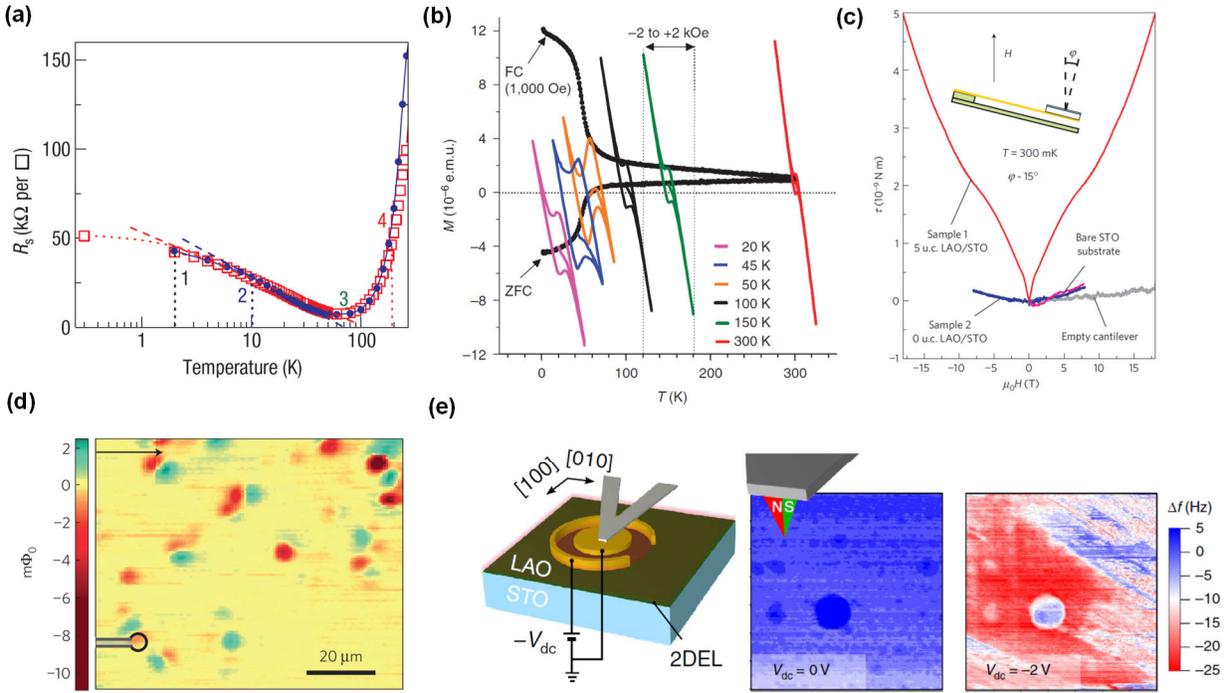

**Figure 4.5 Experimental evidence for ferromagnetism at the LaAlO$_3$/SrTiO$_3$ interface**. (a) Kondo minimum in the temperature dependence of the resistance (Reprinted by permission from Macmillan Publishers Ltd: *Nature Materials* [29], copyright 2007) (b) SQUID magnetometry: hysteresis loops taken at different temperature superimposed on top of the temperature dependence of magnetic moments (Reprinted by permission from Macmillan Publishers Ltd: *Nature Communications* [212], copyright 2011) (c) cantilever-based magnetometry: the magnetic moment induced a torque under external magnetic field, the magnetization of the sample is inferred accordingly (Reprinted by permission from Macmillan Publishers Ltd: *Nature Physics,* [390], copyright 2011) (d) Magnetic patches under scanning SQUID. With a micrometer-sized SQUID on a probe tip, microscopic magnetization can be imaged. Dipole-shaped patches are observed. (Reprinted by permission from Macmillan Publishers Ltd: *Nature Physics* [391], Copyright 2011) (e) Room temperature electrically-controlled ferromagnetism observed with magnetic force microscope (MFM). The magnetism signal is observed only when the interface is insulating. (Adapted from Ref. [213] by permission from the authors).

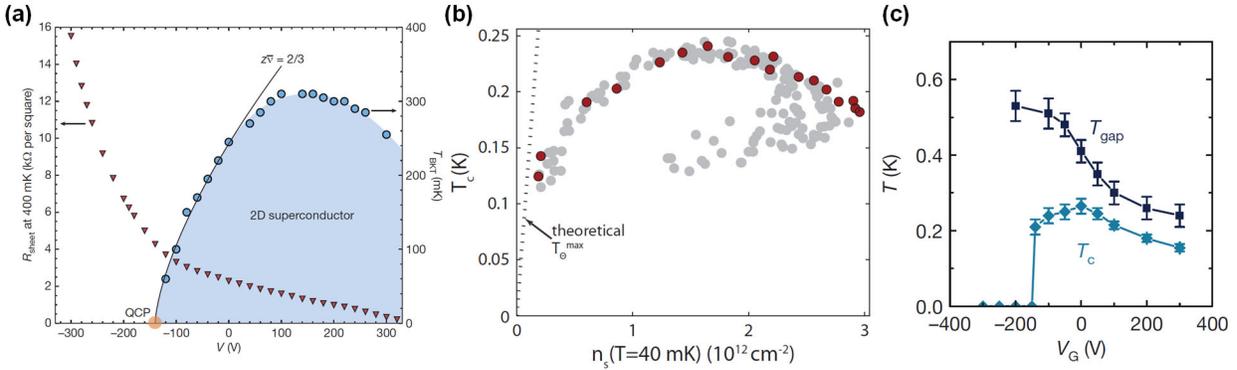

**Figure 4.6 Superconducting dome and pseudogap of $LaAlO_3/SrTiO_3$.** (a) The superconducting transition temperature $T_c$ as a function of the gate voltage, mapped out by Caviglia et al. [31]. (Reprinted by permission from Macmillan Publishers Ltd: *Nature* [31], copyright 2008.) (b) $T_c$ as a function of carrier density inferred from scanning SQUID, by Bert et al. [405]. (Reprinted figure with permission from Bert et al., *Phys. Rev. B* **86**, 060503(R) (2012). Copyright 2012 by the American Physical Society.) (c) The critical temperature for the pseudogap observed in tunneling spectroscopy as a function of gate voltage, by Richter et al. [182]. (Reprinted by permission from Macmillan Publishers Ltd: *Nature* [182], copyright 2013)

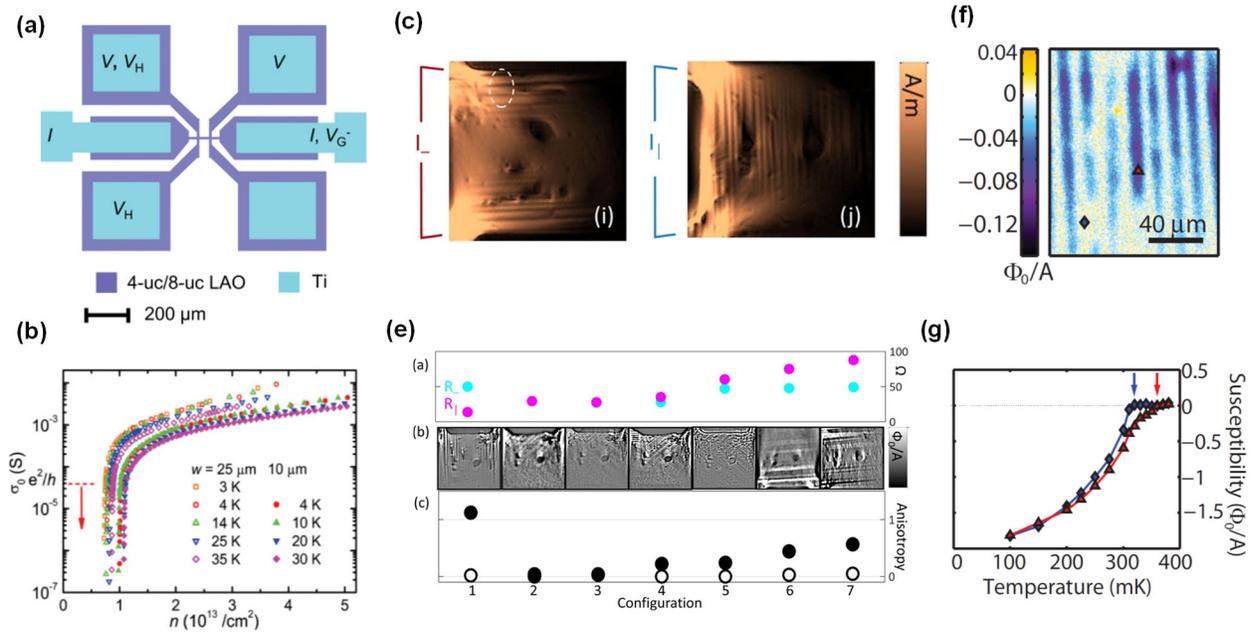

**Figure 5.1 Inhomogeneous transport.** (a) The device in Liao et al. [449] on the metal-insulator-transition controlled by gate. (b) Distinct "knee" of about $\sigma_0 \sim 2 \times 10^{-4}$ S is observed. ((a) and (b): Reprinted figure permission from Liao et al., *Phys. Rev. B* **83**, 075402 (2011). Copyright 2011 by the American Physical Society.) (c) Anisotropy due to tetragonal domain, seen under SQUID. The color scale is the current density. The image on the left is taken when the current is passed perpendicular to the domains while the right with current parallel to the domains. (d) The comparison of the four terminal resistance for two measurement configurations ($R_|$ and $R_\perp$) for different ferroelastic domain patterns. ((c) and (d): Reprinted with permission from *ACS Appl. Mater. Interfaces*, **2016**, 8 (19), pp 12514–12519. Copyright 2016 American Chemical Society.) (f) The superconducting transition temperature is modulated by the ferroelastic domains. The image is taken at $T = 350\ mK$. The color scale is the magnetic susceptibility. Some regions of the sample, for example, the region marked by the red triangle (red curve in (g), with $T_c \sim 360\ mK$), have already become superconducting, while other regions, such as the region marked by green diamond (blue curve in (g), with $T_c \sim 320\ mK$), have not. [452] (g) The temperature dependence of the magnetic susceptibility as a function of the temperature for the two regions (marked red triangle and blue diamond) in (f) [452]. ((e) and (f): Reprinted figure with permission from Noad et al., *Phys. Rev. B* **94**, 174516 (2016). Copyright 2016 by the American Physical Society.)

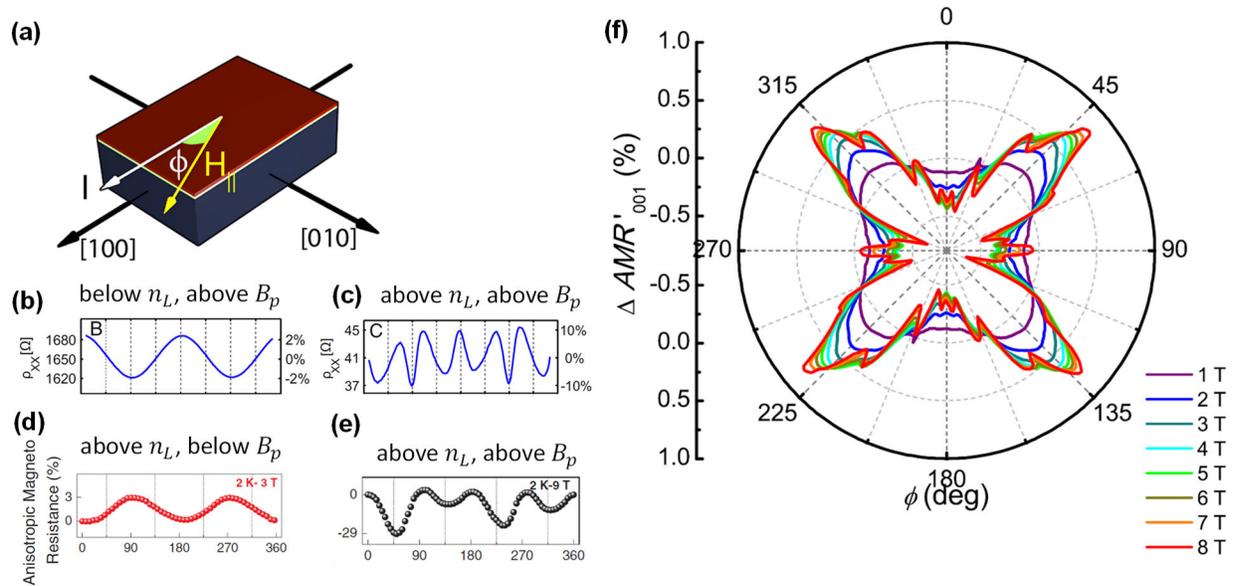

**Figure 5.2 Anisotropic magnetoresistance.** (a) The setup for the in-plane anisotropic magnetoresistance measurement. The anisotropy (excluding the 2-fold Lorentz contribution) only appears when the both the carrier concentration $n_e > n_L$ and $B > B_p$. The transition from regular sinusoidal to irregular or higher harmonics is revealed when the carrier concentration is increased from $n_e < n_L$ (b) to $n_e > n_L$ (c), or when from $B < B_p$ (d) to and $B > B_p$ (e). (f) The complex anisotropy with rich structure reported by Miao et al. [466]. ((b) and (c): Adapted from [384]. Copyright 2013 National Academy of Sciences. (d) and (e): Reprinted figure with permission from Annadi et al., *Physical Review B*, 87, 201102(R) (2013). Copyright 2013 by the American Physical Society. (f): Reprinted from *Applied Physics Letters,* **109**, 26 (2016), with the permission of AIP Publishing.)

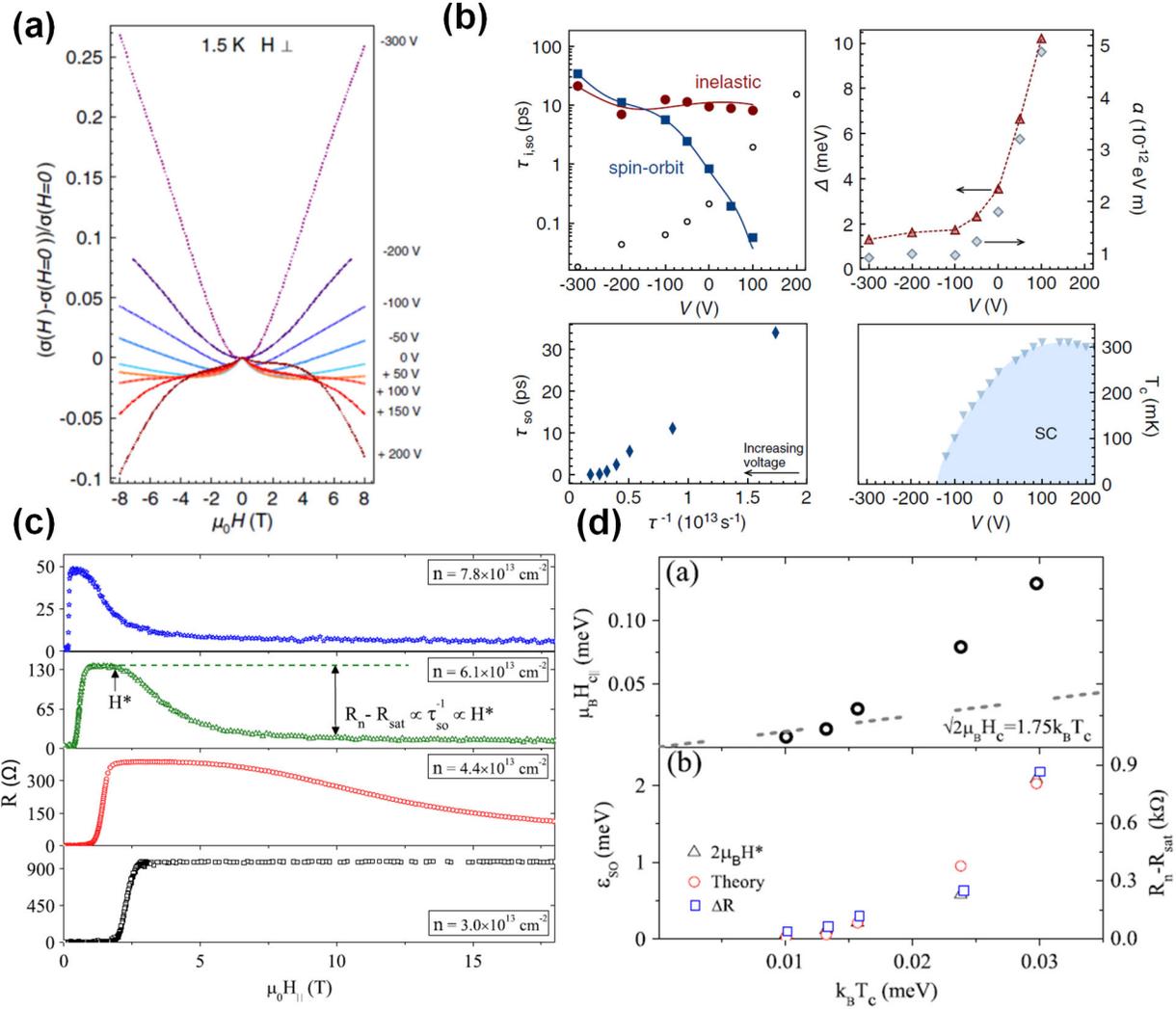

**Figure 5.3 Gate-tunable spin-orbit interaction.** (a) The evolution of the magnetoconductance as a function of gate voltage. The crossover from WL to WAL is observed. [470]. (b) Caviglia et al. [470] then fitted the data to the MF theory, obtaining the spin-orbit strength $\Delta \cong 1 - 10\ meV$. *Left Panels*: various dephasing time from the fit. *Right panels*: The $\Delta$ increases sharply near the gate voltage corresponding to the superconducting transition at lower temperature. ((a) and (b): Reprinted figure with permission from Caviglia et al., *Phys. Rev. Lett.* **104**, 126803 (2010). Copyright 2010 by the American Physical Society.) (c) The evolution of the resistance as a function of in-plane magnetic field at different gate voltage. (d) *Upper*: the evolution of in-plane the upper critical field as the gate voltage is varied. (Plotted against $k_B T_c$). *Lower*, the spin-orbit strength obtained from the violation of the Pauli limit, using $g\mu_B H^* = \varepsilon_{SO}$. ((c) and (d): Reprinted figure with permission from Ben Shalom et al., *Phys. Rev. Lett.* **104**, 126803 (2010). Copyright 2010 by the American Physical Society.)

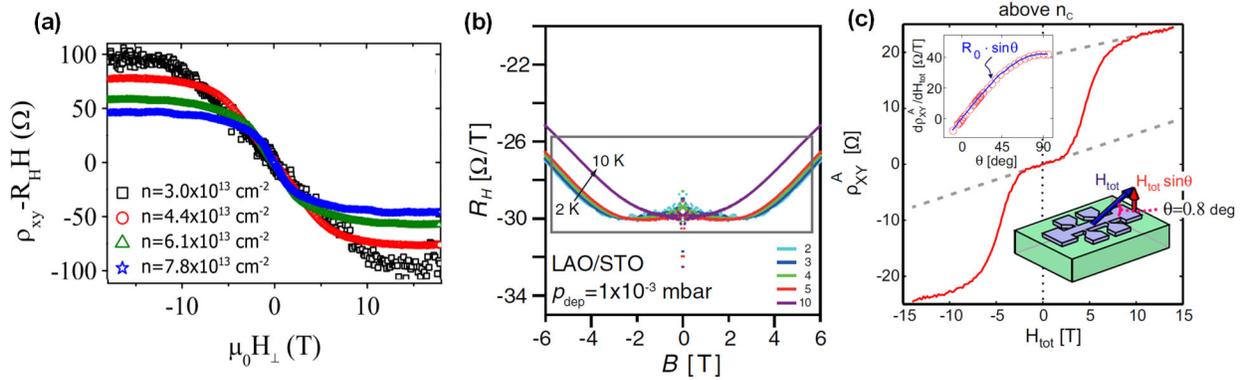

**Figure 5.4 Nonlinear and anomalous Hall effect.** (a) Nonlinearity in the Hall measurement. $R_H$ (the slope of the Hall resistance $R_{xy}$). (Reprinted figure with permission from Ben Shalom et al., *Phys. Rev. Lett.* **104**, 126803 (2010). Copyright 2010 by the American Physical Society.) (b) Hall coefficient $R_H$ as a function of the magnetic field. The small upturn for B ~ -3T to B ~ 3T deviates from the parabolic-like dependence expected for 2-band model. Additional $R_0^{AHE}$ was introduced to fit this deviation [416]. (Adapted from Ref. [416] under Creative Commons 3.0) (c) The evolution of $\rho_{xy}$ as a function of magnetic field $H_{tot}$, 0.8° to the direction of transport. (Adapted from Ref. [384], Copyright 2013 National Academy of Sciences.)

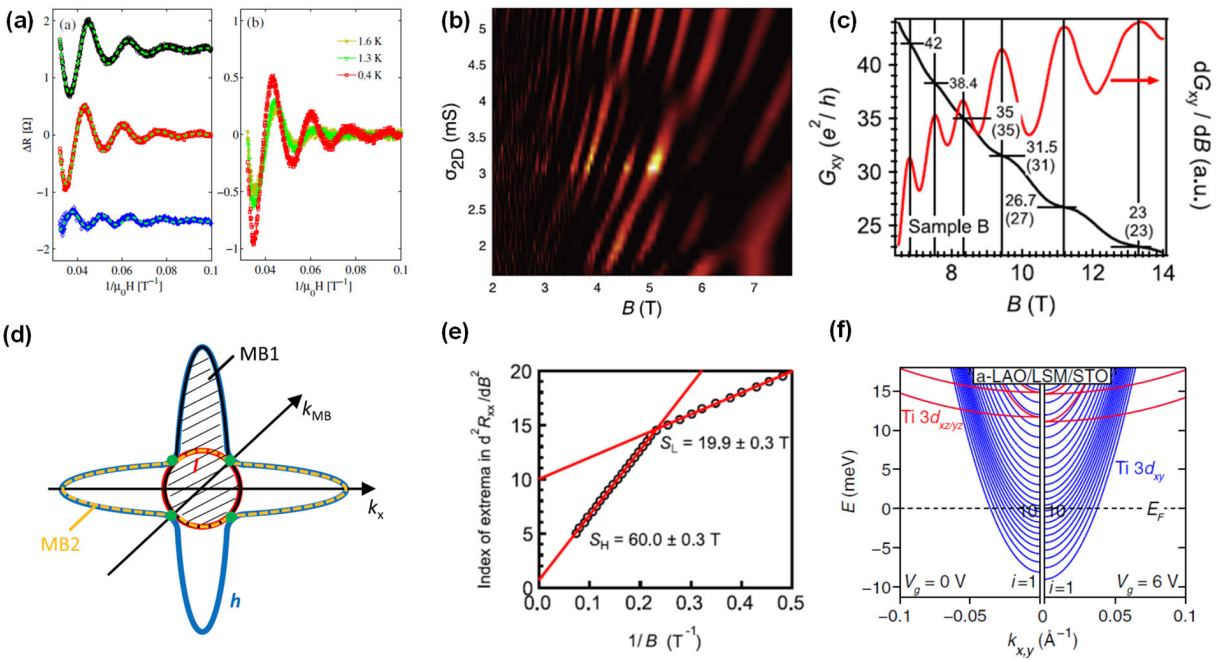

**Figure 5.5 Shubnikov-de Haas (SdH) oscillations.** (a) SdH reported by Ben Shalom et al. [7]. (Reprinted figure with permission from Ben Shalom et al., *Phys. Rev. Lett.* **105**, 206401 (2010). Copyright 2010 by the American Physical Society.) (b) The evolution of the SdH as a function of gate voltage. Both peak-splitting and phase jumps are visible. [10]. (Adapted from Ref. [10] under Creative Commons 3.0) (c) The QHE-like plateaus with associated SdH oscillations reported by Xie et al. [9]. The plateaus have $\Delta v \sim 4$. (d) To explain the $\Delta v \sim 4$, Xie el al. [9] suggested a 4-fold magnetic broken (MB) down orbits, shown as the shaded area. The inner Fermi surface dominates the transport at lower magnetic field ($B < 3.3T$) while the MB orbits dominate at higher magnetic field. ($B > 3.3T$). (e) The SdH frequence changes from $F = 20T$ to $F = 60T$ at about $B = 3.3T$. Thus, the area for the MB orbits have to be 3 times of that of the inner Fermi surface [9]. ((c), (d) and (e): Reprinted from *Solid State Communications*, Volume 197, November 2014, Xie et al., *Quantum longitudinal and Hall transport at the LaAlO3/SrTiO3 interface at low electron densities, Pages 25–29*, Copyright 2014, with permission from Elsevier) (f) The 10-band model proposed by Trier et al. [492] to explain the observed $\Delta v \sim 10, 20$. (Reprinted figure with permission from Trier et al., *Phys. Rev. Lett.* **117**, 096804 (2016). Copyright 2016 by the American Physical Society.)

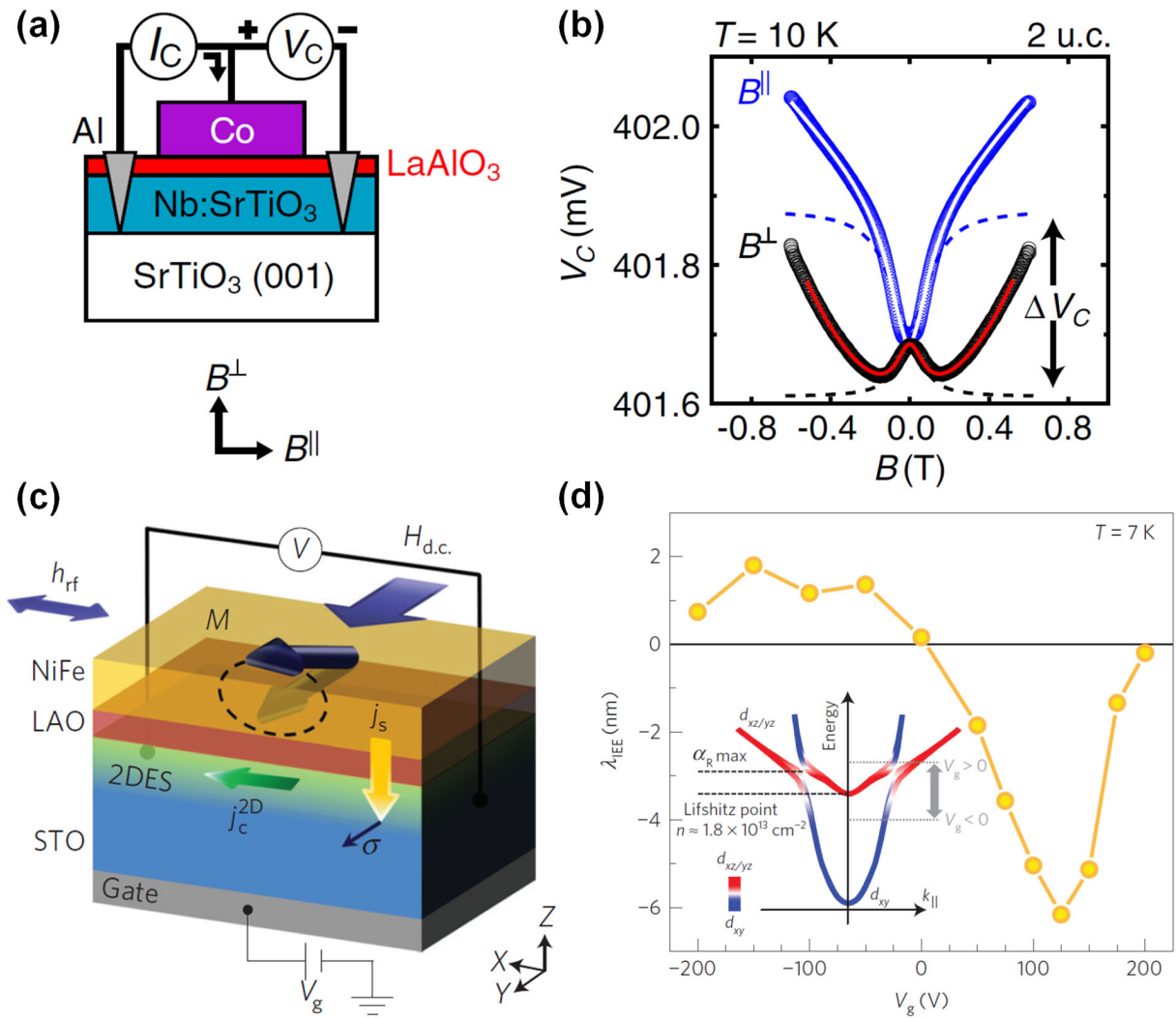

**Figure 5.6 Spintronics effects.** (a) The device schematics for 3-terminal tunneling magnetoresistance measurement [503]. (b) The Lorentzian lineshape of the tunneling magnetoresistance measurement [503]. (Adapted from Ref. [503] under Creative Commons 3.0) (c) The device schematics of Lesne et al. [478]. The spin imbalance at the 2DES is created by spin-pumping technique. This spin imbalance is then

converted to charge by the Rashba effect at the interface (the inverse Edelstein effect (IEE)). (D) The evolution of the spin to charge conversion efficiency $\lambda_{IEE}$ as a function of gate voltage $V_g$. $\lambda_{IEE}$ changes sign at about $V_g = 0$. Lesne et al. [478] suggested that the sign and the magnitude of $\alpha_R$ may be different for $d_{xy}$ and $d_{xz}/d_{yz}$ [478]. (Reprinted by permission from Macmillan Publishers Ltd: *Nature Materials* [478], copyright 2016)

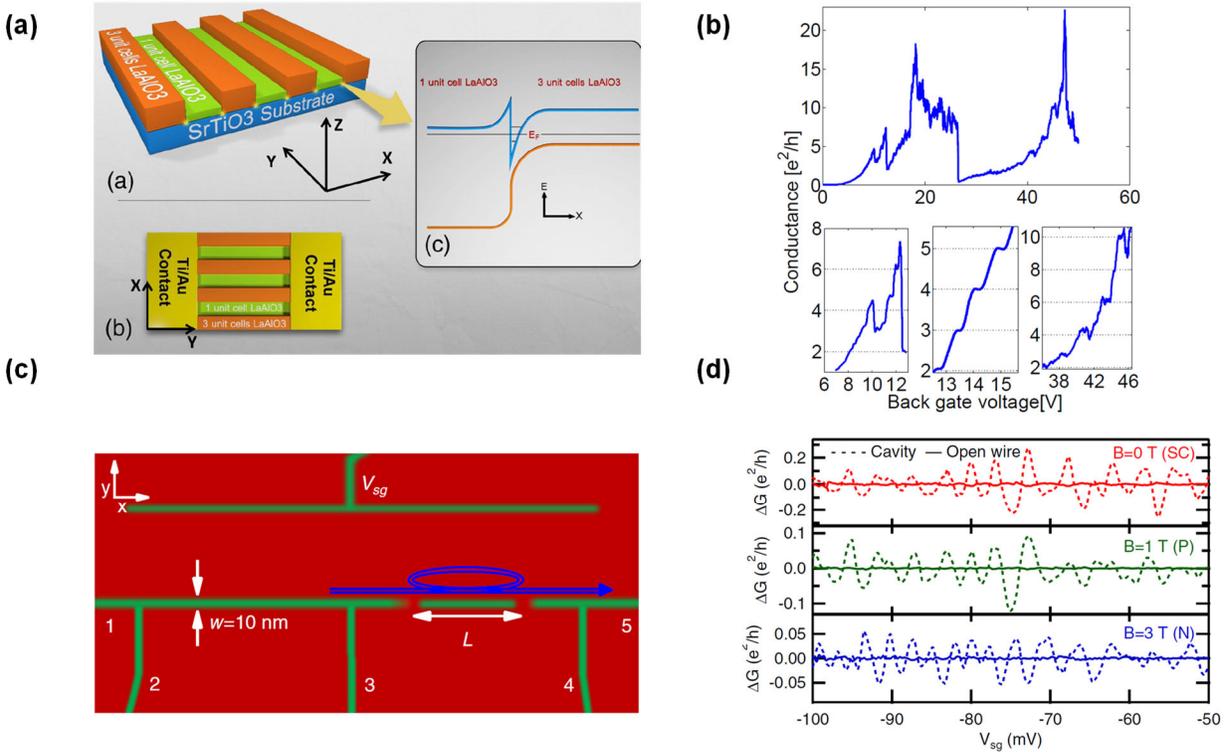

**Figure 6.1 Micron-long ballistic 1D channels.** (a) Ron et al. [32] created parallel quasi-1D channels by covering the SrTiO₃ with alternating 1 u.c. and 3 u.c. of LaAlO₃ overlayers. The transport is proposed to be supported by the edges between the regions covered by 1 u.c. and that covered by 3 u.c.. (b) The resulting conductance as a function of the gate voltage for a device with 28 parallel 4 μm channels. Top panel: the full range. Bottom: the close-ups. (Reprinted figure with permission from Ron et al., *Phys. Rev. Lett.* **112**, 136801 (2014). Copyright 2014 by the American Physical Society.) (c) Tomczyk et al. [33] created ballistic quasi-1D channel up to 4 μm long using c-AFM lithography. (d) The Fabry–Pérot-like pattern for the c-AFM defined quasi-1D channel. (Adapted from Ref. [33] by permission from the authors.)

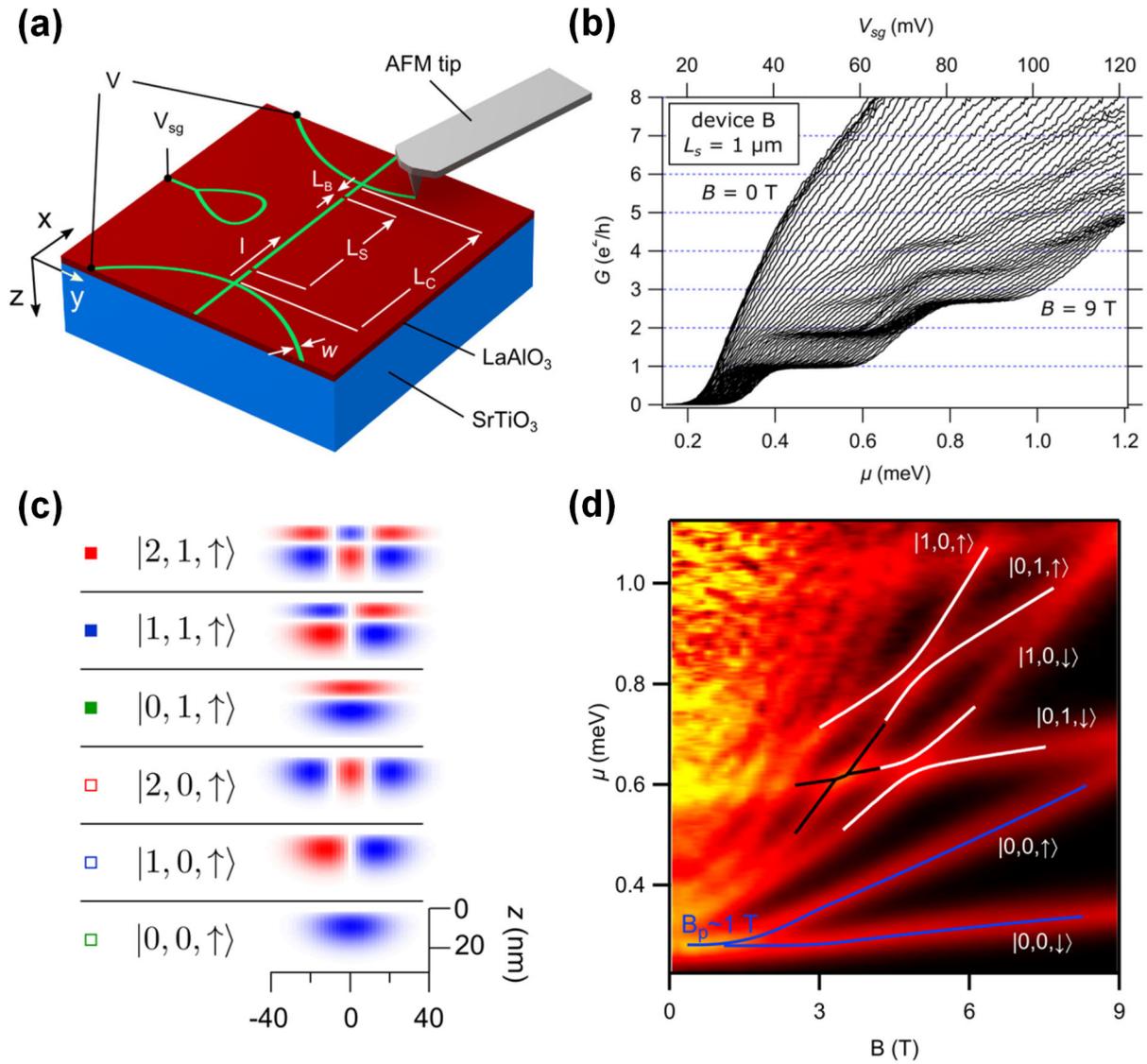

**Figure 6.2 Dissipationless quantum wire.** (a) The device design of the quantum wire. (b) The quantized conductance as a function of the chemical potential µ. (c) The spatial modes and spin state of the quantum wire. (d) The evolution of the modes as a function of the spatial mode and spin state. (Adapted from Ref. [312] by permission from the authors.)

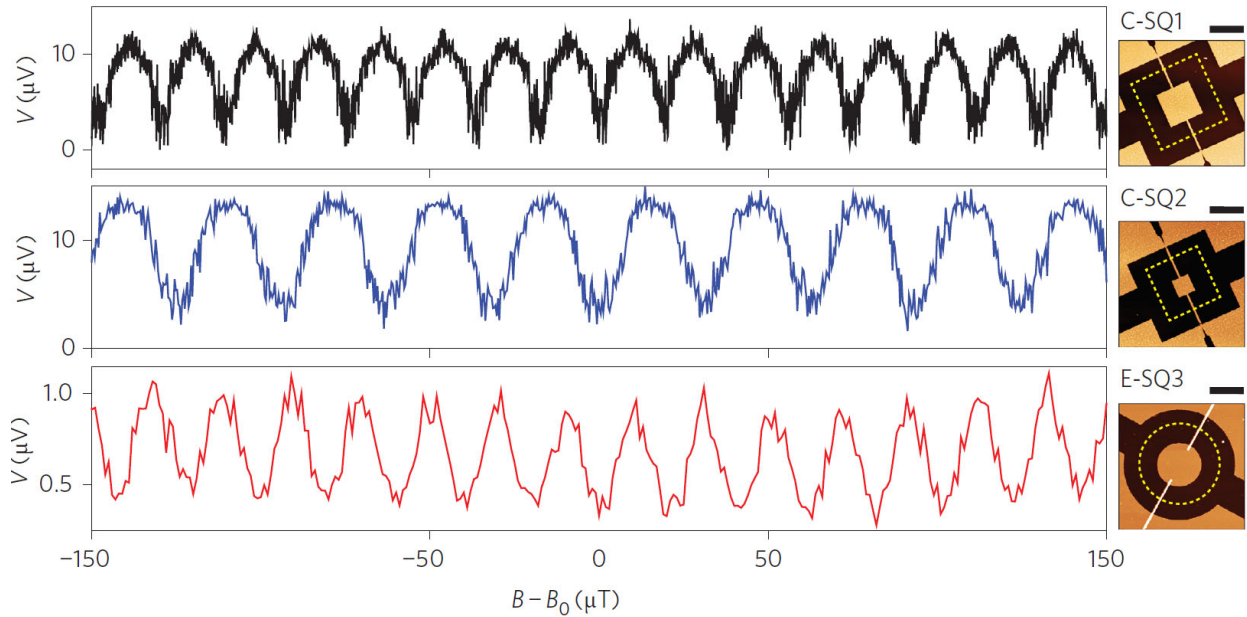

**Figure 6.3 SQUID devices of LaAlO$_3$/SrTiO$_3$.** The oscillation as a function of the magnetic field. C-SQ1 and C-SQ2: devices with constrictions defined by lithography. E-SQ3: device with junctions controlled by electrostatic gating. (Reprinted by permission from Macmillan Publishers Ltd: *Nature Nanotechnology* [508], copyright 2016)

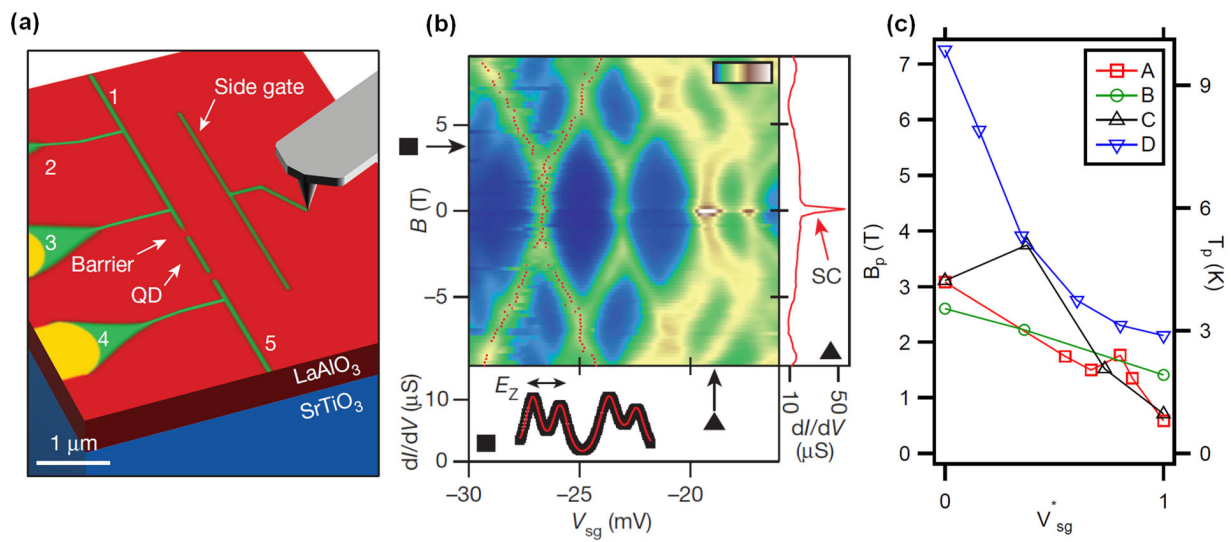

**Figure 6.4 Electronic pairing without superconductivity.** (a) The device design of the SET device. (b) The differential conductance ($dI/dV$) as a function of the sidegate voltage $V_{sg}$. The features correspond to the resonant tunneling of the single ($B > 2T$) or paired electrons ($B < 2T$) through the QD. (c) The dependence of the pairing strength $B_p$ as a function of the normalized sidegate voltage $V_{sg}^*$. $B_p$ decreases

monotonically with increasing $V_{sg}^*$ for all the devices (Four of the devices, A, B, C, and D, are shown). (Adapted and annotated from Ref. [187] by permission from the authors).

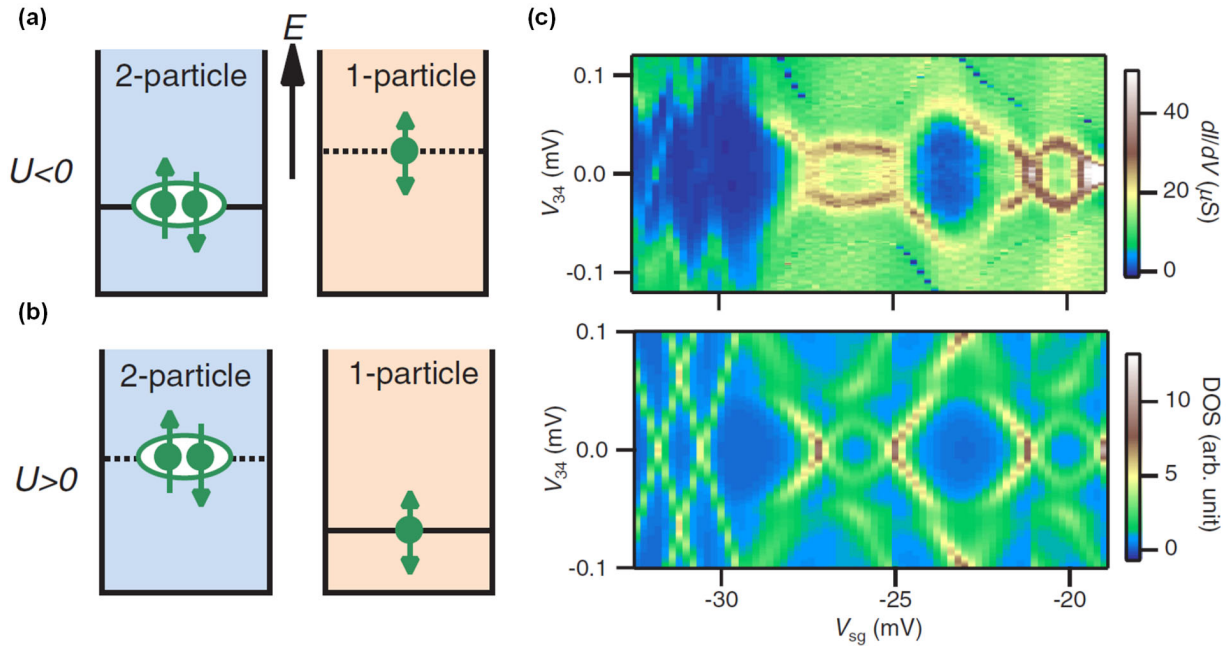

**Figure 6.5 Tunable electron-electron interaction**. (a) and (b), the two-electron ground state and one-electron ground state. (a) When $U < 0$, the energy level for the two-electron ground state is lower than that of one-electron ground state. (b) When $U > 0$, the energy level for the two electron ground state is higher than that of single electron ground state. The transport is governed by Andreev reflections. (c) The differential conductance ($dI/dV$) as a function of the sidegate voltage $V_{sg}$ and the 4-terminal voltage $V_{34}$. In the lower sidegate regime ($V_{sg} < -29\ mV$), the "×" features from pair-tunneling dominate the transport. For higher sidegate voltage ($V_{sg} > -29\ mV$), the loop-like features, attributed to Andreev bound states, dominate the transport. (Adapted from Ref. [517] by permission from the authors.)